\documentclass[prb,aps,nofootinbib,amssymb,twocolumn,superscriptaddress,10pt]{revtex4-2}
\usepackage{project_macros-public}
\usepackage{project_macros-private}
\usepackage{makecell}

\crefname{appendix}{Appendix}{Appendices}
\crefname{equation}{Eq.}{Eqs.}
\crefname{figure}{Fig.}{Figs.}
\crefname{table}{Table}{Tables}
\crefname{section}{Section}{Sections}
\creflabelformat{appendix}{[#2#1#3]}

\makeatletter
\AtBeginDocument{\let\LS@rot\@undefined}
\makeatother

\makeatletter
\renewcommand\onecolumngrid{
\do@columngrid{one}{\@ne}%
\def\set@footnotewidth{\onecolumngrid}
\def\footnoterule{\kern-6pt\hrule width 1.5in\kern6pt}%
}

\newcommand{\ChargeSpinZeroEstimateOneBPV}{4.1}
\newcommand{\ChargeSpinOneEstimateOneBPV}{9.9}

\newcommand{\NeutralSpinZeroEstimateTwoBPV}{2.4}
\newcommand{\NeutralSpinOneEstimateTwoBPV}{1.8}
\newcommand{\ChargeSpinZeroEstimateTwoBPV}{3.2}
\newcommand{\ChargeSpinOneEstimateTwoBPV}{4.0}

\newcommand{\NeutralSpinOneSpinZeroRatioOneBPV}{2.9}
\newcommand{\NeutralSpinOneSpinZeroRatioFlatbandOneBPV}{3.6}

\global\long\def\bsl#1{\boldsymbol{{\bf #1}}}%

\global\long\def\ket#1{\left| #1\right\rangle }%

\global\long\def\up{\uparrow}%

\global\long\def\down{\downarrow}%

\allowdisplaybreaks

\begin{document}
\title{\titlePaper}
\paperAuthors

\author{Miguel Gonçalves}
\affiliation{Princeton Center for Theoretical Science, Princeton University, Princeton NJ 08544, USA}

\author{Juan Felipe Mendez-Valderrama}
\affiliation{Department of Physics, Princeton University, Princeton, New Jersey 08544, USA}

\author{Jonah Herzog-Arbeitman}
\affiliation{Department of Physics, Princeton University, Princeton, New Jersey 08544, USA}
\author{Jiabin Yu}
\affiliation{Department of Physics, University of Florida, Gainesville, FL, USA}

\author{Xiaodong Xu}
\affiliation{Department of Physics, University of Washington, Seattle, Washington 98195, USA}
\affiliation{Department of Materials Science and Engineering, University of Washington, Seattle, Washington 98195, USA}

\author{Di Xiao}
\affiliation{Department of Materials Science and Engineering, University of Washington, Seattle, Washington 98195, USA}
\affiliation{Department of Physics, University of Washington, Seattle, Washington 98195, USA}
\affiliation{Pacific Northwest National Laboratory, Richland, WA, USA}

\author{B. Andrei Bernevig}
\affiliation{Department of Physics, Princeton University, Princeton, New Jersey 08544, USA}
\affiliation{Donostia International Physics Center, P. Manuel de Lardizabal 4, 20018 Donostia-San Sebastian, Spain}
\affiliation{IKERBASQUE, Basque Foundation for Science, Bilbao, Spain}

\author{Nicolas Regnault}
\affiliation{Center for Computational Quantum Physics, Flatiron Institute, 162 5th Avenue, New York, NY 10010, USA}
\affiliation{Department of Physics, Princeton University, Princeton, New Jersey 08544, USA}
\affiliation{Laboratoire de Physique de l'Ecole normale sup\'{e}rieure, ENS, Universit\'{e} PSL, CNRS, Sorbonne Universit\'{e}, Universit\'{e} Paris-Diderot, Sorbonne Paris Cit\'{e}, 75005 Paris, France}

\let\oldaddcontentsline\addcontentsline

\begin{abstract}
Fractionally charged elementary excitations, the quasi-electron and quasi-hole, are one of the hallmarks of the fractional Chern insulator (FCI). In this work, we observe that spontaneous spin polarization in twisted MoTe$_2$ leads to \emph{multiple species} of low-energy quasi-particles distinguished by their spin quantum numbers. We perform large-scale exact diagonalization (ED) calculations to investigate the nature of these excitations and develop a method to extract their fundamental energetic properties. Focusing on $\theta = 3.7^{\degree}$ and filling factor $\nu = -2/3$ relevant to recent experiments, we show that spin-preserving (spinless) charge excitations have smaller gap than spin-flipping (spinful) excitations both with and without band mixing.
This result is in qualitative agreement with the measured magnetic field dependence of the transport gaps. Beyond the spinless and spinful quasi-particle gaps, we extract the full quasi-electron and quasi-hole ``band structure'' and find significant dispersion with emergent magnetic translation symmetry --- a fundamental departure from the immobile excitations of the quantum Hall fluid. Our results establish a framework for computing the properties of novel elementary excitations in FCIs. 

\end{abstract}
\maketitle

\paragraph*{Introduction.---}
Fractional Chern insulators (FCI) \cite{PhysRevLett.106.236804,Sheng2011,PhysRevX.1.021014,tang2011,PhysRevLett.106.236803} are zero-magnetic field analogues of the fractional quantum Hall (FQH) effect that share similar topologically ordered ground states and arise purely due to electron-electron interactions at fractional filling fractions of narrow Chern bands. Recent optical, transport, and orbital magnetization measurements have enabled the observation of FCIs in twisted homobilayer MoTe$_{2}$ at filling fractions $\nu=-2/3,-3/5, -4/7, -5/9$ \cite{Park2023,Zeng2023,Cai2023,PhysRevX.13.031037,Ji2024,Redekop2024,park2025observationhightemperaturedissipationlessfractional,xu2025signaturesunconventionalsuperconductivitynear}. FCIs were also observed in rhombohedral graphene on aligned hBN substrate \cite{Lu2024,Xie2025,Lu2025,aronson2024displacementfieldcontrolledfractionalchern}, demonstrating that these phases can emerge beyond moiré MoTe$_{2}$.
Like their FQH counterparts, FCIs are characterized by fractionalized excitations, fractionally quantized Hall conductivity and topological ground-state degeneracy. Although FCIs echo many features of FQH physics, they also exhibit key differences.

Most notably, FCIs break time reversal symmetry spontaneously by developing a finite spin/valley polarization that then stabilizes topological order at fractional fillings. This spontaneous symmetry breaking provides an enriched set of elementary excitations that can feature distinct interactions and quantum numbers from their FQH analogues \cite{PhysRevResearch.3.L032070,PhysRevLett.131.136501,PhysRevLett.131.136502,PhysRevB.108.245159,PhysRevB.108.085117,PhysRevB.107.L201109,MFCI_0,FTI_Yves,PhysRevLett.132.036501,PhysRevB.109.L121107,shen2024stabilizingfractionalcherninsulators,zaklama2025structurefactortopologicalbound,doi:10.1073/pnas.2316749121,PhysRevLett.133.066601,PhysRevLett.133.186602,PhysRevB.66.075408,PhysRevB.70.241307,PhysRevLett.132.236502,chen2025fractionalcherninsulatorquantum,Jain_2007,PhysRevLett.54.237,Yoshioka84,PhysRevB.33.2903,PhysRevB.55.7702,KPark_1999,PhysRevB.66.155320,PhysRevB.66.075408,PhysRevB.70.241307,PhysRevLett.55.1606,PhysRevLett.65.2916,PhysRevB.37.8476,PhysRevLett.124.156801,PhysRevLett.70.2944,PhysRevLett.127.056801,PhysRevLett.127.056801} where time-reversal is broken explicitly by the magnetic field. Understanding these excitations provides insight into the stability of these interacting topological phases as well as their electromagnetic responses.

\begin{figure}[h!]
     \centering
     \includegraphics[width=\linewidth]{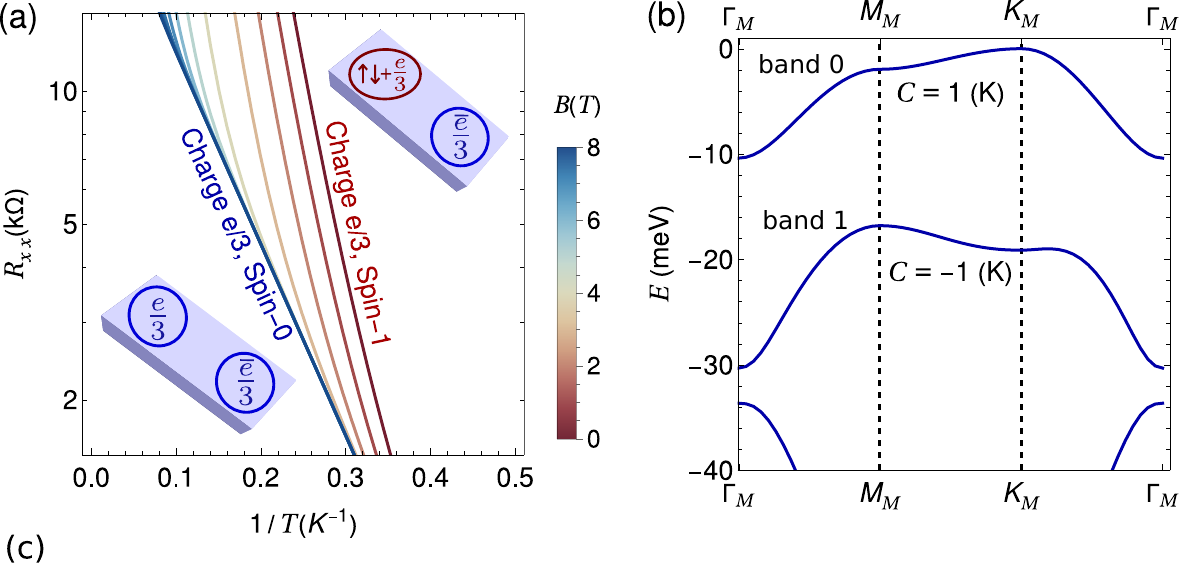}

    \resizebox{0.85\columnwidth}{!}{
    \begin{tabular}{|c|c|c||c|c||c|c|}
    \hline 
    Excitation type & $Q$ & $s$ & \makecell{$\Delta_{\textrm{1BPV}}$ \\ (meV)}   & \makecell{Method \\ (1BPV)} & \makecell{$\Delta_{\textrm{2BPV}}$ \\ (meV)}   & \makecell{Method \\ (2BPV)}  \tabularnewline
    \hline 
    \hline 
    Magnetoroton & 0 & 0 & 2.4 & $N_{s}=39$ & \textcolor{black}{\NeutralSpinZeroEstimateTwoBPV}  & $N_s=21$ \tabularnewline
    \hline 
    Magnon & 0 & 1 & 7.0 & $N_{s}=33$ & \NeutralSpinOneEstimateTwoBPV & All $N_s$ \tabularnewline
    \hline 
    Charge-$e/3$ spin-0 & $e/3$ & 0 & \ChargeSpinZeroEstimateOneBPV & $N_{s}=39$ & \ChargeSpinZeroEstimateTwoBPV & $N_s=21$ \tabularnewline
    \hline 
    Charge-$e/3$ spin-1 & $e/3$ & 1 & \ChargeSpinOneEstimateOneBPV & $N_{s}=30$ & \ChargeSpinOneEstimateTwoBPV & $N_s=21$ \tabularnewline
    \hline 
    \end{tabular}
    }

     \caption{Summary of results. (a) Sketch of the experimental results in Ref.$\,$\cite{park2025observationhightemperaturedissipationlessfractional} for $R_{xx}$ as a function of $1/T$, at different magnetic fields $B$ and $\nu=-2/3$. $R_{xx}$ is thermally activated, satisfying  $R_{xx} = R_{xx}^0 \exp(-\Delta_c/(2 k_B T))$, where $R_{xx}^0$ is a resistance pre-factor and $\Delta_c$ is the transport gap. At $B=0$, $R_{xx}^0$ and $\Delta_c$ are larger than those at large $B$.
    We identify the large-$B$ transport gap as the energy required to create a charge-$e/3$ quasi-electron well separated from a charge-$\bar{e}/3$ quasi-hole (illustrated in the figure, in blue). At $B=0$, we identify the transport gap as the energy required to create a charge-$\bar{e}/3$ quasi-hole well separated from a charge-$e/3$ quasi-electron dressed by a magnon ("$\up \down + \frac{e}{3}$" in the figure text).  (b) Single-particle band structure of moiré MoTe$_{2}$ at $\theta=3.7^{\degree}$, for the parameters in Ref.$\,$\cite{MFCI_I}, for a path along the high symmetry lines in the moiré Brillouin zone. The Chern numbers of the topmost valence bands at $\bf{K}$ are shown (they reverse sign at $\bf{K}'$). 
     (c) Summary of best estimates for the energy gaps for $\epsilon=10/0.9$. For the 1BPV calculation, these correspond to the computed gaps for the largest considered system size (specified in the 'Method (1BPV)' column). For the 2BPV calculations we obtain the estimates either from the largest considered sizes and number of holes in band 1 (specified in the 'Method (2BPV)' column), or from a collapse of all the gap data for different system sizes  as detailed in section "\textit{2BPV Results}" (specified by 'All $N_s$' in the 'Method (2BPV)' column). Note that in this table we only show results for the smallest spin-1 gap, $\Delta_{e/3}^{-1,-}$. }
     \label{fig:1}
\end{figure}

Elementary excitations in FCIs can be probed in transport experiments by measuring the quasi-particle charge gap, $\Delta_c$, which corresponds to the energy required to create a well-separated (unbound) quasi-electron/quasi-hole pair. This gap manifests itself in the exponentially activated temperature dependence of the longitudinal resistance, $R_{xx}\propto \exp(-\Delta_c/2k_B T)$, where $T$ is the temperature and $k_B$ is the Boltzmann constant. In the case of twisted bilayer MoTe$_{2}$, a sizable transport gap of $\Delta_c=2.0 \pm 0.6$meV was observed in Ref.$\,$\cite{Park2023} for the most robust FCI at $\nu=-2/3$, at twist angle $\theta=3.7^{\degree}$ (consistent with predictions from magnetization measurements \cite{Redekop2024} at a higher twist angle $\theta\approx3.89^{\degree}$). 
More recently, Ref.$\,$\cite{park2025observationhightemperaturedissipationlessfractional} 
reported an intriguing behavior for the transport gaps at $\nu=-2/3$, $\theta=3.7^{\degree}$ and variable magnetic field $B$. As sketched in Fig.$\,$\ref{fig:1}(a), a thermal activation with significantly larger longitudinal resistance was observed at $B=0$ compared to larger $B$ (up to $B=8T$), with the largest activation gap of around $4.7$meV being observed at $B=0T$. This gap is substantially larger than the gap reported in Ref.$\,$\cite{Park2023} also at  $B=0T$, likely due to the improved sample quality: disorder is well know to be responsible for decreasing transport gaps compared to their expected clean-limit value in the FQH literature \cite{PhysRevLett.55.1606,PhysRevLett.65.2916,PhysRevB.37.8476,PhysRevLett.70.2944,PhysRevLett.124.156801,PhysRevLett.127.056801}.  As $B$ is increased, $R_{xx}$ decreases and the temperature dependence of $R_{xx}$ saturates at $4T \lesssim  B <  8T$, being associated with a smaller activation gap $\Delta_c\approx 1.7$meV. 
A possible explanation introduced in Ref.$\,$\cite{park2025observationhightemperaturedissipationlessfractional} is that at and near $B=0$, the resistance, $R_{xx}$, is primarily dominated by the thermally activated behavior of electronic quasi-particles with a spin-flip relative to the fully polarized FCI ground-state. \footnote{Because the spin-flip quasi-particles have a larger charge gap than those with the same spin as the FCI ground state, they can dominate transport at $B = 0$ only if their resistance pre-factor is significantly larger, as proposed in  Ref.~\cite{park2025observationhightemperaturedissipationlessfractional}. 
This effect, although crucial for describing the experiment, is beyond the scope of the present work.}  At larger $B$, these excitations are suppressed due to the Zeeman effect and instead quasi-particles with the same spin as the FCI ground state provide the most significant contribution to $R_{xx}$.

Early theoretical studies of spinful FQH quasi-particle excitations --- those involving spin flips relative to a polarized ground-state --- have shown that they can be competitive with polarized quasi-particle excitations 
\cite{PhysRevB.33.2221,PhysRevB.36.5454,PhysRevB.41.10862,PhysRevLett.57.130,PhysRevB.64.125310,PhysRevB.80.045407,PhysRevLett.105.176802} and are likely relevant to understand experiments with tilted magnetic fields \cite{PhysRevLett.62.1540,PhysRevLett.62.1536,PhysRevB.41.7910,PhysRevB.45.3418,PhysRevLett.121.226801}. Nevertheless, these studies have been limited and a comprehensive investigation of these quasi-particles is missing for FCIs.

In this letter, for the first time, we perform direct calculations of quasi-particle charge gaps in a FCI, and compare the results for spinless and spinful charge and neutral excitations, respectively with the same spin as the polarized FCI ground-state, or with relative spin-flips. We carry out a thorough study using exact diagonalization (ED), determining the robustness of our results to system size and band-mixing effects.  We focus on electron filling $\nu=-2/3$ and $\theta=3.7^{\degree}$ with the main goal of understanding the microscopic origin of the hierarchy of gaps observed in Ref.$\,$\cite{park2025observationhightemperaturedissipationlessfractional}. 
In Fig.$\,$\ref{fig:1}(c) we summarize our results for our best estimates for the different computed excitation gaps. We also obtain the bandwidths of the spinless quasi-hole and quasi-electron dispersions, that we estimate respectively to be $1.1\,$meV and $3\,$meV.

\paragraph*{Model and Methods.---}
To carry out an exhaustive study of the excitations in MoTe$_2$, we perform ED calculations on the single-particle first-harmonic continuum model \cite{PhysRevLett.122.086402} which provides a good description of the electronic bands at $\theta=3.7^{\circ}$ with the parameters obtained in Ref.$\,$\cite{MFCI_I} (see Appendix \ref{sec:Model_definition} for details).
The resulting band structure is shown in Fig.$\,$\ref{fig:1}(b). To model the interactions, we take the double-gated screened Coulomb interaction between holes \cite{PhysRevB.108.085117,MFCI_0}, with gate distance $\xi=20$nm  and work with the full Hamiltonian in the hole basis. Unless otherwise stated, for all the presented results, we take an electron filling $\nu=-2/3$, $\theta=3.7^{\degree}$ and dieletric constant $\epsilon = 10/0.9$.

We performed calculations by projecting the many-body Hamiltonian into (i) the topmost valence band to perform one band per valley (1BPV) ED calculations, allowing to reach the largest possible system sizes and (ii) the two topmost valence bands, that we will refer to as band 0 (topmost) and band 1 (second topmost), to perform two bands per valley (2BPV) ED calculations in order to account for effects of band mixing that were previously shown to be crucial to fruitfully capture the physics of twisted MoTe$_{2}$ \cite{MFCI_0,FTI_Yves,PhysRevB.109.L121107,doi:10.1073/pnas.2316749121}.

In the 2BPV ED calculations  we will employ the "band maximum" technique introduced in Ref.$\,$\cite{MFCI_IV}. This technique limits the number of holes allowed in the remote band (band 1) while leaving the occupation of band 0 unrestricted. This approach enables a controlled inclusion of band mixing effects by tuning maximum number of holes allowed in band 1, $N_{\text{band}1}$.
In order to minimize finite-size effects, we typically employ tilted boundary conditions \cite{PhysRevLett.111.126802, PhysRevB.90.245401}. We identify different system sizes with the integers $(N_x,N_y,n_{11},n_{12},n_{21},n_{22})$, where $N_x$ and $N_y$ are the number of unit cells along each moiré lattice vector and the remaining integers characterize different types of tilted boundary conditions (see Appendix \ref{sec:1BPV} for details).

\paragraph*{Charge gaps.---}

Due to the topological order of the FCI ground states, we anticipate that the lowest-energy charge excitations will correspond to quasi-electrons and quasi-holes with fractional charge. 
Inspired by standard FQH methods - where quasi-particles are introduced in ED calculations by changing the magnetic flux - we define analogous excitations in FCIs by appropriately varying the number of particles and sites in the system. We now present the general theory (see Appendix. \ref{sec:gaps} for details). 

The fundamental quantity computed in ED is the spectrum in each symmetry sector. We define $E_{\bf{k}}(N_h,S_{z},N_s)$ as the ground-state energy for a system with $N_h$ holes on $N_s$ sites in the spin $S_z$ (that defines the number of holes in valleys $\bf{K}$ and $\bf{K}'$, respectively $N_h^{\up} = 2 S_z$ and  $N_h^{\down} =N_h - 2 S_z$)  and many-body momentum $\bf{k}$ sector. To determine the ground-state at filling $\nu_h = p/q$, we take $N_h=pN$ and $N_s=qN$, where $q N \rightarrow \infty$ in the thermodynamic limit. We can then add or remove $p'$ holes and $q'$ sites to the system and define the energy dispersions of excitations with charge $e^*=\bar{e}\frac{p' q - q' p}{q}$, spin $\Delta S_z\equiv \frac{e^*}{2 \bar{e}}-s$ (where $s$ is the number of spin-flips with respect to the fully polarized ground-state)  and momentum $\bf{k}$ relative to a zero-momentum FCI ground-state as

\begin{align}
    \epsilon_{e^*,\nu_h}^s({\bf k},N)&=\frac{e^*}{|e^*|}  \big(E_{{\bf k}}(pN\!+p',S_{z}^{\textrm{max}}\!\!-s,qN\!+q')  -E_0 \big) \, ,
    \label{eq:eps_general_main}
\end{align}

\noindent where  $\bar{e}=|e|$ is the hole charge with $e<0$, we have $p,q,p',q',s \in \mathbb{Z}$ and 
$E_0=\langle E_{{\bf k} = {\bf k}_{FCI}}(pN,S_z^{\textrm{max}},qN) \rangle_{\Psi_{FCI}}$ is the average energy among the $q$ quasi-degenerate FCI ground-states, with $S_z^{\textrm{max}} = pN/2$. Note that for brevity, we omit the explicit dependence $e^*=e^*(\nu_h=p/q,p',q')$.  In Eq.$\,$\ref{eq:eps_general_main} we adopted a "band picture" for the excitations, that allows us to define quasi-particle charge gaps by subtracting the quasi-hole and quasi-electron dispersions, through

\begin{align}
    \Delta_{e^{*},\nu_h}^{s,s'}=  \underset{\bf{k},\bf{k}'}{\min} \, \frac{e^*}{|e^*|} \big( \epsilon_{e^*,\nu_h}^s({\bf k},N)-\epsilon_{-e^*,\nu_h}^{s'}({\bf k}',N) \big) 
    \, ,
    \label{eq:charge_gap_general_main}
\end{align}

\noindent where $\epsilon_{-e^*,\nu_h}^{s'}({\bf k},N)$ can simply be obtained from Eq.$\,$\ref{eq:eps_general_main} by replacing $p',q'\rightarrow-p',-q'$. The charge gaps in Eq.$\,$\ref{eq:charge_gap_general_main} correspond to the energy required to create well separated quasi-particles with opposite charges $e^*=\pm \bar{e} \frac{p' q - q' p}{q}$ with a net spin difference $\Delta S_z = -(s+s')$ with respect to the ground-state.  Importantly, reliable extraction of these charge gaps requires that  $p' \ll pN$ and $q' \ll qN$, ensuring that the quasi-particles remain dilute and that there are no severe finite-size effects. We focus on filling $\nu_h=2/3$ ($p=2,q=3$), so we will drop the dependence on $\nu_h$ in Eqs.$\,$\ref{eq:eps_general_main} and \ref{eq:charge_gap_general_main} from now on for brevity. At this filling, the minimal (charge-$\pm \bar{e}/3$) gaps can only be obtained by choosing $p'=q'=\pm 1$, that is by adding or removing simultaneously both one hole and one flux.\footnote{Note that for the more familiar case of the Laughlin state at $\nu_h=1/3$, adding (removing) one flux while keeping the electron number unchanged, i.e. choosing $p'=0, q'=1$ ($p'=0, q'=-1$), creates a charge $\bar{e}/3$ quasi-hole ($e/3$ quasi-electron) excitation.} In Appendix \ref{sec:gaps}, we explain this counting with a connection to the composite fermion picture in FQH.
Additionally, since the ground-state is fully polarized  \cite{MFCI_0,prepar}, we only study the excitations that involve the smallest amount of spin flips, $s,s'=0,1$. In particular, we study charge-$e/3$ gaps with different net spin difference relative to the ground-state: spin-0 gap $\Delta_{e/3}^{0}\equiv \Delta_{e/3,2/3}^{0,0}$, spin-1 gaps $\Delta_{e/3}^{-1,-} \equiv \Delta_{e/3,2/3}^{1,0}$ and $\Delta_{e/3}^{-1,+} \equiv \Delta_{e/3,2/3}^{0,1}$, and spin-2 gap $\Delta_{e/3}^{-2} \equiv \Delta_{e/3,2/3}^{1,1}$. Note that  $\Delta_{e/3}^{-1,-}$ is the energy required to create a magnon-dressed quasi-electron well separated from a quasi-hole, while $\Delta_{e/3}^{-1,+}$ is the energy required to create a quasi-electron well separated from a magnon-dressed quasi-hole. Excitations that differ by their total spin quantum number can be distinguished experimentally by their coupling to the magnetic field via the Zeeman effect. The Zeeman energy differences from the ground-state   are, respectively, $\Delta \epsilon_z = 0,g \mu_B B,2g \mu_B B$ for the gaps with $\Delta S_z=0,-1,-2$, where $\mu_B=e\hbar/(2mc)$ is the Bohr magneton and $g$ is the effective g-factor.

\paragraph*{Neutral Gaps.---}
 In the previous section, we discussed charge gaps, defined by adding the energies required to create a quasi-electron and quasi-hole independently, without any binding energy. Neutral gaps, on the other hand, are the energies required to create bound quasi-electron/quasi-hole pairs. They are defined as excitation energies at a fixed number of holes and number of unit cells. We define the spin-0 neutral gap (or more commonly known as the FCI or many-body gap), $\Delta_N$, and the spin-1 neutral gap, $\Delta_N^s$, as:

\begin{align}
    \Delta_N &= \min\nolimits_{\mathbf{k}}^{\prime} E_{\mathbf{k}}(pN, S_z^{\text{max}}, qN) - E_0 
    \label{eq:Delta_N_main} \\
    \Delta_N^s &= \min_{\bf{k}} E_{ \bf{k} }(pN,S_z^{\textrm{max}}-1,qN)-E_0
    \label{eq:Delta_Ns_main}
\end{align}

\noindent where $\min'_{\bf{k}}  E_{\bf{k}}$ is defined as the minimum many-body energy excluding the energies of the FCI ground-states.
Eqs.$\,$\ref{eq:Delta_N_main},\ref{eq:Delta_Ns_main} correspond respectively to the energies required to excite magnetoroton \cite{PhysRevB.33.2481,PhysRevB.90.045114,wolf2024intrabandcollectiveexcitationsfractional,shen2025magnetorotonsmoirefractionalchern,long2025spectramagnetorotonchiralgraviton,lu2024interactiondrivenrotoncondensationc} or magnon-like collective modes.

\begin{figure}[h!]
     \centering
     \includegraphics[width=1.0\linewidth]{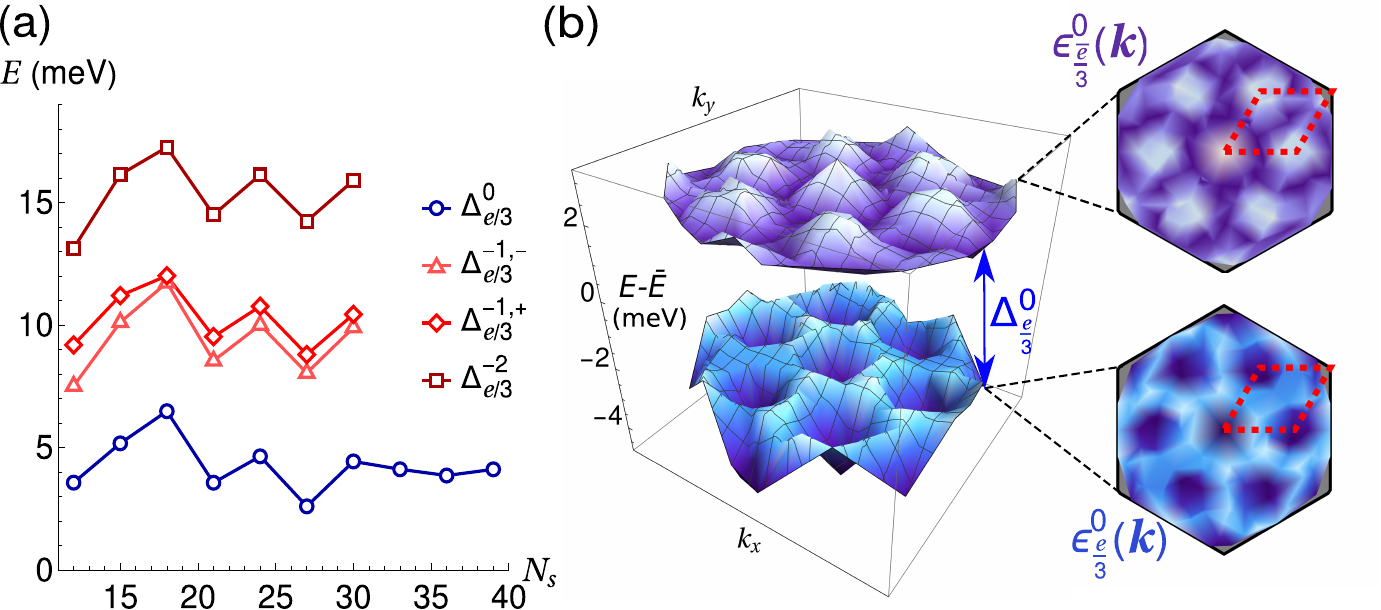}
     \caption{ 1BPV results for quasi-particle charge gaps. (a) Charge $e/3$ gaps computed through Eq.$\,$\ref{eq:charge_gap_general_main} for different system sizes.  The calculations of each charge gap involve systems with $N_s$ and $N_s\pm1$ sites (see Appendix \ref{sec:1BPV} for details on all the used system sizes). (b) Quasi-hole and quasi-electron dispersions for the spin-0 charge excitations. The results were obtained by interpolating the data combined from all the system sizes with $N_s \geq 20$. In order to avoid fluctuations in the dispersion due to the variations in the gap for different sizes, we made energy shifts in all the data such that  $ \min_{\bf k} \epsilon_{\bar{e}/3}^0({\bf k},N) =  \min_{\bf k} \epsilon_{\bar{e}/3}^0({\bf k},N_{\textrm{max}})$ (that is, the minimum over $\bf{k}$ of the quasi-hole energy at any $N$ is chosen to be same as the mininum over $\bf{k}$ at $N=N_{\textrm{max}}$) and $ \max_{\bf k} \epsilon_{e/3}^0({\bf k},N) =  \max_{\bf k} \epsilon_{e/3}^0({\bf k},N_{\textrm{max}})$ (the maximum over $\bf{k}$ of the quasi-electron energy at any $N$ becomes the same as the maximum over $\bf{k}$ at $N=N_{\textrm{max}}$), where $N_{\textrm{max}}=39$ is the largest system size. We also shifted all energies  by  $\bar{E} = \big( \min_{\bf k} \epsilon_{\bar{e}/3}^0({\bf k},N_{\textrm{max}}) 
 + \max_{\bf k} \epsilon_{e/3}^0({\bf k},N_{\textrm{max}}) \big)/2$ in order to center the energy gap around $E=0$. The approximate emergent Brillouin zone is highlighted in the density plots by dashed lines.}
     \label{fig:2}
\end{figure}

\begin{figure}[h!]
     \centering
     \includegraphics[width=1.0\linewidth]{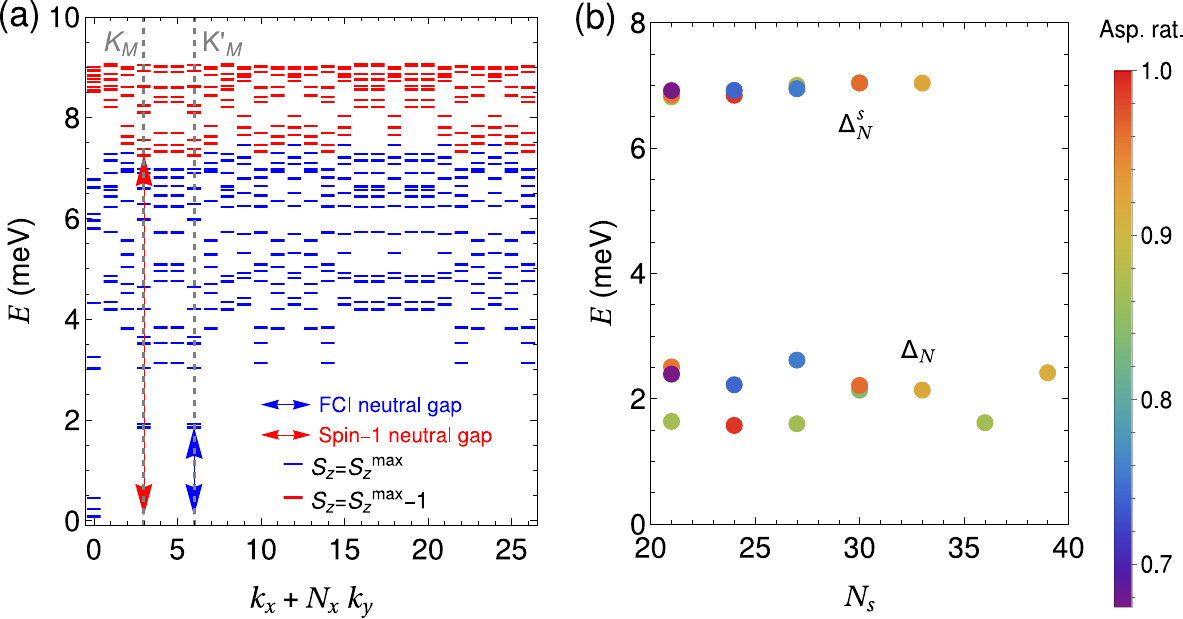}
     \caption{ 1BPV results for neutral gaps. (a) Energy spectrum in the $S_z=S_z^{\textrm{max}}$ (lowest 12 eigenvalues per $\bf{k}$-sector) and $S_z=S_z^{\textrm{max}}-1$ (lowest 8 eigenvalues) sectors, for the 27-site lattice defined by $(N_x,N_y,n_{11},n_{12},n_{21},n_{22})=(9,3,1,1,0,1)$. The spin-0 (FCI) and spin-1 neutral gaps are depicted respectively by the blue and red arrows, with the minimal gap occurring at $\bf{k}=\bf{K}_M,\bf{K'}_M$ in both cases.  
      (b) Spin-0 ($\Delta_N$) and spin-1 ($\Delta_N^s$) neutral gaps for different system sizes.  The different points are colored by aspect ratio.   }
     \label{fig:3}
\end{figure}

\paragraph*{1BPV Results.---}
We start by presenting 1BPV results on the charge and neutral gaps shown respectively in Figs.$\,$\ref{fig:2} and \ref{fig:3}. In Fig.$\,$\ref{fig:2} we present finite-size scaling results for all the charge-$e/3$ gaps. The spin-0 gaps show a good convergence with system size, with a relative variation of only $6\%$  across the three largest sizes. Among the spinful charge excitations, the smallest gap is $\Delta_{e/3}^{-1,-}$ and corresponds to the energy to create a quasi-hole well separated from a magnon-dressed quasi-electron , as illustrated in Fig.$\,$\ref{fig:1}(a). 
Since it has the smallest spinful gap, we identify this excitation as the most relevant for the thermal activation observed in experiment near $B=0$. We note however that $\Delta_{e/3}^{-1,+}$ is also very close in energy and may also provide a strong contribution. The spin-2 gap $\Delta_{e/3}^{-2}$, on the other hand, is significantly larger than the spin-1 gaps (approximately by $6\,$meV). The finite-size fluctuations are comparable between spin-0 and spinful gaps, suggesting that finite-size effects are under control for $N_s \geq 30$. 

For the largest system size we used in calculations of spinful gaps ($N_s=30$) we find $\Delta_{e/3}^{-1,-}/\Delta_{e/3}^{0}\approx 2.2$. In Appendix \ref{sec:1BPV}, we also show results in the flat band limit where the single-particle energy dispersion is neglected and the interaction is the only energy scale. In this case, we find $\Delta_{e/3}^{-1,-}/\Delta_{e/3}^{0}=2.5$, indicating that the ratio is not very sensitive to interaction strength for the interaction range that we consider. 
In Appendix \ref{sec:1BPV}, we also present results for quasi-particle gaps of charge-$e$ and charge-$2e/3$ excitations, with the latter being a target of significant recent interest \cite{xu2025dynamicsclustersanyonsfractional,gattu2025molecularanyonsfractionalquantum,zhang2025holonmetalchargedensitywavechiral,nosov2025anyonsuperconductivityplateautransitions,shi2025anyondelocalizationtransitionsdisordered}. These show stronger finite-size effects than the charge-$e/3$ gaps here presented. This is expected since adding charges $2e/3$ or $e$ is equivalent to creating respectively two or three fractionalized quasi-electrons, which can experience strong mutual repulsion in finite systems. While we find $\Delta_{2e/3}^{s}\approx 2\Delta_{e/3}^{s}$ for the largest studied $N_s$, we always observe $\Delta_{e}^{s} > 3\Delta_{e/3}^{s}$, with a strong $N_s$ dependence for the range of studied system sizes. 
This highlights the importance of computing charge-$e/3$ gaps directly, rather than estimating them from the standard charge-1 gaps $\Delta_{e}^{s}$ , which are more susceptible to finite size effects (see Appendix \ref{sec:1BPV}).

In Fig.$\,$\ref{fig:2}(b) we show the dispersion of the spin-0 quasi-electron and quasi-hole excitations computed through Eq.$\,$\ref{eq:eps_general_main}. Our procedure is to compute $\epsilon_{e*}^0(\mbf{k},N)$ on many different lattices $N_s = 20,\dots, 37$ in order to sample different meshes in the Brillouin zone. The resulting energy dispersions for each system size are shown in Appendix~\ref{sec:charge_excitations_2o3}.
 We then combine the data across all $\bf{k}$ points obtained from different system sizes and perform a linear interpolation to obtain the plots in  Fig.$\,$\ref{fig:2}(b) (see the caption for more details). Remarkably, we find that the data all collapses to a smooth band structure, demonstrating that finite size effects are under control. Importantly, the spectra reveal an emergent approximate periodicity depicted by the red dashed lines in Fig.$\,$\ref{fig:2}(b), with reciprocal lattice vectors three times smaller than the moiré reciprocal lattice vectors in correspondence with a $3\times 3$ enlarged unit cell. The most natural interpretation of this data is as a Hofstadter spectrum, where magnetic translation symmetry at $\phi/2\pi = 1/3$ flux per unit cell enforces additional periodicity in momentum space. While a many-body derivation of this claim is beyond the scope of this paper, it can be expected from simple symmetry considerations. (1) In twisted MoTe$_{2}$, the parent Chern band  can be thought of as a ``Hofstadter'' band at $\phi = 2\pi$ flux \cite{2024PhRvB.110k5114K,2024PhRvB.110c5130S} where magnetic translations by lattice vectors commute. Since the orbital coupling to the magnetic field is proportional to the quasi-particle charge, a quasi-particle with charge $\pm e/3$ is expected to feel $\phi_{e^*}=2\pi/3$ flux. This explains the increased periodicity, since magnetic translations require $\phi_{e^*}$-periodicity across the Brillouin zone \cite{bernevig2013topological,PhysRevB.79.195132} (see Appendix \ref{sec:charge_excitations_2o3} for details). (2) In the FQH system, the dispersion of the quasi-electrons and quasi-holes is flat in the thermodynamic    limit due to the extensive center-of-mass degeneracy enforced by magnetic translation symmetry (see Appendix \ref{sec:charge_excitations_2o3} for details). However, in a FCI, the lattice breaks continuous translation down to a discrete subgroup, leading to modulations in the electrostatic potential and magnetization created by the non-uniform density and circulating current in the ground state. This periodicity, in the presence of $1/3$ flux, generically leads to a Hofstadter spectrum with nonzero dispersion. However, this heuristic explanation obfuscates the fact that no magnetic flux appears in the microscopic Hamiltonian. The approximate Hofstadter periodicity is evidence that the quasi-particle obeys a projective representation of the translation group\cite{2002PhRvB..65p5113W}, motivated by our numerical observations.

Turning to the neutral excitations, Fig.$\,$\ref{fig:3} shows the 1BPV results for the spin-0 and spin-1 neutral gaps. Fig.$\,$\ref{fig:3}(a) shows a representative example of the energy spectra in the spin sectors $S_z=S_z^{\textrm{max}}$ and $S_z=S_z^{\textrm{max}}-1$, where it can be seen that $\Delta_N^s$ is significantly larger than $\Delta_N$. The finite-size scaling shown in Fig.$\,$\ref{fig:3}(b) indicates that $\Delta_N^s$ is well converged - with relative variations below $1.3 \%$ for the three larger system sizes - while $\Delta_N$  shows more pronounced finite-size fluctuations. 
For the best estimates, we obtain $\Delta_N^s/\Delta_N=\NeutralSpinOneSpinZeroRatioOneBPV$. In the flatband limit (Appendix \ref{sec:1BPV}), we obtain $\Delta_N^s/\Delta_N=\NeutralSpinOneSpinZeroRatioFlatbandOneBPV$, showing  that the hierarchy is robust to band dispersion and interaction strength at the level of 1BPV calculations. 
Finally, in Fig.$\,$\ref{fig:3}(a) we can also identify collective modes, whose detailed study is left for a separate work.

\begin{figure}[h!]
     \centering
     \includegraphics[width=1.0\linewidth]{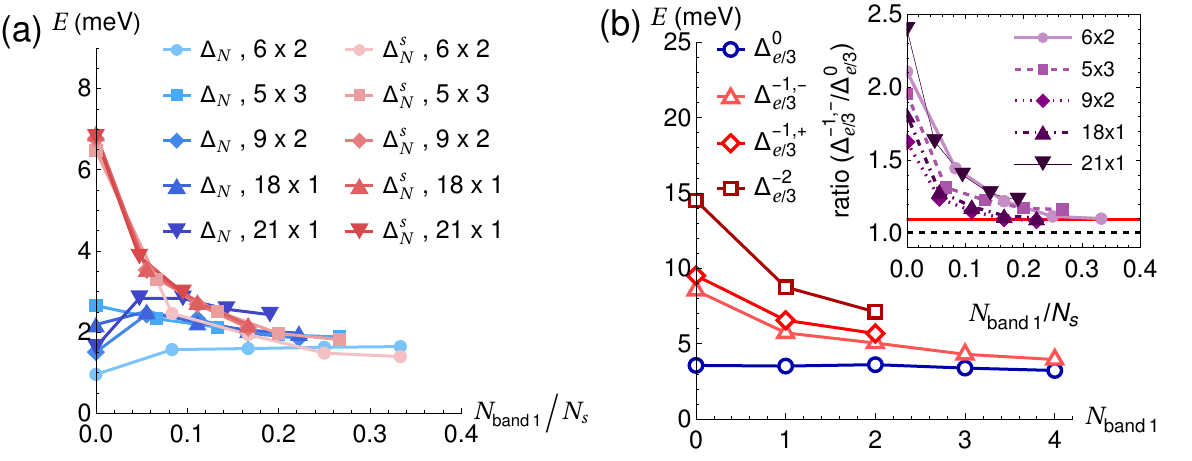}
     \caption{ Band mixing effects on neutral and charge gaps. (a) Attempt of data collapse for the spin-0 ($\Delta_N$) and spin-1 ($\Delta_N^s$) neutral gap data as a function of $N_{\textrm{band1}}/N_s$, for all the considered system sizes. (b) Charge $e/3$ gaps as function of $N_{\textrm{band}1}$, for the 21-site system defined by $(N_x,N_y,n_{11},n_{12},n_{21},n_{22})=(21,1,1,-5,0,1)$. Inset: Ratio $\Delta_{e/3}^{-1,-}/\Delta_{e/3}^{0}$ as a function of density of holes in band 1, for different system sizes. The full red line indicates the 12-site full 2BPV result. }
     \label{fig:4}
\end{figure}

\paragraph*{2BPV Results.---}
We now address the effects of band mixing on the charge and neutral gaps by performing 2BPV calculations, summarized in Figs.$\,$\ref{fig:4}(a,b). The key observation is that the spin-0 neutral and charge gaps remain relatively unaffected by band mixing, while spinful gaps undergo a substantial reduction (as expected from the impact of band mixing on magnetic properties \cite{MFCI_0,FTI_Yves}). 
In Fig.$\,$\ref{fig:4}(a) we attempt a collapse of all the data for the neutral gaps as a function of $N_\textrm{band1}/N_s$, for all the considered system sizes.
For the spin-1 neutral gap, we find excellent data collapse for $N_s \geq 15$, while the data for $N_s=12$ exhibits significant deviations.  We therefore take the point with largest $N_{\textrm{band1}}/N_s$ (obtained for $N_s=15$) as the best estimate for $\Delta_N^s$, shown in Fig.$\,$\ref{fig:1}(c), identifying this procedure by 'All $N_s$' in Fig.$\,$\ref{fig:1}(c). 
For the spin-0 gap, no good quality data collapse is observed. For the best estimate of the spin-0 neutral gap in Fig.$\,$\ref{fig:1}(c) we therefore consider the result for the largest system size $N_s=21$ and largest $N_{\textrm{band1}}$(=4), which is  converged in $N_{\textrm{band1}}$ (with only a $5\%$ relative difference between the two largest $N_{\textrm{band1}}$).  
From these results, we conclude that band mixing brings the spin-0 and spin-1 neutral gaps to comparable values, contrasting with the large ratio observed in 1BPV calculations.

A similar trend holds for the charge gaps, as shown in Fig.$\,$\ref{fig:4}(b): the spin-0 charge gap is essentially unaffected, while the smallest spinful charge gap  $\Delta_{e/3}^{-1,-}$ is strongly reduced, becoming comparable - though still slightly larger than - the spin-0 gap.
The inset of Fig.$\,$\ref{fig:4}(b), shows that although no clear scaling collapse of the ratio $\Delta_{e/3}^{-1,-}/\Delta_{e/3}^{0}$ exists for the different system sizes, the asymptotic values at large enough $N_{\textrm{band1}}$ are similar. 
For the best estimates of the 2BPV spin-0 and spin-1 gaps shown in Fig.~\ref{fig:1}(c), we use the results obtained at $N_{\textrm{band1}} = 4$ for the $N_s = 21$ system. This choice is motivated not only because $N_s = 21$ is the largest system size accessible in our 2BPV calculations (for which it is possible to converge in $N_{\textrm{band1}}$), but also due to its demonstrated reliability: for the spin-0 gap, the 1BPV result for $N_s=21$  differs by only $13\%$ from the value obtained at $N_s = 39$, the largest system size considered overall, as shown in Fig.$\,$\ref{fig:2}. We note however that the most demanding calculation to obtain the spin-1  gap $\Delta_{e/3}^{-1,-}$ ($N_s=20,N_{\textrm{band1}}=4$) involves an Hilbert space dimension of $10^9$. Therefore we restrict the calculation to the momentum-sectors of the five lowest-energy states for $N_{\textrm{band1}}=4$. This approximation is well justified because  (i) we observe that the momentum sector of the lowest-energy state remains unchanged for all cases where $N_{\textrm{band1}} = 4$ calculations are feasible (see Appendix \ref{sec:2BPV}) and (ii) we observed that the energetic hierarchy of the five lowest-energy states obtained for $N_{\textrm{band1}}=3$ was unchanged for $N_{\textrm{band1}}=4$.
We also find that band mixing significantly affects the dispersion of the charged excitations, with the general trend of narrowing it, as shown in Appendix \ref{sec:spectra_charge_excitations_bmixing}. 
However, unlike the 1BPV results shown in Fig.$,$\ref{fig:2}(b), a satisfactory collapse of the data is not possible due to strong finite-size effects.
 
We finally note that spinful gaps could be underestimated by not properly including short-range  repulsion at the atomic scale. At the moiré scale, this interaction is suppressed with a factor $(a/a_M)^2$, where $a$ and $a_M$ are respectively the atomic and moiré lattice constants, but can still be significant (see Appendix \ref{sec:Hubbard} for details ). However, reasonable modification of the interaction by adding a Hubbard term compatible with the magnetic properties at e.g. $\nu=-4/3$, only increases the spin-1 gap by at most $30\%$ (see Appendix \ref{sec:Hubbard} for a detailed discussion).  
This modest correction suggests that the spinless and spinful charge gaps can be affected differently by
additional physical effects not included in our model, with disorder likely being the leading contributor.

\paragraph*{Discussion.---}

Our results qualitatively reproduce the key experimental hierarchy: the smallest spinful charge gap is consistently  larger than the spinless (spin-0) charge gap both in our 1BPV and 2BPV calculations.
However, we find that band mixing significantly reduces the spinful gap, bringing it close to the spin-0 gap (although still remaining larger). For 1BPV, our best estimate for the ratio is $\Delta_{e/3}^{-1,-}/\Delta_{e/3}^{0}=2.4$, decreasing to $1.2$ in the 2BPV calculations.
The 2BPV ratio is therefore substantially smaller than the experimental value reported in Ref.$\,$\cite{park2025observationhightemperaturedissipationlessfractional} --- $\Delta_{e/3}^{-1,-}/\Delta_{e/3}^{0} \approx 2.8$ --- pointing to the likely importance of additional physical ingredients, such as disorder or additional short-range interactions.

Quantitatively, our best estimates for the spin-0 quasi-particle charge gap - $\Delta_{e/3}^{0}=\ChargeSpinZeroEstimateOneBPV \,$meV (39-site, no 1BPV) and $\Delta_{e/3}^{0}=\ChargeSpinZeroEstimateTwoBPV \,$meV (21-site, 2BPV) - are larger but of the order of the experimental value of $\Delta_{e/3}^{0}\approx1.7$meV observed in Ref.$\,$\cite{park2025observationhightemperaturedissipationlessfractional}. This level of discrepancy, exceeding $50\%$, is comparable to that observed between the most accurate numerical calculations of FQH gaps and experimental measurements in GaAs heterostructures \cite{PhysRevLett.127.056801}. For the spin-1 charge gap, our best estimates are $\ChargeSpinOneEstimateOneBPV \,$meV (30-site, 1BPV) and $\ChargeSpinOneEstimateTwoBPV \,$meV (21-site, 2BPV), with the 2BPV result being closer, but slightly smaller than the experimental value in Ref.$\,$\cite{park2025observationhightemperaturedissipationlessfractional} of $4.7$meV.
 We note, however, that the precise value of these gaps can be tuned by varying the dielectric constant, not known precisely, and that we fixed in this work. Rather than aiming for perfect quantitative agreement, our goal is to demonstrate that theoretical estimates of the charge gaps can approximate the experimentally observed values using realistic, \emph{ab initio}-derived model parameters. 

Finally, our work highlights several important directions for future study. First, the good convergence of quasi-particle charge gaps suggests that the methods introduced here may serve as a foundation for extracting transport gaps in future theoretical studies of FCIs, enabling direct comparison with experiments. Second, we underscore the physical importance of spinful quasi-particles, which, while often overlooked in favor of spin-polarized excitations, can be energetically competitive and may even dominate transport in experimentally relevant regimes. Our results strongly motivate further theoretical and experimental exploration of spinful excitations in FCIs, especially in materials with small Zeeman energy scales and strong multi-band effects. Finally, we have introduced the notion of the quasi-particle band structure which makes concrete the idea of an emergent, weakly-interacting, and \emph{mobile} charged particle. In so doing, we have quantitatively revealed a fundamental distinction between FCI and FQH fluids. While the quasi-particles in the FQH have a quenched kinetic energy due to continuous magnetic translation symmetry, those of the FCI are dispersive. They may give rise to exotic phases of matter that replace the FQH-like plateaus  \cite{PhysRevLett.51.605} if disorder can be reduced below the quasi-particle bandwidth.

\begin{acknowledgments}
We thank Zlatko Papic, Yves Kwan, Zhao Liu and Valentin Cr\'epel for fruitful discussions.

B.A.B. was supported by the Gordon and Betty Moore Foundation through Grant No.~GBMF8685 towards the Princeton theory program, the Gordon and Betty Moore Foundation’s EPiQS Initiative (Grant No.~GBMF11070), the Office of Naval Research (ONR Grant No.~N00014-20-1-2303), the Global Collaborative Network Grant at Princeton University, the Simons Investigator Grant No.~404513, the BSF Israel US foundation No.~2018226, the NSF-MERSEC (Grant No.~MERSEC DMR 2011750), the Simons Collaboration on New Frontiers in Superconductivity (SFI-MPS-NFS-00006741-01), and the Schmidt Foundation at the Princeton University. This project was partially supported by the European Research Council (ERC) under the European Union’s Horizon 2020 Research and Innovation Programme (Grant Agreement No. 101020833). The Flatiron Institute is a division of the Simons Foundation. J. Y.'s work was supported by startup funds at University of Florida.
X. X. and D. X. are supported by DE-SC0012509.

Part of the simulations presented in this article were performed on computational resources managed and supported by Princeton Research Computing, a consortium of groups including the Princeton Institute for Computational Science and Engineering (PICSciE) and Research Computing at Princeton University.

\end{acknowledgments}

\renewcommand{\addcontentsline}[3]{}
\bibliographystyle{apsrev4-2}
\bibliography{refs}

\begin{thebibliography}{120}
\bibitem [{pre()}]{prepar}%
  \BibitemOpen
  \href@noop {} {\bibinfo  {journal} {The full many-body spectra across all
  spin sectors --- demonstrating that the ground state is in the fully
  polarized sector for the same model parameters used in this work --- will be
  presented in a separate manuscript}\ }\BibitemShut {NoStop}%
\end{thebibliography}
\let\addcontentsline\oldaddcontentsline

\renewcommand{\thetable}{S\arabic{table}}
\renewcommand{\thefigure}{S\arabic{figure}}
\renewcommand{\theequation}{S\arabic{section}.\arabic{equation}}
\onecolumngrid
\pagebreak
\thispagestyle{empty}

\cleardoublepage

\begin{center}
	\textbf{\large Supplementary Information for ''\titlePaper{}``}\\[.2cm]
\end{center}

\appendix
\renewcommand{\thesection}{\Roman{section}}
\tableofcontents
\let\oldaddcontentsline\addcontentsline
\newpage

\section{Review of the single-particle and interacting model}

\label{sec:Model_definition}

In this section we briefly review the model for twisted MoTe2 used in this work. For a more detailed overview, the interested reader is referred to Refs.$\,$\cite{MFCI_0,MFCI_I}.

The symmetry group of AA stacked moiré MoTe2 is generated by the moiré lattice translations, rotational symmetries $C_3$, $C_{2y}$, and time-reversal symmetry $\mathcal{T}$. 

The single-particle continuum model can be written in real-space as \cite{PhysRevLett.122.086402}

\begin{equation}
H_{\eta}^{0}({\bf r})=\int d^{2}r\,\left(c_{\eta,b,{\bf r}}^{\dagger}\,c_{\eta,t,{\bf r}}^{\dagger}\right)\left(\begin{array}{cc}
\frac{\hbar^{2}\nabla^{2}}{2m^{*}}+V_{\eta,b}({\bf r}) & t_{\eta}({\bf r})\\
t_{\eta}^{*}({\bf r}) & \frac{\hbar^{2}\nabla^{2}}{2m^{*}}+V_{\eta,t}({\bf r})
\end{array}\right)\left(\begin{array}{c}
c_{\eta,b,{\bf r}}\\
c_{\eta,t,{\bf r}}
\end{array}\right)
\end{equation}
where the coarse-grained continuum operator $c_{\eta,\ell,{\bf r}}^{\dagger}$
creates an electron at position ${\bf r}$, layer $\ell$ and valley
$\eta=\pm1$ (respectively $\pm{\bf K}$ valleys). Under the first
harmonic approximation, we have 

\begin{equation}
V_{\eta,\ell}({\bf r})=2V\sum_{i=1}^{3}\cos({\bf g}_{i}\cdot{\bf r}-(-1)^{\ell}\psi)
\end{equation}

\begin{equation}
t_{\eta}({\bf r})=w\sum_{i=1}^{3}e^{-{\rm i}\eta{\bf q}_{i}\cdot{\bf r}}\,,
\end{equation}
where $(-1)^{\ell}=1,-1$ respectively for $\ell=b,t$, ${\bf g}_{i}=C_{3}^{i-1}{\bf b}_{M,1}$
with ${\bf b}_{M,1}=\frac{4\pi}{\sqrt{3}a_{M}}\big(1,0\big)^{T}$
and ${\bf b}_{M,2}=C_{6}{\bf b}_{M,1}$ the moiré reciprocal lattice
vectors and $a_{M}=\frac{a_{0}}{2\sin(\theta/2)}$ the moiré lattice
constant ($a_{0}=3.52\text{\AA})$; ${\bf q}_{1}=\frac{8\pi}{3a_{0}}\sin(\theta/2)(0,1)^{T},{\bf q}_{j+1}=C_{3}^{j}{\bf q}_{1}\,(j=1,2)$.
We choose the model parameters extracted for the first-harmonic model in Ref.$\,$\cite{MFCI_I} from
fitting the DFT band structure at $\theta=3.89^{\circ}$: effective
mass $m_{*}=0.6m_{e}$ (with $m_{e}$ the electron mass), interlayer
tunneling amplitude $w=-18.8$meV, moiré potential strength $V=16.5$meV
and phase $\psi=-105.9^{\circ}$. 

Introducing the plane-wave Fourier basis,

\begin{equation}
c_{{\bf k},{\bf Q},\eta}^{\dagger}=\frac{1}{\sqrt{\Omega_{tot}}}\int d^{2}{\bf r}e^{i({\bf k}-{\bf Q})\cdot{\bf r}}c_{{\bf r}l\eta}^{\dagger}
\end{equation}
where  $\Omega_{tot}$ is the total area of the moiré sample, ${\bf k}$
is defined in the moiré Brillouin zone and ${\bf Q}\in\mathcal{Q}_{l}^{\eta}=\{{\bf G}_{M}+\eta(-)^{\ell}{\bf q}_{1}\}$,
where ${\bf G}_{M}$ are moiré reciprocal lattice vectors, so that

\begin{equation}
c_{{\bf r}l\eta}^{\dagger}=\frac{1}{\sqrt{\Omega_{tot}}}\sum_{{\bf Q}\in\mathcal{Q}_{l}^{\eta}}\sum_{{\bf k}\in{\rm MBZ}}e^{-i({\bf k}-{\bf Q})\cdot{\bf r}}c_{{\bf k},{\bf Q},\eta}^{\dagger}\,.
\end{equation}

Defining $\mathcal{Q}=\{{\bf G}_{M}+{\bf q}_{1}\}\cup\{{\bf G}_{M}-{\bf q}_{1}\}$,
we can write the single-particle Hamiltonian in the plane-wave basis
as 

\begin{equation}
H_{\eta}^{0}({\bf r})=\sum_{{\bf k}}\sum_{{\bf Q},{\bf Q}'\in\mathcal{Q}}c_{{\bf k},{\bf Q},\eta}^{\dagger}\,h_{{\bf Q}{\bf Q}'}^{\eta}({\bf k})\,c_{{\bf k},{\bf Q}',\eta'}^{\dagger}\label{eq:H_SP_k}
\end{equation}
where 

\begin{equation}
\begin{array}{cc}
h_{{\bf Q}{\bf Q}'}^{\eta}({\bf k}) & =-\frac{\hbar^{2}({\bf k}-{\bf Q})^{2}}{2m^{*}}\delta_{{\bf Q},{\bf Q}'}+w\sum_{i=1}^{3}[\delta_{{\bf Q}+{\bf q}_{i}.{\bf Q}'}+\delta_{{\bf Q}.{\bf Q}'+{\bf q}_{i}}]\\
 & +V\sum_{i=1}^{3}[e^{-{\rm i}\eta(-1)^{{\bf Q}}\psi}\delta_{{\bf Q}+{\bf g}_{i}.{\bf Q}'}+e^{{\rm i}\eta(-1)^{{\bf Q}}\psi}\delta_{{\bf Q}-{\bf g}_{i}.{\bf Q}'}]
\end{array}\label{eq:H_QQp}
\end{equation}
\noindent and  $(-1)^{\bf Q}=\pm 1$ respectively for ${\bf Q }\in \{{\bf G}_{M}\pm{\bf q}_{1}\}$.

For the interaction Hamiltonian, we will consider the Coulomb interaction
between holes. The total Hamiltonian in the hole basis reads

\begin{equation}
H=\int d^{2}r\,\tilde{c}_{\eta,\ell,{\bf r}}^{\dagger}[-H_{\eta}^{0}({\bf r})]_{\ell\ell'}\tilde{c}_{\eta,\ell',{\bf r}}+\frac{1}{2}\sum_{\ell\eta,\ell'\eta'}\int d^{2}r\textrm{ }d^{2}r'\,V({\bf r}-{\bf r}')\tilde{c}_{\eta,\ell,{\bf r}}^{\dagger}\tilde{c}_{\eta',\ell',{\bf r}'}^{\dagger}\tilde{c}_{\eta',\ell',{\bf r}'}\tilde{c}_{\eta,\ell,{\bf r}}
\end{equation}
where $\tilde{c}_{\eta,\ell,{\bf r}}^{\dagger}$ creates a hole
at position ${\bf r}$, layer $\ell$ and valley $\eta=\pm1$, $V({\bf r})$
is the double-gated Coulomb potential given by

\begin{equation}
V({\bf r})=\int\frac{d^{2}p}{(2\pi)^{2}}\,V({\bf p})e^{i{\bf p}\cdot{\bf r}}=V_{\xi}\sum_{n=-\infty}^{\infty}\frac{(-1)^{n}}{\sqrt{({\bf r}/\xi)^{2}+n^{2}}}\,\,\,\,,V({\bf p})=\pi\xi^{2}V_{\xi}\frac{\tanh(\xi|{\bf p}|/2)}{\xi|{\bf p}|/2}\,\,\,,V_{\xi}=\frac{e^{2}}{4\pi\epsilon\epsilon_{0}\xi}
\end{equation}
where $\xi$ is the gate distance that we take $\xi=20nm$, $e<0$
is the electron charge, $\epsilon_{0}$ is the vacuum permittivity
and $\epsilon$ is the dielectric constant.

In order to make the ED calculations manageable, we finally project
the total Hamiltonian into a finite number of bands defined by the
operators 

\begin{equation}
\gamma_{\eta,m,{\bf k}}^{\dagger}=\sum_{{\bf Q}}[U_{\eta,m,{\bf k}}]_{{\bf Q}}\tilde{c}_{\eta,{\bf k},{\bf Q}}^{\dagger}
\label{eq:band_operators}
\end{equation}
where $\{U_{\eta,m,{\bf k}}\}$ are the eigenvectors of the Hamiltonian
matrix in Eq.$\,$\ref{eq:H_QQp}. The projected Hamiltonian is therefore
given by 

\begin{equation}
H=-\sum_{{\bf k}\in{\rm MBZ},\eta,m}E_{\eta,m}({\bf k})\gamma_{\eta,m,{\bf k}}^{\dagger}\gamma_{\eta,m,{\bf k}}+\frac{1}{2\Omega_{\textrm{tot}}}\sum_{\eta\eta'}\sum_{m,m',n',n}\sum_{{\bf k},{\bf k}',{\bf q}\in{\rm MBZ}}V_{\eta\eta',mm'n'n}({\bf k},{\bf k}',{\bf q})\gamma_{\eta,m,{\bf k}+{\bf q}}^{\dagger}\gamma_{\eta',m',{\bf k}'-{\bf q}}^{\dagger}\gamma_{\eta',n',{\bf k}'}\gamma_{\eta,n,{\bf k}}
\label{eq:H_full_band_basis}
\end{equation}
where $E_{\eta,m}({\bf k})$ are the energy bands obtained by diagonalizing
the electron-basis Hamiltonian in Eq.$\,$\ref{eq:H_SP_k} and the
interaction matrix elements are given by 

\begin{equation}
V_{\eta\eta',mm'n'n}({\bf k},{\bf k}',{\bf q})=\sum_{{\bf b}\in\{{\bf G}_{M}\}}V({\bf q}+{\bf b})M_{\eta,mn}({\bf k},{\bf q+{\bf b}})M_{\eta',m'n'}({\bf k}',-{\bf q-{\bf b}})
\end{equation}
where we defined the form factors 

\begin{equation}
M_{\eta,mn}({\bf k},{\bf q}+{\bf b})=\sum_{{\bf Q}}[U_{\eta,m,{\bf k}+{\bf q}}^{*}]_{{\bf Q}-{\bf b}}[U_{\eta,n,{\bf k}}]_{{\bf Q}}\,,
\end{equation}
and used the periodic convention $[U_{\eta,m,{\bf k}+{\bf q}+{\bf b}}^{*}]_{{\bf Q}}=[U_{\eta,m,{\bf k}+{\bf q}}^{*}]_{{\bf Q}-{\bf b}}$.
The form factors satisfy $M_{mn}^{(\eta)*}({\bf k}-{\bf q},{\bf q+{\bf b}})=M_{nm}^{(\eta)}({\bf k},-{\bf q-{\bf b}})$
and the interaction matrix elements the properties $V_{\eta'\eta,m'mnn'}({\bf k}',{\bf k},-{\bf q})=V_{\eta\eta',mm'n'n}({\bf k},{\bf k}',{\bf q})$
and $V_{\eta\eta',nn'm'm}^{*}({\bf k}+{\bf q},{\bf k}'-{\bf q},-{\bf q})=V_{\eta\eta',mm'n'n}({\bf k},{\bf k}',{\bf q})$,
that respectively follow from fermionic statistics and hermiticity.

Unless otherwise stated, for all the presented results, we take an electron filling $\nu=-2/3$, twist angle $\theta=3.7^{\degree}$ and define the interaction strength $u=10/\epsilon$ that we choose as $u=0.9$.

\clearpage

\section{Definition of charge and neutral gaps}

\label{sec:gaps}

In this section, we provide details on the calculation for the two types of excitation gaps analyzed in this work—the charge gap and the neutral gap—each of which probes a distinct physical process and links to different experimental observables. The neutral gap is the minimum excitation energy obtained at fixed particle number (equivalently, fixed hole number in our hole representation; see Sec. \ref{sec:Model_definition}). Because the net charge is unchanged, this gap measures the cost of creating the lowest-energy collective mode formed by a bound quasi-electron–quasi-hole pair. In the fractional quantum Hall (FQH) effect, the canonical spin-conserving neutral mode is the magnetoroton \cite{PhysRevB.33.2481,Jain_2007,GirvinYang2019}, which have analogs in FCI \cite{PhysRevB.90.045114,wolf2024intrabandcollectiveexcitationsfractional,shen2025magnetorotonsmoirefractionalchern,long2025spectramagnetorotonchiralgraviton,lu2024interactiondrivenrotoncondensationc}. Spinful neutral excitations, including spin waves \cite{PhysRevLett.70.3983,PhysRevLett.73.3568} and spin-flip rotons \cite{PhysRevB.60.13702,PhysRevB.63.201310} have also been explored both theoretically and experimentally in FQH systems.  

Generically, we can define the charge gap $\Delta_c$ as the energy required to create a \emph{spatially} separated quasi-electron and quasi-hole excitations that may have any combination of quantum numbers --that is, an unbound pair of charge excitations that can move independently through the sample \cite{Jain_2007,GirvinYang2019}. Because these defects carry opposite charges once they are pulled apart, they are the lowest-energy excitations able to transport real charge over macroscopic distances (note that at this stage we are not specifying the quantum numbers of the lowest energy excitations, that will depend on the particular system). Consequently $\Delta_c$ sets the Arrhenius scale for the dc longitudinal conductivity, $\sigma_{xx} \propto e^{-\Delta_c/(2 k_B T)}$, where $T$ is the temperature and $k_B$ is the Boltzmann constant. When the Hall conductivity is quantized, at small temperatures we have $\sigma_{xy} \gg \sigma_{xx}$, which implies that $\rho_{xx} \propto \sigma_{xx}$, where $\rho_{xx}$ is the longitudinal resistivity, and therefore that $\rho_{xx} \propto e^{-\Delta_c/(2 k_B T)} $. Therefore, charge gaps are the key energy scales that determine the onset of 
dissipation in experiments measuring longitudinal resistance.

In the FQH regime, it is well established that the neutral and charge gaps typically track each other closely--differing by no more than a few percent--because the binding energy at the magnetoroton mininum is small compared to the large momentum particle-hole excitation energy \cite{PhysRevB.66.075408,PhysRevB.70.241307} . By contrast, in fractional Chern insulators the (moir\'e) lattice-scale structure can decouple these two gaps significantly \cite{PhysRevLett.132.236502,chen2025fractionalcherninsulatorquantum, shi2025effectsberrycurvatureideal,kousa2025theorymagnetorotonbandsmoire, paul2025shininglightcollectivemodes}. 
In particular, quasi-particle charge gaps have been extensively computed for FQH states \cite{Jain_2007,PhysRevLett.54.237,Yoshioka84,PhysRevB.34.2670,PhysRevB.34.5639,PhysRevB.33.2903,PhysRevB.55.7702,KPark_1999,PhysRevB.66.075408,PhysRevB.66.155320,PhysRevB.66.075408,PhysRevB.70.241307,PhysRevLett.100.166803} and compared with experimental results \cite{PhysRevLett.55.1606,PhysRevLett.65.2916,PhysRevB.37.8476,PhysRevLett.124.156801,PhysRevLett.70.2944,PhysRevLett.127.056801}. Despite inclusion of Landau level mixing and finite-width corrections, theoretical predictions typically overestimate experimental gaps, with discrepancies reaching over $50\%$ at $\nu = 1/3$ for two-dimensional electron systems confined in quantum wells \cite{PhysRevLett.127.056801}.
The discrepancies are typically attributed to disorder which creates localized in-gap states and induces spatial fluctuations in the potential landscape that can reduce the activation energy observed in transport measurements. 
Some early studies have also focused on characterizing the FQH quasi-particle excitations with one or more spin-flips relative to the polarized ground-state, showing that they can have lower energy than quasi-particle excitations with the same spin as the ground-state \cite{PhysRevB.33.2221,PhysRevB.36.5454,PhysRevB.41.10862,PhysRevLett.57.130,PhysRevB.64.125310,PhysRevB.80.045407,PhysRevLett.105.176802}. These are of direct relevance to experiments with tilted magnetic fields \cite{PhysRevLett.62.1540,PhysRevLett.62.1536,PhysRevB.41.7910,PhysRevB.45.3418,PhysRevLett.121.226801}, where the Zeeman energy can be tuned at fixed fillings.

While there have been extensive studies of spinless and spinful charged excitations in the context of FQH, the great majority of works in FCIs is focused on the calculation of neutral gaps and the study of neutral excitations \cite{PhysRevResearch.3.L032070,PhysRevLett.131.136501,PhysRevLett.131.136502,PhysRevB.108.245159,PhysRevB.108.085117,PhysRevB.107.L201109,MFCI_0,FTI_Yves,PhysRevLett.132.036501,PhysRevB.109.L121107,shen2024stabilizingfractionalcherninsulators,zaklama2025structurefactortopologicalbound,doi:10.1073/pnas.2316749121,PhysRevLett.133.066601}.
In the context of FCI, quasi-electron and quasi-hole excitations have been studied before \cite{PhysRevB.91.045126,PhysRevB.87.205136,PhysRevB.99.045136} with focus on their characterization, including e.g. their charge distribution, size and statistics. However, there has not been a previous attempt of a direct calculation of the quasi-particle charge gap, to our knowledge . 
In the following subsection, we introduce general definitions for energy dispersions and gaps associated with spinless and spinful charged excitations. These definitions can be directly applied both in the context of FCI and FQH. We will make the connection with FQH, where appropriate.

\subsection{Charge gaps}

\subsubsection{General definitions}

We first focus on the charge gaps. We will start by providing general definitions for a system with valley/spin $U(1)$ symmetry and with a fully polarized ground-state, and particularize later to $\nu=-2/3$. 
Start by defining $E_{\bf{k}}(pN,S_{z},N_s)$ as the ground-state energy for a system with $pN$ holes on $N_s$ unit cells in the spin $S_z$ and many-body momentum $\bf{k}$ sector. We want to probe excitations around a general hole filling $\nu_h \equiv -\nu = (N p)/( qN) = p/q$, where  $qN\rightarrow \infty$ in the thermodynamic limit. Consider now the effect of changing the number of holes and unit cells by $p'$ and $q'$ respectively. In the case of a spin-polarized ground state, it is also useful to consider the situation where $s$ holes with the same spin as the ground-state are removed and $s$ holes with the opposite spin polarization are added . Naturally, the quantities introduced above are all integers, $p,q,p',q',s \in \mathbb{Z}$. We can now check what is the effect of these changes on the hole filling fraction $\nu_h$. In particular, after the addition of particles and unit cells, the filling fraction $\nu_h =p/q$ can change to a new value $\nu_h^\prime$ which we determine by first calculating the changes in filling fraction for each spin sector  $\nu_{h,\sigma}\rightarrow\nu_{h,\sigma}^\prime$, where $\sigma=\uparrow,\downarrow$ and in our notation we assume that $\sigma=\uparrow$ corresponds to the spin polarization of the ground state . With these definitions we find that the changes in filling factor for each spin sector will be given by:

\begin{align}
    \Delta\nu_{h,\uparrow}&=\nu'_{h,\uparrow}-\nu_{h,\uparrow}=\frac{pN+p'-s}{qN+q'}-\frac{p}{q}=\frac{1}{N_{s}}\Bigg(\frac{p'q-q'p}{q}-s\Bigg) \\
    \Delta\nu_{h,\downarrow}&=\nu'_{h,\downarrow}-\nu_{h,\downarrow}=\frac{s}{N_{s}}.
\end{align}
\noindent where we identified $N_s=qN+q'$ as the total number of unit cells increased by $q'$ compared to the ground-state.
The total number of excited holes and difference in spin polarization with respect to the ground-state at filling $\nu_h=p/q$ is then given by
\begin{align}
    \delta N_h = N_s\Delta\nu_h&=N_s(\Delta\nu_{h,\uparrow}+\Delta\nu_{h,\downarrow})=\frac{e^{*}}{\bar{e}}\\
    \Delta S_{z}&=N_s\Big(\frac{\Delta\nu_{h,\uparrow}-\Delta\nu_{h,\downarrow}}{2}\Big)=\frac{e^{*}}{2 \bar{e}}-s
\end{align}
where we introduced the definition of the effective charge $e^*\equiv\bar{e}\frac{p' q - q' p}{q}$ ($\bar{e}=|e|$ is the hole charge, with $e<0$), the extra factor of $1/2$ in the definition of $\Delta S_{z}$ follows from the fact that we consider spin $1/2$ fermions. Note that as long as $\frac{q}{pq'} \notin \mathbb{Z}$ there is a unique value of $q'$ and $p'$ for each $e^*$ for fixed $p,q\in \mathbb{Z}$ since $p',q'\in \mathbb{Z}$. The cases that we consider below in which the charge, $e^*$, is fractional have a unique assignment of $q'$ and $p'$, and thus we can identify the values $p',q'$ and $s$ by only specifying $e^*$ and $s$. In cases where $e^* \in \mathbb{Z}$, we always chose $p'$ to be an integer and $q'=0$ so that for all cases below we need only to specify $\nu_h,s,e^*$. Furthermore, with the prescription that $p$ and $q$ are co-prime integers, the whole set $p,q,p',q',s$ is completely specified by $\nu_h,s,e^*$. We then find that excitations that can carry both spin and fractional charge can be added to the system by changing the number of spin up and spin down holes and the number of unit cells . The energy dispersion of these charge excitations can be computed as follows:
\begin{equation}
    \epsilon_{e^*,\nu_h}^s({\bf k},N)=\textrm{sgn}(e^*) \big(E_{{\bf k}}(pN+p',S_{z}^{\textrm{max}}-s,qN+q')-E_0 \big) \, ,
    \label{eq:eps_general}
\end{equation}
\noindent where $E_0=\langle E_{{\bf k} = {\bf k}_{FCI}}(pN,S_z^{\textrm{max}},qN) \rangle_{\Psi_{FCI}}$ is the average energy among the $q$ quasi-degenerate FCI ground-states on the torus, and $S_{z}^{\textrm{max}}=pN/2$ is the maximal $S_z$ for a system with $pN$ holes. We emphasize that the excitation charge $e^*$ is a function of $p/q,p'$ and $q'$, but we omitted the dependence for brevity. We note that to define the energy dispersions in Eq.$\,$\ref{eq:eps_general}, we adopt a "band picture" of excitations.  
This convention allows us to define charge gaps with different quantum numbers by subtracting the quasi-hole and quasi-particle dispersions, through:
\begin{equation}
    \Delta_{e^{*},\nu_h}^{s,s'}= \textrm{min}_{\bf{k},\bf{k}'} \, \big[ \textrm{sgn}(e^*)\big( \epsilon_{e^*,\nu_h}^s({\bf k},N)-\epsilon_{-e^*,\nu_h}^{s'}({\bf k}',N) \big) \big]
    \label{eq:charge_gap_general}
\end{equation}

\noindent where $\epsilon_{-e^*}^{s'}({\bf k},N)$ can simply be obtained from Eq.$\,$\ref{eq:eps_general} by replacing $p',q'\rightarrow-p',-q'$. The charge gaps in Eq.$\,$\ref{eq:charge_gap_general} correspond to the energy required to create well separated quasi-particles with opposite charges $e^*=\pm \bar{e}\frac{p' q - q' p}{q}$, giving rise to a net spin difference $\Delta S_z = -(s+s')$ with respect to the ground-state. Importantly, we note that the extraction of these charge gaps is only reliable for finite-size systems in the limit where the excitation is dilute, that is,  $p' \ll pN$ and $q' \ll qN$. We now simplify our notation to streamline the discussion to follow. First, since we work exclusively at filling  $\nu_h=2/3$ ($p=2,q=3$), we will omit the $\nu_h$ from the specification of the dispersions and charge gaps in Eqs.$\,$\ref{eq:eps_general} and \ref{eq:charge_gap_general} from this point on. We also identify cases  where there is a unique combination $s,s'$ that gives rise to a particular $\Delta S_z$. Overall, we have the following definitions:
\begin{align}
    \Delta_{e^{*}}^{0} &\equiv  \Delta_{e^{*},\nu_h=2/3}^{s=0,s'=0},\\
    \Delta_{e^{*}}^{-1,-} &\equiv  \Delta_{e^{*},\nu_h=2/3}^{s=1,s'=0},\\    
    \Delta_{e^{*}}^{-1,+} &\equiv  \Delta_{e^{*},\nu_h=2/3}^{s=0,s'=1},\\
    \Delta_{e^{*}}^{-2} &\equiv  \Delta_{e^{*},\nu_h=2/3}^{s=1,s'=1}\, ,
\end{align} 
\noindent that satisfy the relation $(\Delta_{e^*}^{-1,+}-\Delta_{e^*}^{0}) + (\Delta_{e^*}^{-1,-}-\Delta_{e^*}^{0}) = (\Delta_{e^*}^{-2}-\Delta_{e^*}^{0})$.
We will consider charge excitations with charges $e^*=e/3,2e/3,e$ and total spin $N_s\Delta S_z = 0,1,2$.
To simplify the notation, in the expressions that follow we will use the definition 
\begin{equation}
    E(pN,S_{z},N_s) \equiv \min_{\bf{k}} E_{\bf{k}}(pN,S_{z},N_s)
\end{equation}

The charge-$e$ gaps corresponding to set $p'=1$, $q'=0$ and $s,s'=0,1$ in Eq.$\,$\ref{eq:charge_gap_general}, are explicitly calculated as follows:
\begin{eqnarray}
\Delta_{e}^{0}=E(2N-1,S_{z}^{\textrm{max}},3N)+E(2N+1,S_{z}^{\textrm{max}},3N)-2E_0
\label{eq:charge_gaps_e1}\\
\Delta_{e}^{-1,-}=E(2N-1,S_{z}^{\textrm{max}}-1,3N)+E(2N+1,S_{z}^{\textrm{max}},3N)-2E_0
\label{eq:charge_gaps_e2}
\\
\Delta_{e}^{-1,+}=E(2N-1,S_{z}^{\textrm{max}},3N)+E(2N+1,S_{z}^{\textrm{max}}-1,3N)-2E_0 \label{eq:charge_gaps_e3}\\
\Delta_{e}^{-2}=E(2N+1,S_{z}^{\textrm{max}}-1,3N)+E(2N+1,S_{z}^{\textrm{max}}-1,3N)-2E_0.
\label{eq:charge_gaps_e4}
\end{eqnarray}
An important consequence of the spin of each excitation is that different Zeeman energies will contribute to the different gaps under a magnetic field $B$. The Zeeman energies are respectively $\Delta \epsilon_z = 0,g \mu_B B,2g \mu_B B$ for the gaps with $N_s \Delta S_z=0,-1,-2$, where $\mu_B=e\hbar/(2mc)$ is the Bohr magneton and $g$ is the effective g-factor.

We next move on to define the different charge gaps with fractional charge. These involve changing the number of unit cells in the system by $\pm 1$ and potentially also changing the number of particles. The charge $e/3$ gaps amount to take $p'=1$, $q'=1$, $s,s'=0,1$ in  Eq.$\,$\ref{eq:charge_gap_general} and are given by

\begin{eqnarray}
\Delta_{e/3}^{0}=E(2N-1,S_{z}^{\textrm{max}},3N-1)+E(2N+1,S_{z}^{\textrm{max}},3N+1)-2E_0
\label{eq:charge_gaps_eOver3_1}\\
\Delta_{e/3}^{-1,-}=E(2N-1,S_{z}^{\textrm{max}}-1,3N-1)+E(2N+1,S_{z}^{\textrm{max}},3N+1)-2E_0
\label{eq:charge_gaps_eOver3_2}\\
\Delta_{e/3}^{-1,+}=E(2N-1,S_{z}^{\textrm{max}},3N-1)+E(2N+1,S_{z}^{\textrm{max}}-1,3N+1)-2E_0
\label{eq:charge_gaps_eOver3_3}\\
\Delta_{e/3}^{-2}=E(2N-1,S_{z}^{\textrm{max}}-1,3N-1)+E(2N+1,S_{z}^{\textrm{max}}-1,3N+1)-2E_0
\label{eq:charge_gaps_eOver3_4}
\end{eqnarray}

Finally, the charge $2e/3$ gaps correspond to choosing $p'=0$, $q'=1$, $s,s'=0,1$ in  Eq.$\,$\ref{eq:charge_gap_general}. Explicitly, they are given by

\begin{eqnarray}
\Delta_{2e/3}^{0}=E(2N,S_{z}^{\textrm{max}},3N+1)+E(2N,S_{z}^{\textrm{max}},3N-1)-2E_0
\label{eq:charge_gaps_2eOver3_1}\\
\Delta_{2e/3}^{-1,-}=E(2N,S_{z}^{\textrm{max}}-1,3N+1)+E(2N,S_{z}^{\textrm{max}},3N-1)-2E_0
\label{eq:charge_gaps_2eOver3_2}\\
\Delta_{2e/3}^{-1,+}=E(2N,S_{z}^{\textrm{max}},3N+1)+E(2N,S_{z}^{\textrm{max}}-1,3N-1)-2E_0
\label{eq:charge_gaps_2eOver3_3}\\
\Delta_{2e/3}^{-2}=E(2N,S_{z}^{\textrm{max}}-1,3N+1)+E(2N,S_{z}^{\textrm{max}}-1,3N-1)-2E_0
\label{eq:charge_gaps_2eOver3_4}
\end{eqnarray}

\begin{figure}[h!]
\centering{}%
\includegraphics[width=0.85\columnwidth]{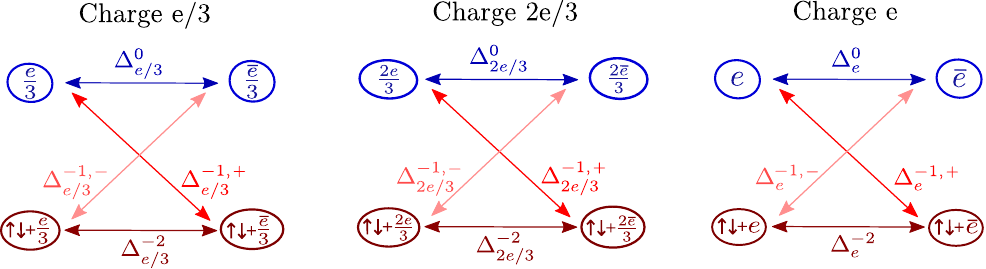}\caption{ Schematic of all the computed charge gaps and involved excitations. The arrows indicate the types of excitations involved in the corresponding energy gaps.
Using the same notations than Fig.$\,$\ref{fig:1}(a) of the main text, $\up \down + x$ is the combination of charge excitation ($x=e/3, 2e/3, e$) and a magnon-like excitation. The arrows, shown in different colors, represent the excitations associated with each gap. Each gap is labeled near its corresponding arrow using the same color. }\label{fig:gaps_summary_schematic}
\end{figure}

We provide a comprehensive schematic of all the computed charge gaps in Fig.$\,$\ref{fig:gaps_summary_schematic}.

\subsubsection{Connection with FQH and composite fermion picture} 
The definitions introduced in this section are general and can be applied to any system hosting integer or fractionalized excitations. In order to explore them in a more familiar context, here we take the FQH case and briefly make the connection to the composite fermion (CF) picture \cite{Jain_2007}. 
Assuming that each electron is attached by two fluxes, the magnetic fields felt by the electrons and composite fermions, that we denote respectively by $B$ and $B^*$, are related through \cite{Jain_2007}

\begin{align}
    B = B^* + 2\rho \phi_0 \\
    \leftrightarrow N^*_{\phi}=b^*(N_{\phi}-2pN)
    \label{eq:B_Bstar}
\end{align}

\noindent where $\phi_0=h c/\bar{e}$ is the flux quantum and we use here a notation consistent with the previous definitions, where $pN$ is the number of particles (electrons in this example, instead of holes in the previous definitions). $N_{\phi}=BA/\phi_0$ and $N^*_{\phi}=|B^*|A/\phi_0$ are respectively the number of flux quanta felt by the electrons and composite fermions and $b^*=\textrm{sgn}(B^*)$. 

The electron and CF fillings are given by $\nu=\rho \phi_0/B$ and $\nu^*=\rho \phi_0/|B^*|$. From Eq.$\,$\ref{eq:B_Bstar}, we obtain 

\begin{equation}
    \nu^* = \frac{b^*}{\nu^{-1}-2}
    \label{eq:nu_nustar}
\end{equation}
\noindent where $b^*=\textrm{sgn}(B^*)$. Choosing an integer filling for the composite fermions $\nu_0^*=p$ corresponds to an electronic fractional filling  $\nu_0=p/q$, with $q=b^*+2p$. We emphasize that here we are restricted to Jain's filling fractions specified by $p$ and $b^*$, so $q=q(b^*,p)=b^*+2p$.

We now inspect how changing the number of fluxes felt by the electrons by $q'$ and the number of electrons by $p'$ impacts the filling fraction of the composite fermions, $\nu^*$. The number of particles and number of fluxes are respectively changed to $pN+p'$ and $N'_{\phi}=qN+q'$, so that the electronic filling is changed to $\nu=(pN+p')/(qN+q')$. This implies that

\begin{equation}
    \Delta \nu^* = \nu^*-\nu_0^* = \nu^*-p  = \frac{b^*(pN+p')}{b^*N-2p'+q'}=\frac{p'q-q'p}{b^* N_{\phi}^{'*}}
    \label{eq:nustar_general_composite}
\end{equation}

\noindent where we identified $N_{\phi}^{'*}=b^*\big(N_{\phi}-2(pN+p')\big)=b^*(qN+q'-2pN-2p')=b^*(b^*N+q'-2p')$. The number of excited composite fermions relative to $\nu_0^*=p$ filled $\Lambda$-levels  -- the effective Landau levels that composite fermions occupy \cite{Jain_2007} --  is simply given by $\delta N^{'*} =  N_{\phi}^* \Delta \nu^* = b^*(p'q-q'p)$. It is important to appreciate that we can excite a non-trivial number of composite fermions $\delta N^{'*}$ that is not trivially equal to the number of added or removed electrons, but is also a function of the number of added or removed fluxes. The reason for this non-trivial behavior is that the degeneracy of the $\Lambda$ levels also depends on $p'$ and $q'$. In other words, the total number of CF that fully fill $p$ $\Lambda$ levels is a function of $p'$ and $q'$, leaving a leftover charge of CF that is not simply equal to the number of added/removed electrons. 
In the same way, we have for the electrons

\begin{equation}
    \Delta \nu = \nu-\nu_0= \frac{p'q-q'p}{q N'_{\phi}}
    \label{eq:nu_general_composite}
\end{equation}

\noindent from which we can see that in the electrons, the fraction of charge $e^*/\bar{e}$ defined in the previous section is excited. 

Eqs.$\,$\ref{eq:nustar_general_composite} and \ref{eq:nu_general_composite} imply that adding $q'$ fluxes and $p'$ particles with respect to the filling $\nu=p/q$ corresponds to excite an integer charge of CF quasi-particles $e^*_{cf} = e \, b^*(p'q-q'p)$ and a fraction of this charge $b^* e^*_{cf}/q$ of electronic quasi-particles. 

We now take specific examples. 

\paragraph*{\textbf{Spinless excitations at $\nu=1/3$.---}} We first consider excitations with respect to the standard case of a polarized FQH state at $\nu=1/3$. This state corresponds to  choosing $p=1$ and $b^*=1$ ($B^*$ pointing in the ${\bf e}_z$ direction). In this case, adding or removing a single flux ($p'=0,q'=\pm 1$) corresponds to adding a charge $e^*_{cf}=\mp e$, that is, creating a single CF quasi-hole or quasi-electron. The minimal (charge-$e/3$) excitations around $\nu=1/3$ can therefore be obtained simply by changing the number of fluxes.

\begin{figure}[h!]
\centering{}%
\includegraphics[width=0.8\columnwidth]{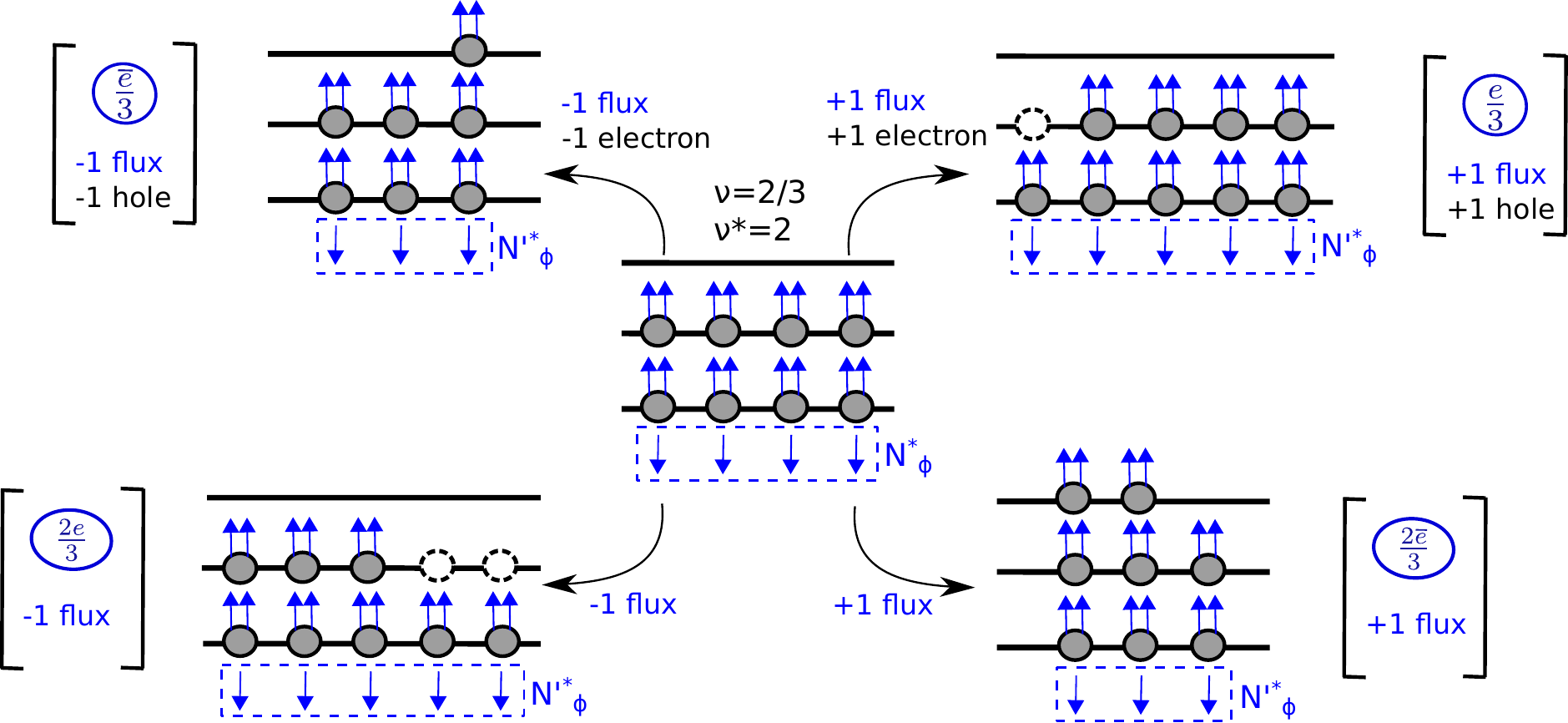}\caption{Illustration of the CF picture of spinless excitations at electronic filling $\nu=2/3$. The corresponding CF filling is $\nu^*=2$, that is, two fully filled CF $\Lambda$-levels (central illustration). We take this illustrative example with $8$ electrons, that is, $N=4$.   Positive and negative fluxes are respectively represented by up and down arrows, in blue. Since the total number of fluxes at $\nu=2/3$ is $N_{\phi}=3N=12$ and  each electron captures two fluxes, that is, $4N=16$ fluxes in total, $N^*_{\phi}=N=4$ unattached negative fluxes remain. The CFs feel a negative magnetic field $B^*=-N^*_{\phi} \phi_0/A \,{\bf e_z}$ due to these negative unattached fluxes, responsible for the creation of the $\Lambda$-levels. Adding or removing a single flux (or equivalently removing or adding a negative flux) with respect to the $\nu^*=2$ case in the central illustration changes the $\Lambda$-level degeneracy and is equivalent to respectively creating two quasi-electrons or quasi-holes (left and right bottom illustrations). The minimal excitations, a single quasi-hole or quasi-electron are created respectively by adding or removing both a single flux and single electron (left and right top illustrations). For each CF excitation, we indicate the corresponding hole excitation in MoTe$_2$ --- as depicted in Fig.~\ref{fig:gaps_summary_schematic} --- within square brackets.  }\label{fig:CF_analogy_2o3}
\end{figure}

\paragraph*{\textbf{Spinless excitations at $\nu=2/3$.---}}  We now consider the example of a polarized FQH state at $\nu=2/3$, to understand why changing only the number of fluxes is not sufficient to obtain the minimal (charge-$e/3$) excitations. We note that in a Landau level setup, the ground-state at $\nu=2/3$ is an unpolarized spin-singlet at small enough Zeeman energy \cite{PhysRevB.30.7320,PhysRevB.40.3487,Jain_2007,PhysRevB.45.3418,PhysRevLett.82.3665,Verdene2007}. For this example, we will however consider a spin-polarized FQH state, which is more closely related to the fully polarized FCI observed at $\nu_h=2/3$ for twisted MoTe2. The spin-polarized FQH state can be reached by increasing the Zeeman energy at fixed filling by tilting the magnetic field \cite{PhysRevLett.62.1540,PhysRevLett.62.1536,PhysRevB.41.7910,PhysRevB.45.3418,PhysRevLett.121.226801}.  Focusing on this state, it corresponds to choosing $p=2$ and $b^*=-1$ ($B^*$ pointing in the $-{\bf e}_z$ direction). In this case, adding or removing a single flux ($p'=0,q'=\pm 1$) corresponds to exciting a CF charge $e^*_{cf}=\pm 2e$, that is, exciting two CF quasi-particles into the next empty $\Lambda$ level or creating two empty orbitals in the highest filled $\Lambda$ level, as illustrated in Fig.$\,$\ref{fig:CF_analogy_2o3}. In order to create a single empty orbital in the highest filled $\Lambda$ level, we need to add a single flux and a single electron ($p'=q'=1$), so that $e^*_{cf}=\bar{e}$.  On the other hand, to excite a single CF quasi-electron to the next $\Lambda$ level, we need to remove a single flux and a single electron  ($p'=q'=-1$), so that $e^*_{cf}=e$. The example of $\nu=2/3$ is illustrated in Fig.$\,$\ref{fig:CF_analogy_2o3}.

\paragraph*{\textbf{Spinful excitations, general case.---}}  In the case of spinful excitations, the composite fermions occupy opposite-spin $\Lambda$ levels. In particular, assuming again that we add $p'$ electrons and $q'$ fluxes, for spin-$s$ excitations the number of electrons in the spin-up sector simply changes from $pN+p'$ to $pN+p'-s$, and the missing $s$ electrons populate the spin-down sector. The number of fluxes felt by the CF, on the other hand, is not changed since it depends on the total number of particles, which is still $pN+p'$. Therefore, in the spin-up sector, the equivalent of Eq.$\,$\ref{eq:nustar_general_composite} is  

\begin{equation}
    \Delta \nu_{\up}^* = \nu_{\up}^*-p  = \frac{b^*(p'q-q'p)-s}{ N_{\phi}^{'*}}
    \label{eq:nustar_general_composite_spinful}
\end{equation}

\noindent which implies that a number $\delta N_{\up}^{'*}=b^*(p'q-q'p)-s$ CF is excited in the spin-up sector with respect to $p$ filled $\Lambda$-levels, while $\delta N_{\down}^{'*}=s$ CF are excited in the spin-down sector. In other words, $s$ CF are excited from the spin-up sector into the spin-down sector.

\begin{figure}[h!]
\centering{}%
\includegraphics[width=1\columnwidth]{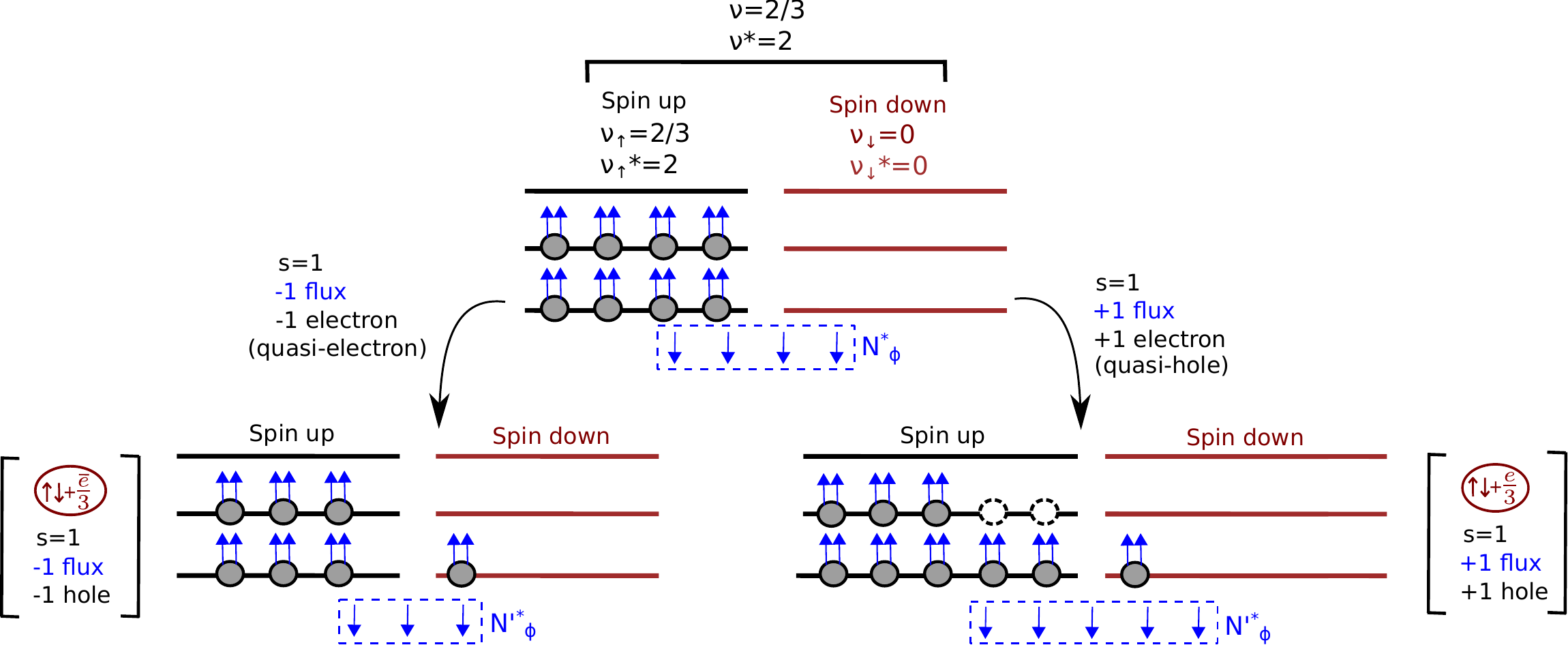}\caption{ Illustration of the CF picture of spin-$1$ excitations at electronic filling $\nu=2/3$. The spin-up (spin-down) CF $\Lambda$ levels are shown respectively in black (red). Positive and negative fluxes are respectively represented by up and down arrows, in blue.  The excitations involving adding or removing both a flux and an electron, accompanied by a spin-flip, are the simplest CF spinful excitations. In particular, they involve filling the opposite spin $\Lambda$ level with a single CF. For each CF excitation, we indicate the corresponding hole excitation in MoTe$_2$ --- as depicted in Fig.~\ref{fig:gaps_summary_schematic} --- within square brackets.}\label{fig:CF_analogy_2o3_spinful}
\end{figure}

\paragraph*{\textbf{Spin-1 excitations at $\nu=2/3$.---}} Taking the example of $\nu=2/3$ with $s=1$, in the quasi-hole channel ($p'=q'=1$) we have $e^{*\up}_{cf} = 2\bar{e}$ and $e^{*\down}_{cf}=e$. In the quasi-electron channel ($p'=q'=-1$), we have instead $e^{*\up}_{cf} = 0$ and $e^{*\down}_{cf}=e$. These are the minimal spinful excitations for the composite fermions, that we sketch in Fig.$\,$\ref{fig:CF_analogy_2o3_spinful}. The interested reader is referred to  Refs.$\,$\cite{Morf1987,PhysRevB.49.5085,PhysRevB.64.125310,onoda2001,Jain_2007,PhysRevB.80.045407} for detailed studies of different kinds of spinful excitations in FQH. 

\subsection{Neutral gaps}
 We will focus on two types of neutral gaps:
\begin{itemize}
    \item  The spin-0  neutral gap $\Delta_N$ (more commonly known as the FCI or many-body gap), defined as the excitation energy in the fully polarized sector $S_z = S_z^{\textrm{max}}$;
    \item The spin-1 neutral gap $\Delta_N^s$, defined as the excitation energy to the ground-state in the sector with one spin-flip with respect to the polarized sector $S_z = S_z^{\textrm{max}}-1$. 
\end{itemize}

Explicitly, we define these neutral gaps as 

\begin{equation}
    \Delta_N = \min\nolimits_{\mathbf{k}}^{\prime}\Big( E_{\mathbf{k}}(pN, S_z^{\text{max}}, qN) - \bar{E}_{\text{FCI}} \Big)
    \label{eq:Delta_N}
\end{equation}

\begin{equation}
    \Delta_N^s = \min_{\bf{k}} \Big( E_{ \bf{k} }(pN,S_z^{\textrm{max}}-1,qN)-E_0 \Big)
    \label{eq:Delta_N_s}
\end{equation}

\noindent where $\min'_{\bf{k}}  E_{\bf{k}}$ is defined as the mininum many-body energy excluding the energies of the FCI ground-states.

\begin{figure}[h!]
\centering{}%
\includegraphics[width=0.4\columnwidth]{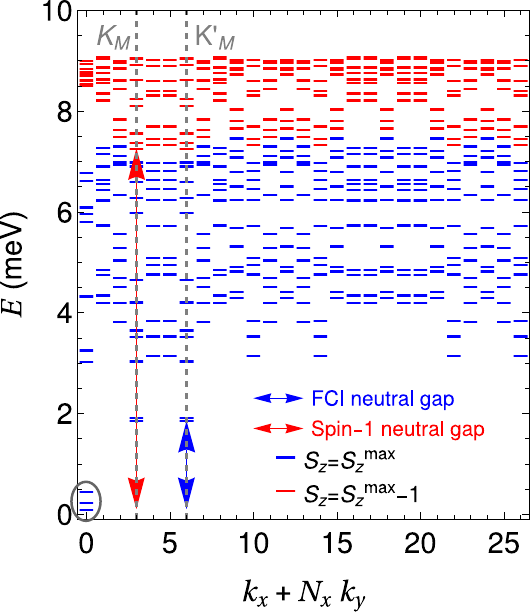}\caption{ Illustration of neutral FCI and spin-1 gaps from exact diagonalization spectra obtained for spin sectors $S_z = S_z^{\textrm{max}}$ and $S_z = S_z^{\textrm{max}}-1$, respectively. Results for $u=10/\epsilon=0.9$ and the 27 unit cell lattice specified by  $(N_x,N_y,n_{11},n_{12},n_{21},n_{22})=(9,3,1,1,0,1)$ (see Sec.$\,$\ref{sec:1BPV} for the full definition of this specification). The lowest energy is shifted to be zero. The quasi-degenerate FCI ground-states are highlighted by the gray circle and the moiré Dirac points $\bsl{K}_M$ and $\bsl{K}'_M$ by the gray dashed lines. We only show the 12 and 8 lowest eigenvalues for each momentum sector for $S_z = S_z^{\textrm{max}}$ and $S_z = S_z^{\textrm{max}}-1$.  }\label{fig:spectra_27site}
\end{figure}

The spin-0 and spin-1 neutral gaps are illustrated in Fig.$\,$\ref{fig:spectra_27site}, which shows the ED spectrum for MoTe$_2$, for a 27 unit cell system .

\subsection{FCI vs. FQH excitation gaps}

We conclude this section by comparing the excitation gaps computed in this work for MoTe$_2$ at filling $\nu = -2/3$ with the corresponding gaps for the spin-polarized FQH state at $\nu = 1/3$ in the well-studied spin-$SU(2)$ symmetric lowest Landau level (LLL) case (at zero Zeeman energy), using unscreened (pure) Coulomb interaction. We recall that we use a screened Coulomb interaction for MoTe$_2$. However, for the gate distance used in this work ($\xi=20\,nm$) we expect the results not to differ significantly from those obtained with a pure Coulomb interaction. Detailed results and discussion on the excitation gap calculations in the FCI case are provided in Sections~\ref{sec:1BPV} and~\ref{sec:2BPV}. Here, we summarize only the best estimates for each excitation gap, which are also reported in Fig.~\ref{fig:1}(c) of the main text (see also the main text for additional details on the methods to obtain the best estimates). A comparison between different excitation gaps in the FCI and FQH cases is provided in Tab.$\,$\ref{tab:FCI_vs_FQH}. 

It is important to emphasize that there is a key difference between the FCI and FQH cases here considered: the FQH state has full spin-$SU(2)$ symmetry, while the FCI state in MoTe$_2$ retains only spin-$U(1)$  symmetry due to the ising-SOC of the system. As a result, the magnon excitation is gapless in the FQH case, but generically gapped for MoTe$_2$.

The magnetoroton gap is $[20-40] \%$ smaller than the spin-0 charge gap in the FQH case, while for the FCI it is $59 \%$ and $25 \%$ smaller respectively for our one band per valley (1BPV) and two band per valley (2BPV) calculations. 
The ratio between the binding energy of a quasi-electron/quasi-hole pair with the same spin as the ground state and the spinless charge gap (i.e., the energy of the pair when it is well-separated) is therefore substantially larger in the FCI case than in the FQH case, within the 1BPV calculation. In contrast, this ratio becomes comparable to the LLL result in the 2BPV calculation.

In contrast, the hierarchy between the spin-1 and spin-0 charge gaps is reversed in the FQH and FCI cases. For the FQH state, the spin-1 charge gap to create a magnon-dressed quasi-electron well separated from a quasi-hole with the same spin as the ground-state is approximately $25\%$–$37\%$ smaller than the spin-0 charge gap. In the FCI case, on the other hand, the spin-1 charge gap exceeds the spin-0 gap both for the 1BPV and 2BPV calculations, as detailed in the main text.

\begin{table}[]
    \centering

    \resizebox{0.8\columnwidth}{!}{
    \begin{tabular}{|c|c|c||c|c|c|}
    \hline 
    Excitation type & $Q$ & $s$ & \makecell{FCI, $\Delta_{\textrm{1BPV}}$ \\ (meV)}    & \makecell{FCI, $\Delta_{\textrm{2BPV}}$ \\ (meV)} & \makecell{FQH, LLL ($\nu=1/3$) \\ ($e^2/(\epsilon l_B)$)}   \tabularnewline
    \hline 
    \hline 
    Magnetoroton & 0 & 0 & 1.6 & \textcolor{black}{\NeutralSpinZeroEstimateTwoBPV}
    & $ [0.06-0.08]$ \cite{PhysRevLett.54.237,PhysRevB.33.2481,PhysRevB.34.2670, PhysRevB.95.075201,Majumder2009}
    \tabularnewline
    \hline 
    Magnon & 0 & 1 & 7.0 &  \NeutralSpinOneEstimateTwoBPV 
    & $0$
    \tabularnewline
    \hline 
    Charge-$e/3$ spin-0 & $e/3$ & 0 & \ChargeSpinZeroEstimateOneBPV & \ChargeSpinZeroEstimateTwoBPV  
    & $\approx 0.10$ \cite{PhysRevLett.54.237, PhysRevB.34.2670, PhysRevB.66.075408,PhysRevLett.100.166803}
    \tabularnewline
    \hline 
    \makecell{Charge-$e/3$ spin-1 \\ (qh + magnon-dressed qp)} & $e/3$ & 1 & \ChargeSpinOneEstimateOneBPV & \ChargeSpinOneEstimateTwoBPV 
    & $ [ 0.063-0.075]  $ \cite{Morf1987, PhysRevB.36.5454, PhysRevB.64.125310}
    \tabularnewline
    \hline 
    \end{tabular}
    
    }
\label{tab:FCI_vs_FQH}
\caption{ 
Comparison of our best estimates for different excitation gaps in MoTe$_2$ at filling $\nu = -2/3$, twist angle $\theta = 3.7^{\degree}$, and interaction strength $u = 0.9$, with corresponding excitation gaps in the FQH state at $\nu = 1/3$ in the lowest Landau level. The FQH results were obtained in the zero Zeeman energy limit using a pure (unscreened) Coulomb interaction, without considering neither Landau level mixing nor finite-width corrections. All the references from which the FQH results were extracted are indicated in the table. For all the references, spherical geometry was used, except for Ref.$\,$\cite{PhysRevB.95.075201}, that also considered the torus geometry.
}

\end{table}

\clearpage

\section{1BPV Results}

\subsection{Neutral and charge gaps}

\label{sec:1BPV}

In this section we present detailed results for the neutral and charge gaps at $\nu=-2/3$ using one band per valley (1BPV) ED calculations, i.e. using the Hamiltonian in Eq.$\,$\ref{eq:H_full_band_basis} projected to the topmost valence band for each valley. To minimize  finite-size effects we use tilted boundary conditions \cite{PhysRevLett.111.126802, PhysRevB.90.245401} with momentum meshes that have the best aspect ratios  --- that is, as close as possible to the aspect ratio of momentum meshes that preserve the symmetries of the model, as detailed below. Specifically, the first harmonic model of MoTe$_2$ here considered has $C_3$ and inversion symmetries \cite{MFCI_I}, which together imply $C_6$ symmetry.

We consider momentum meshes defined by the two reciprocal lattice vectors $\bsl{f}_{1}$ and $\bsl f_{2}$ given by

\begin{equation}
\begin{array}{cc}
     \bsl{f}_{1} & =n_{11}\bsl{b}_{M,1}+n_{12}\bsl{b}_{M,2}  \\
     \bsl f_{2}& =n_{21}\bsl b_{M,1}+n_{22}\bsl b_{M,2} \, .
\end{array}
\end{equation}

\noindent where $\bsl{b}_{M,1}$ and $\bsl{b}_{M,2}$ are the moiré reciprocal lattice vectors and
$n_{11},n_{12},n_{21},n_{22}$ are integers satisfying $|n_{11}n_{22}-n_{12}n_{21}|=1$. In other words, we perform an $SL(2,\mathbb{Z})$ transformation of the infinite lattice, and then choose periodic boundary conditions.  We take momentum meshes of $N_{x},N_{y}$ momenta along the directions $\bsl f_{1}$ and $\bsl f_{2}$ respectively, that is,

\begin{equation}
    \bsl{k} = \frac{k_x}{N_x} \bsl{f}_1 + \frac{k_y}{N_y} \bsl{f}_2 \, ,
\end{equation}

\noindent corresponding to a system with $N_s=N_x N_y$ unit cells. In order to define the aspect ratio, we first define the real-space lattice vectors $\bsl{T}_1$  and $\bsl{T}_2$ dual of $\bsl{f}_1$ and $\bsl{f}_2$, that is, satisfying $\bsl{T}_i \cdot \bsl{f}_j = 2\pi \delta_{i,j}$. The aspect ratio is then defined as

\begin{equation}
\mathrm{AR} = 
\min_{\substack{M \in SL(2,\mathbb{Z}) \\ (\boldsymbol{T}_1',\, \boldsymbol{T}_2')^\top = M\, (\boldsymbol{T}_1,\, \boldsymbol{T}_2)^\top}}
\frac{
\min\left(\|\boldsymbol{T}_1'\|,\, \|\boldsymbol{T}_2'\|\right) \cdot \sin \theta(\boldsymbol{T}_1',\, \boldsymbol{T}_2')
}{
\max\left(\|\boldsymbol{T}_1'\|,\, \|\boldsymbol{T}_2'\|\right)
}
\,, \label{eq:AR}
\end{equation}

\noindent
where \( \theta(\boldsymbol{T}_1', \boldsymbol{T}_2') \) is the angle between the two vectors. Eq.~\eqref{eq:AR} computes the aspect ratio by considering all lattice vectors related to \( (\boldsymbol{T}_1, \boldsymbol{T}_2) \) via \( SL(2,\mathbb{Z}) \) transformations and selecting the ones that form the most isotropic parallelogram — that is, the one with the optimal (largest) aspect ratio. We note that with this definition, $C_6$ symmetric lattices have an aspect ratio of AR=$\sqrt{3}/2$, so in our calculations we aim to choose momentum meshes with aspect ratios close to this value.

\begin{figure}[h!]
\centering{}%
\includegraphics[width=0.85\columnwidth]{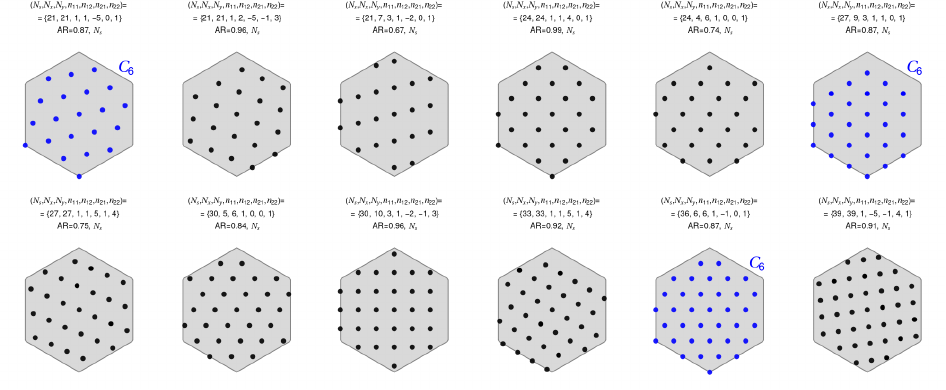}\caption{Momentum meshes used for the 1BPV calculations of the neutral gaps, and corresponding aspect ratio (AR). The $C_6$ symmetric meshes are highlighted in blue. }\label{fig:meshes_neutral_1BPV}
\end{figure}

Using these conventions, we performed extensive numerical calculations for different system sizes in order to properly account for the finite-size dependence. All the momentum-space meshes we used for the calculations are shown in Figs.$\,$\ref{fig:meshes_neutral_1BPV}-\ref{fig:meshes_qp_charge_1BPV_2}.

\begin{figure}[h!]
\centering{}%
\includegraphics[width=0.725\columnwidth]{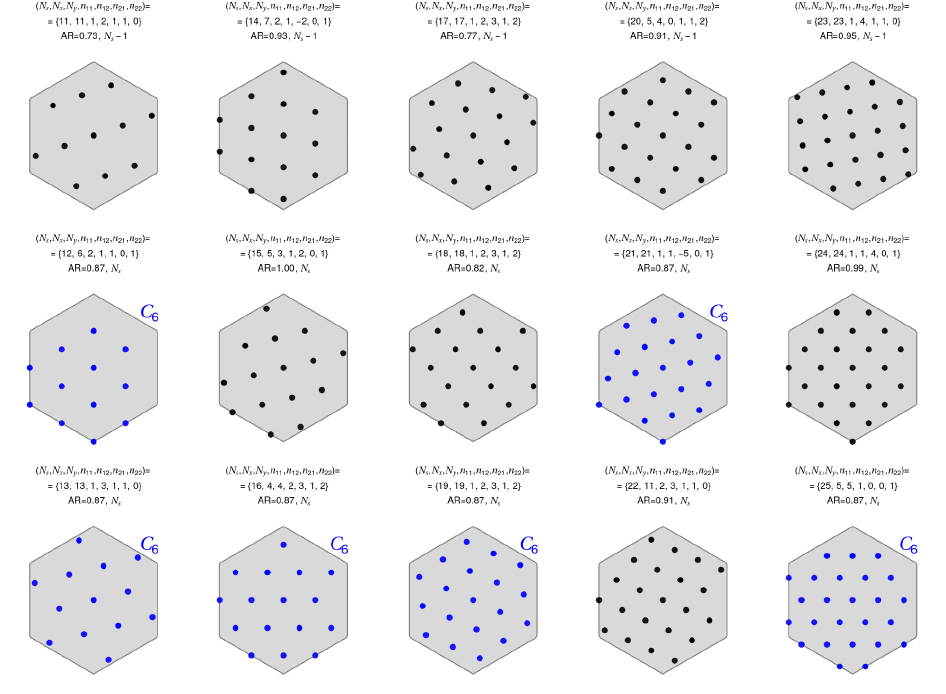}\caption{Momentum meshes used for the 1BPV calculations of the quasi-particle charge gaps, and corresponding aspect ratio (AR). The calculations for the charge $e/3$ and $2e/3$ quasi-particle charge gaps involve three different system sizes, with $N_s-1,N_s$ and $N_s+1$ unit cells, see Eqs.$\,$\ref{eq:charge_gaps_eOver3_1}-\ref{eq:charge_gaps_2eOver3_4}. These are respectively shown from the top to bottom rows, while each column corresponds to the sizes used in a given calculation. For the charge-$e$ gaps, only the number of particles is varied (see Eqs.$\,$\ref{eq:charge_gaps_e1}-\ref{eq:charge_gaps_e4}) and therefore only the sizes in the middle row were used. The $C_6$ symmetric meshes are highlighted in blue. The remaining used lattices are shown in Fig.$\,$\ref{fig:meshes_qp_charge_1BPV_2}.}\label{fig:meshes_qp_charge_1BPV_1}
\end{figure}

\begin{figure}[h!]
\centering{}%
\includegraphics[width=0.575\columnwidth]{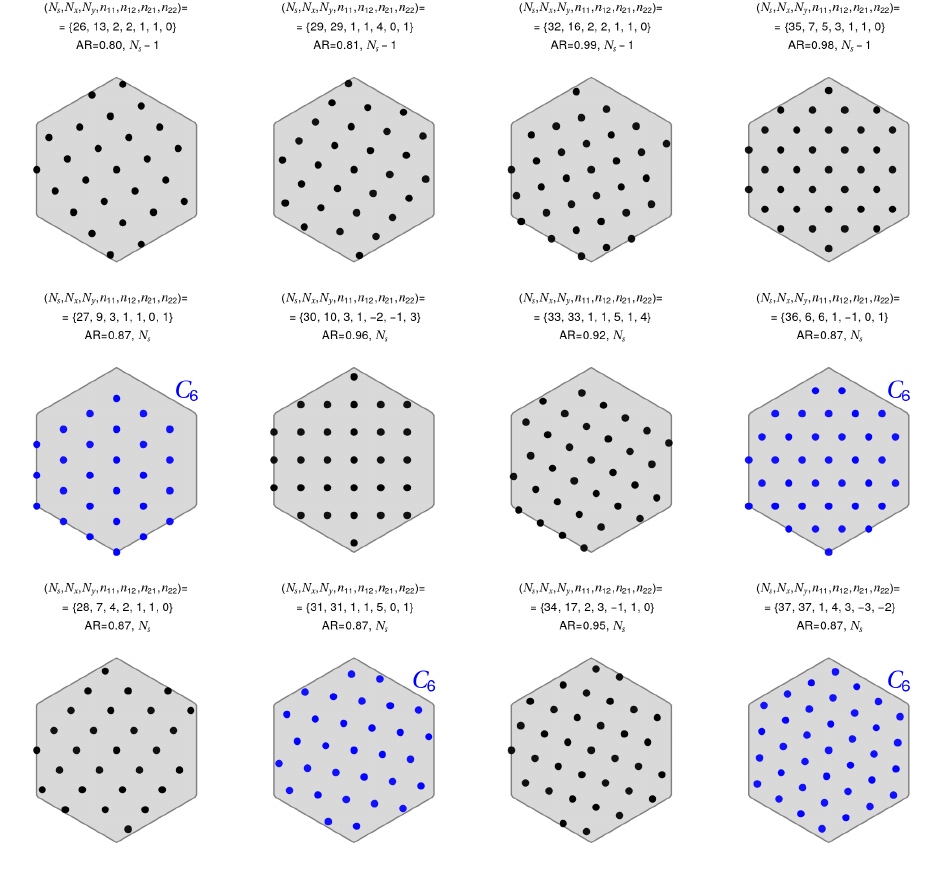}\caption{Continuation of Fig.$\,$\ref{fig:meshes_qp_charge_1BPV_1}, showing the remaining momentum meshes used for the calculation of quasi-particle charge gaps.}\label{fig:meshes_qp_charge_1BPV_2}
\end{figure}

\clearpage

To study the competition between all the energy gaps defined in Sec.$\,$\ref{sec:gaps}, we focus on two cases:
\begin{itemize}
    \item A reasonable value of the dielectric constant, fixing $u=10/\epsilon=0.9$;
    \item Flatband limit, where we neglect the energy band dispersion, i.e. only consider the two-body term in Eq.$\,$\ref{eq:H_full_band_basis}, and take $u=1$.
\end{itemize}

\begin{figure}[h!]
\centering{}%
\includegraphics[width=\columnwidth]{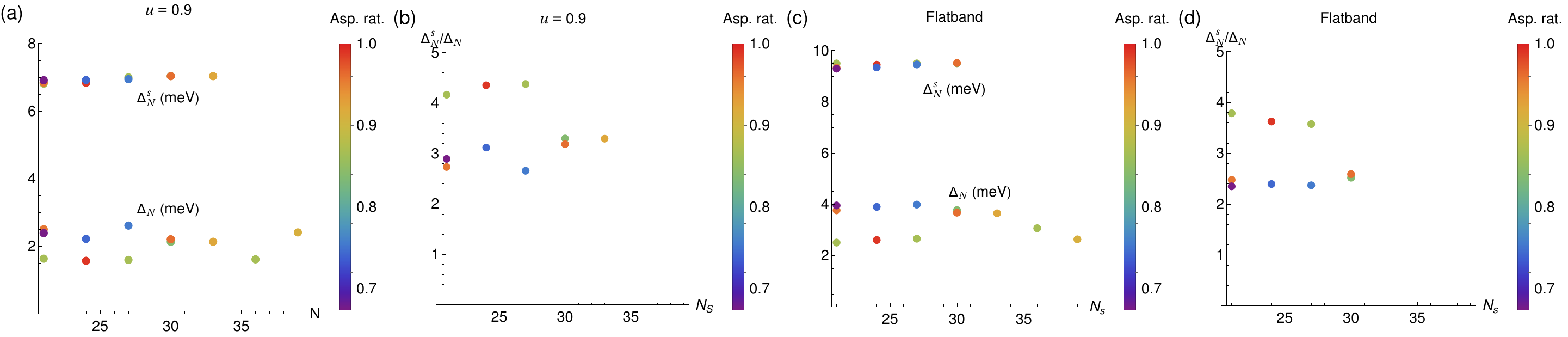}\caption{Finite-size scaling results for the neutral gaps, using 1BPV ED. Results for $u=10/\epsilon=0.9$ (a-b) and for the flatband limit (c-d). All the momentum meshes used for these calculations are shown in Fig.$\,$\ref{fig:meshes_neutral_1BPV}.}\label{fig:gaps_neutral_u0p9_1BPV}
\end{figure}

\begin{figure}[h!]
\centering{}%
\includegraphics[width=0.85\columnwidth]{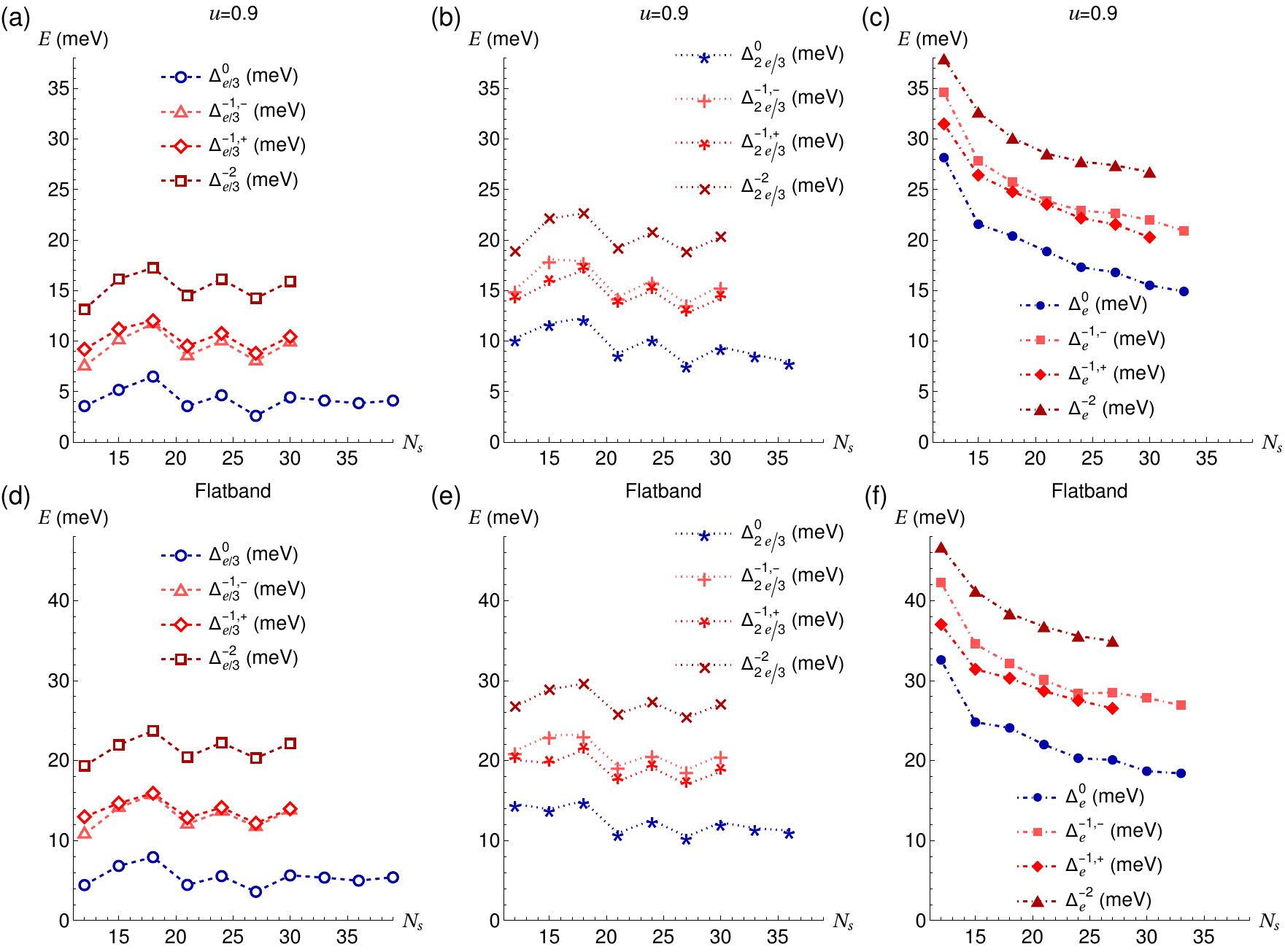}\caption{Finite-size scaling results for all the charge gaps with charge $e^*=e/3,2e/3,e$, using 1BPV ED. Results for $u=0.9$ (a-c) and for the flatband limit (d-f). All the momentum meshes used for these calculations are shown in Figs.$\,$\ref{fig:meshes_qp_charge_1BPV_1} and $\,$\ref{fig:meshes_qp_charge_1BPV_2}.}\label{fig:gaps_charge_u0p9_1BPV}
\end{figure}

We summarize our 1BPV results in Figs.$\,$\ref{fig:gaps_neutral_u0p9_1BPV} and \ref{fig:gaps_charge_u0p9_1BPV}, where we show the finite-size scaling results for all the neutral and charge gaps. For reference, in Tab.$\,$\ref{tab:dimH_1BPV}, we show the maximum system sizes and Hilbert space sizes used for the main calculations shown in the main text and in this section.

\begin{table}[h]
\centering
\begin{tabular}{|c|c|c|c|c|c|c|}
\hline
& $\Delta_N$ & $\Delta_N^s$ & $(e/3,0)$ &   $(\bar{e}/3,0)$ &   $(e/3,1)$ & $(\bar{e}/3,1)$ \\
\hline
dimH & $2 \times 10^8$ &   $3.5 \times 10^8$    &    $1.4\times 10^8$      &    $3\times 10^8$                &   $3.5\times 10^7$       &     $8.5\times 10^7$     \\
\hline
$N_s$ & 39 &    33    &   38       &    40      &      29    &   31                \\
\hline
\end{tabular}
\caption{Largest Hilbert space sizes (dimH) and corresponding  $N_s$ used in the 1BPV calculations, for the neutral gaps and different types of charge excitations defined by $(e^*,s)$ (see Eq.$\,$\ref{eq:eps_general}).}
\label{tab:dimH_1BPV}
\end{table}

In Fig.$\,$\ref{fig:gaps_neutral_u0p9_1BPV}(a,b) we show the finite-size scaling results for the spin-0 and spin-1 neutral gaps, for $u=0.9$. The spin-1 neutral gap, $\Delta_N^s$, quickly becomes system size independent with relatively small fluctuations below $1.3\%$ across the largest three sizes, as evidenced in Fig.$\,$\ref{fig:gaps_neutral_u0p9_1BPV}(a). In contrast, the spin-0 neutral gap $\Delta_N$ exhibits larger fluctuations as a function of system size and aspect ratio. However, the overall hierarchy 
$\Delta_N^s > \Delta_N$ is robust as shown in Fig.$\,$\ref{fig:gaps_neutral_u0p9_1BPV}(b). For the best estimates of these gaps obtained --- $N_s=39$ for $\Delta_N$ and $N_s=33$ for $\Delta_N^s$ ---  shown in Fig.$\,$\ref{fig:1}(c) in the main text --- we obtain $\Delta_N^s/\Delta_N=\NeutralSpinOneSpinZeroRatioOneBPV$. 

To assess the robustness of these results against changes in the band dispersion, we perform calculations in the flat-band limit, where the kinetic energy is completely quenched. The results of these calculations are shown in Fig.$\,$\ref{fig:gaps_neutral_u0p9_1BPV}(c,d).  Note that we cannot directly compare the values of the gaps between the $u=0.9$ case and the flat-band case, since in the latter the only remaining energy scale being set by interactions. However, we can compare the gap ratios. In the flat-band limit, the ratio of the best estimates for the spin-0 and spin-1 neutral gaps ($N_s=39$ for $\Delta_N$ and the $N_s=30$ size with better aspect ratio for $\Delta_N^s$) changes to $\Delta_N^s / \Delta_N = \NeutralSpinOneSpinZeroRatioFlatbandOneBPV$, remaining large and therefore not being very sensitive to changes of the dispersion. Importantly, we emphasize that, in this Appendix, we consider the occupation of the remote bands to remain "frozen".
Multiband effects are systematically studied in Appendix$\,$\ref{sec:2BPV} below, where we find that the gap ratio is substantially changed with respect to this 1BPV calculation.

\begin{figure}[h!]
\centering{}%
\includegraphics[width=\columnwidth]{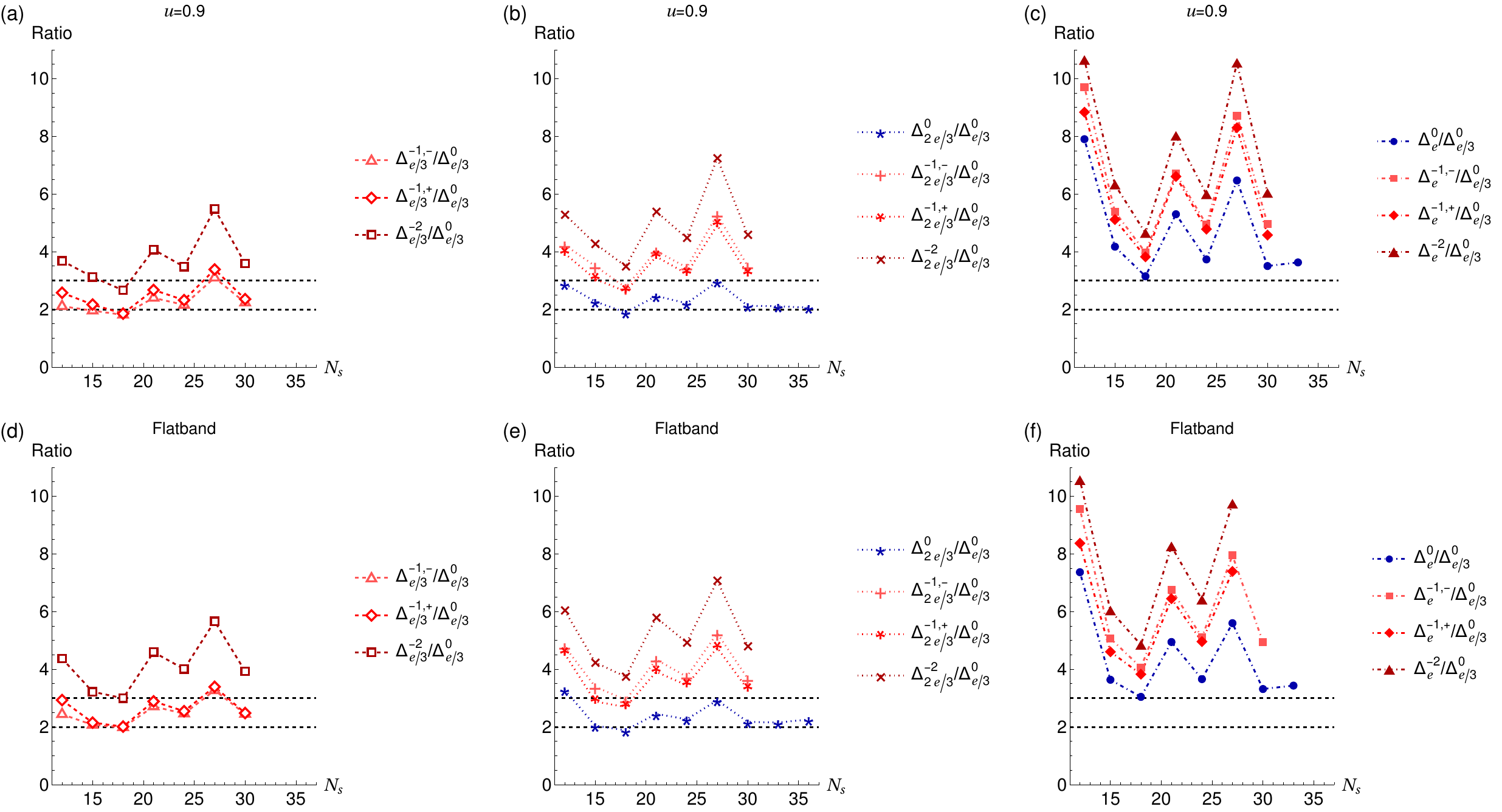}\caption{Ratios between all the gaps and the smallest gap $\Delta_{e/3}^0$, for $u=0.9$ (a-c) and in the flatband limit (d-f). All the momentum meshes used for these calculations are shown in Figs.$\,$\ref{fig:meshes_qp_charge_1BPV_1} and $\,$\ref{fig:meshes_qp_charge_1BPV_2}.}\label{fig:gaps_charge_u0p9_1BPV_ratios}
\end{figure}

 We present finite-size scaling results for charge gaps with quasi-particle charge $e^* = e/3$, $2e/3$, and $e$ in the $u=0.9$ and flat-band cases in Fig.~\ref{fig:gaps_charge_u0p9_1BPV}. The charge-$e/3$ gaps shown in Figs.~\ref{fig:gaps_charge_u0p9_1BPV}(a,d) exhibit
good convergence as a function of the the system size. For the spin-0 gap, $\Delta_{e/3}^0$, a relative variation of only $6\%$ and $7\%$ is observed across the three largest sizes respectively for $u=0.9$ and in the flat-band case. 
The charge-$2e/3$ and charge-$e$ gaps shown in Figs.~\ref{fig:gaps_charge_u0p9_1BPV}(b-c,e-f) exhibit stronger finite-size effects, with the worst case scenario occurring for the charge-$e$ gaps that consistently decrease with $N_s$ for the available system sizes.
This is expected since the charge-$2e/3$ and charge-$e$ gaps correspond to adding respectively two or three quasi-particles, which are harder to accommodate on a given (not large enough) system size compared to a single excitation. 
In Fig.~\ref{fig:gaps_charge_u0p9_1BPV_ratios} we show the ratios between all the gaps and the smallest charge gap $\Delta_{e/3}^0$. Interestingly,  we can see in  this figure that $ \Delta_{2e/3}^0/ \Delta_{e/3}^0 \approx 2$ for the largest studied system sizes (while this is not the case for the spinful gaps since it is not possible to reach system sizes as large as for the spinless case). This implies that for the larger systems, the  Coulomb repulsion between the two quasi-particles is already residual. This is however not the case for the charge-$e$ gaps, for which we observe $ \Delta_{e}^0/ \Delta_{e/3}^0 > 3$ even for the larger sizes, as expected for a repulsive interaction.
In terms of the charge-$e/3$ spinful gaps, we observe that $\Delta_{e/3}^{-1,-}$ - the energy required to create a quasi-hole well separated
from a magnon-dressed quasi-electron - is the smallest. It is however very close in energy to $\Delta_{e/3}^{-1,+}$, the energy required to create a quasi-electron well separated from a magnon-dressed quasi-hole. The spin-2 gap, $\Delta_{e/3}^{-2}$ is significantly larger (approximately $6$meV for $u=0.9$), suggesting that the spin-1 gaps are the ones providing the largest contribution to the thermal activation of $R_{xx}$  at $B = 0$ in experiment \cite{park2025observationhightemperaturedissipationlessfractional}, with the strongest contribution coming from $\Delta_{e/3}^{-1,-}$. Note that, as previously discussed, this requires that the resistance pre-factor $R_{xx}^0$ in the thermal activation expression  $R_{xx} = R_{xx}^0 \exp(-\Delta_c/(2 k_B T))$ is significantly larger for the spin-1 gap than for the spinless gap, given that the latter is smaller.  

We finally estimate the binding energy between the magnon and the quasi-electron and quasi-hole, for the magnon-dressed excitations, that can be respectively defined as 

\begin{equation}
    \Delta_b^{(\textrm{qe} + \up \down)} = \Delta_{e/3}^{-1, -}-(\Delta_{e/3}^{0}+\Delta_N^s)\, ,
    \label{eq:binding_qe_magnon_1BPV}
\end{equation}

\begin{equation}
    \Delta_b^{(\textrm{qh} + \up \down)} = \Delta_{e/3}^{-1, +}-(\Delta_{e/3}^{0}+\Delta_N^s)\, .
    \label{eq:binding_qh_magnon_1BPV}
\end{equation}

Using the charge gaps for $N_s=30$, we have for $u=0.9$ (flatband limit) that $\Delta_b^{(\textrm{qe} + \up \down)} \approx 1.57$meV ($1.25$meV) and $\Delta_b^{(\textrm{qh} + \up \down)} \approx 1.03$meV ($1.10$meV) - and therefore significant binding energies between the magnon and quasi-electron/quasi-hole.

\clearpage

\subsection{Spectra of spin-0 and spin-1 neutral excitations }

\label{sec:S-matrix}

\begin{figure}[h!]
\begin{centering}
\includegraphics[width=\columnwidth]{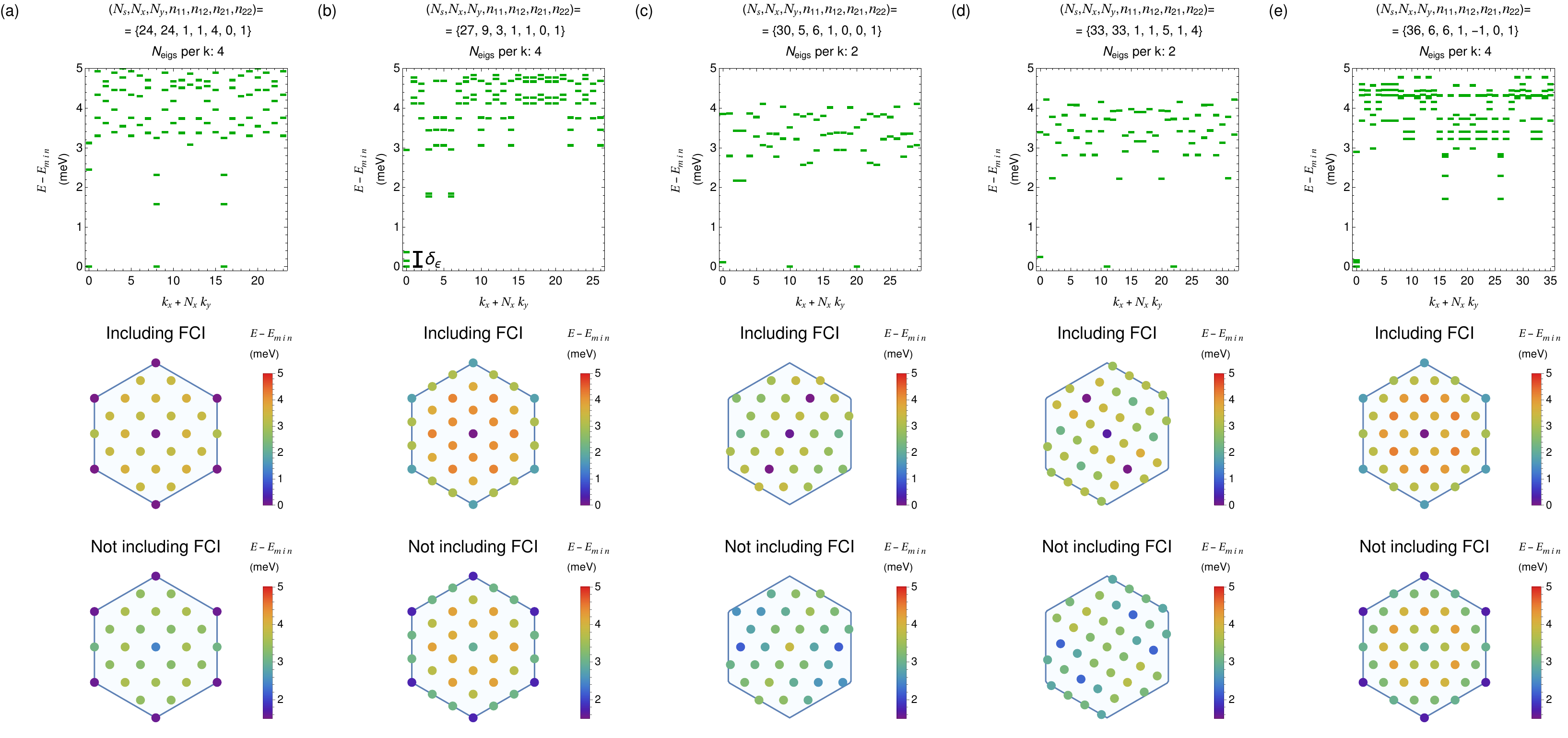}\caption{1BPV ED spectrum of neutral excitations in the polarized sector for $u=0.9$  and different system sizes. Each column, labeled by (a-e), contains data for a given system size specified in the top figure. The figures in the middle and bottom rows show the lowest energy eigenvalues in the moiré Brillouin zone, respectively either including or not the FCI ground-states. By including the  quasi-degenerate FCI ground states, we can identify their respective many-body momenta. Excluding the ground states, as in the bottom row, gives access to the dispersion of the magnetoroton collective modes. The number of computed eigenvalues is also indicated as "$N_{\textrm{eigs}}$ per k" (note that not all are shown for the chosen energy range in (a,b) ).  
 \label{fig:neutral_spectra_u0p9}}
\par\end{centering}
\end{figure}

\begin{figure}[h!]
\centering{}%
\includegraphics[width=0.7\columnwidth]{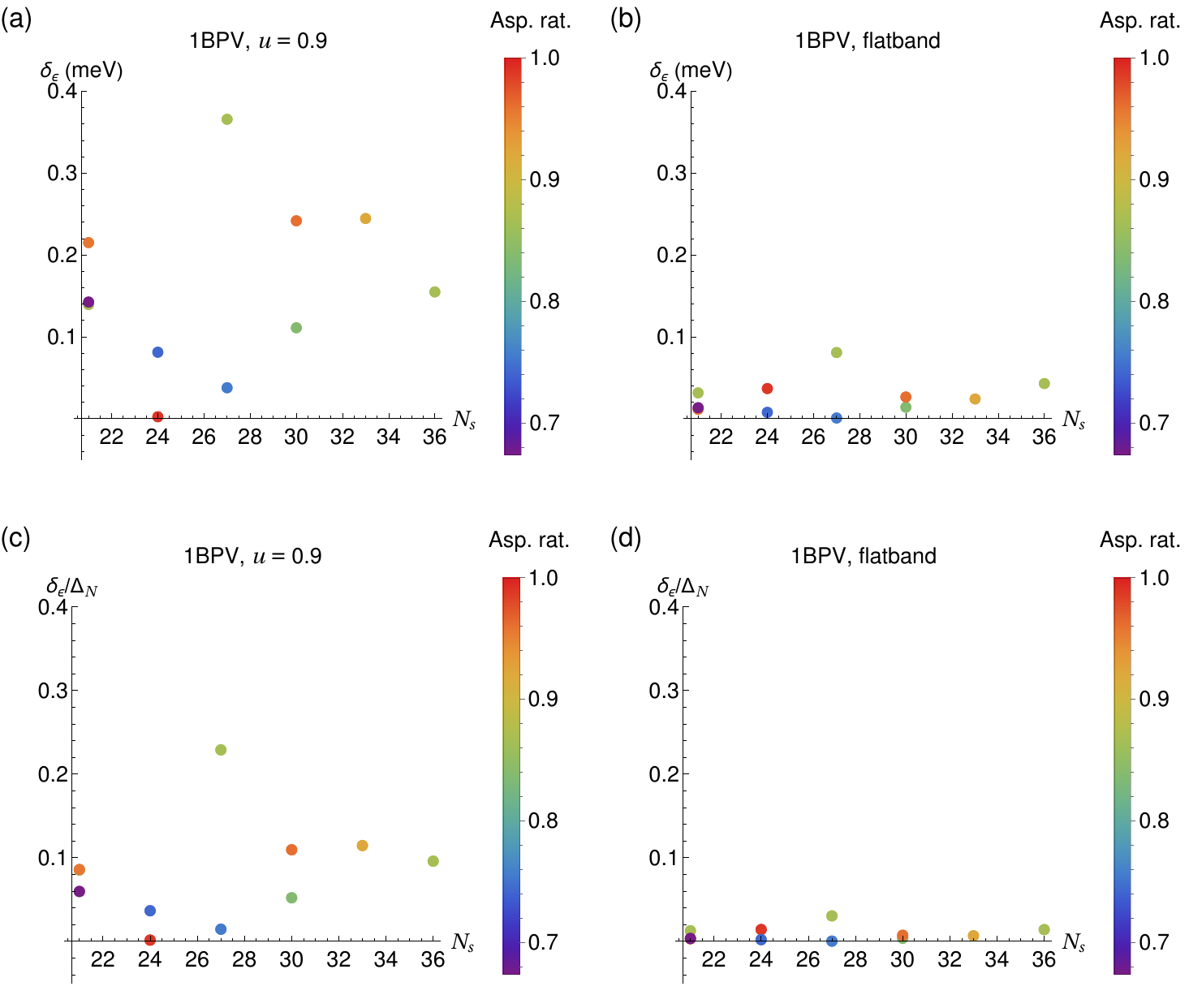}\caption{Finite-size spread of the quasi-degenerate FCI ground-states, defined as  $\delta_{\epsilon} = \textrm{max}(\{E_{FCI}\})-\textrm{min}(\{E_{FCI}\})$, where $\{E_{FCI}\}$ is the set of quasi-degenerate FCI energies. Results for all the system sizes used for the calculations in Fig.$\,$\ref{fig:gaps_neutral_u0p9_1BPV}, for (a) $u=0.9$; (b) flatband limit. In (c,d) we show the ratio between the spreads and the corresponding spin-0 neutral gap respectively for $u=0.9$ and the flatband limit.  The data points are colored according to the corresponding aspect ratio. }\label{fig:spread_1BPV}
\end{figure}

In this section, we provide more detailed results on the neutral spectra for spin-polarized and spin-1 excitations. The results for the energy spectrum in the spin-polarized sector are shown in Fig.$\,$\ref{fig:neutral_spectra_u0p9} where each column corresponds to a different system size. For all the system sizes, we observe three quasi-degenerate FCI ground states that are well separated from the magnetoroton collective mode by a clear FCI neutral gap. To evaluate the robustness of the FCI ground states, we inspect the finite-size spread of the quasi-degenerate ground states on the torus, defined as
\begin{equation}
\delta_{\epsilon}=\textrm{max}(\{E_{FCI}\})-\textrm{min}(\{E_{FCI}\}) \, ,
\label{eq:FCI_spread}
\end{equation}

\noindent where $\{E_{FCI}\}$ is the set of (three) quasi-degenerate energies. The spreads, $\delta_\epsilon$, for all the system sizes used for the calculations in Fig.$\,$\ref{fig:gaps_neutral_u0p9_1BPV} are shown in Fig.$\,$\ref{fig:spread_1BPV}.  For both the $u=0.9$ and for the flat band case, we find that $\delta_{\epsilon}$  reaches at most $20\%$ of the gap for the 27 unit-cell lattice, see Fig.$\,$\ref{fig:spread_1BPV}(c,d). Stronger finite-size effects are expected for this lattice size given that counting enforces the FCI ground-states to occur all a the same momentum sector, thus these states experience residual level repulsion which vanishes in the thermodynamic limit. For the remaining system sizes, we have $\delta_\epsilon/\Delta_N \lesssim 0.1$. Overall, we find that $\delta_{\epsilon}$ is small for the interaction and band dispersion parameters introduced above in Appendix \ref{sec:Model_definition} for all system sizes studied below, thus supporting that finite-size effects on the ground state are under control. 

For each system size, the dispersion as a function of momenta for the lowest energy eigenvalues is shown in the middle and bottom panels of Fig.$\,$\ref{fig:neutral_spectra_u0p9}(a-e). For clarity we show the dispersion of the lowest energy eigenvalues excluding the FCI ground state, which corresponds to the the magnetoroton modes \cite{PhysRevB.33.2481,PhysRevB.90.045114,wolf2024intrabandcollectiveexcitationsfractional,shen2025magnetorotonsmoirefractionalchern,long2025spectramagnetorotonchiralgraviton,lu2024interactiondrivenrotoncondensationc}. For or 27 and 36 site $C_6$-symmetric lattices, we find the magnetoroton minima minima at $\bsl{K}_M$ and $\bsl{K}_M'$, alluding to the incipient instability toward charge-density wave order when the magnetoroton gap collapses.

\begin{figure}[h!]
\centering{}\includegraphics[width=1\columnwidth]{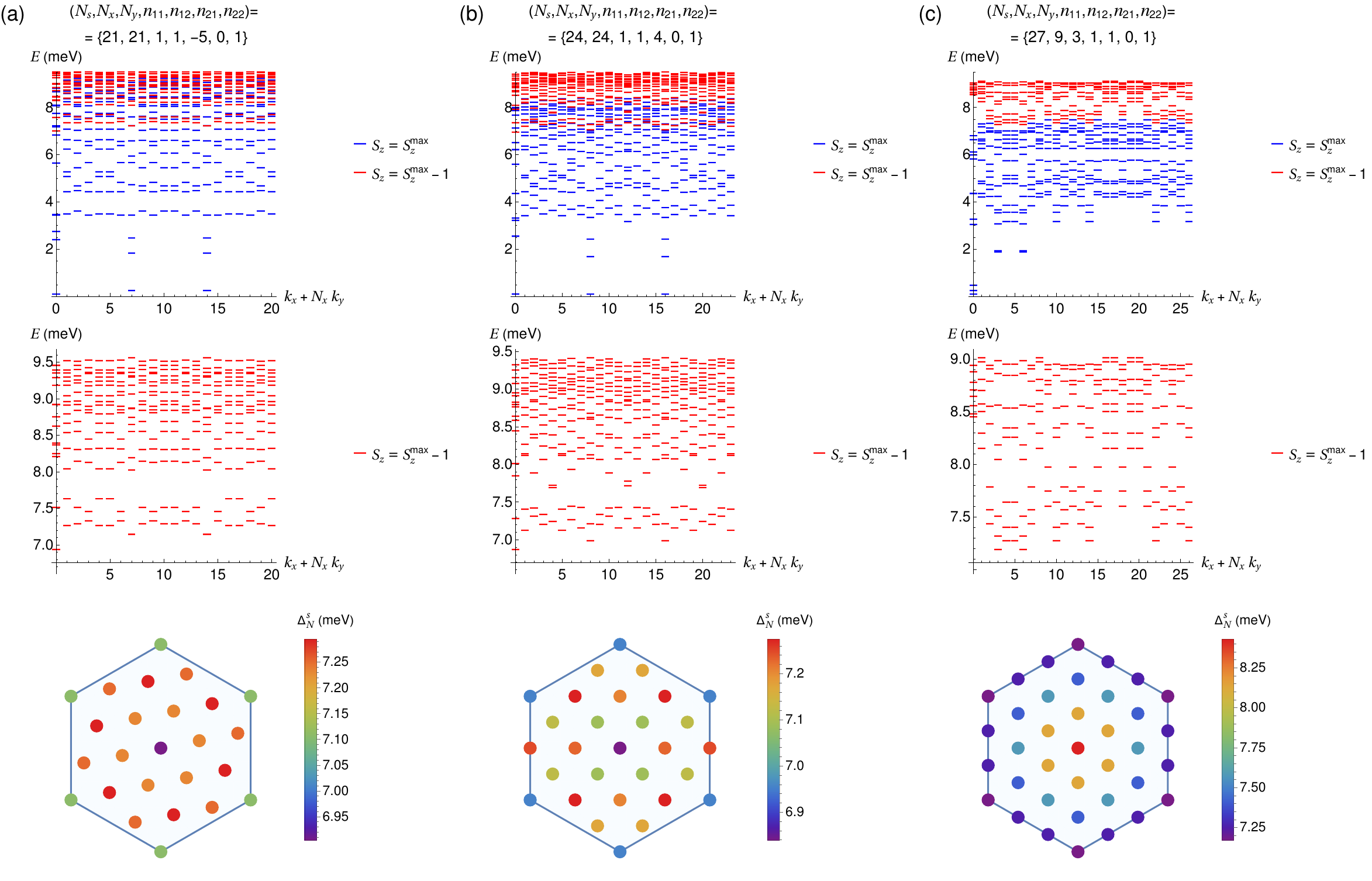}\caption{Neutral spectra for spin-0 (blue) vs. spin-1 (red) excitations, for  $u=0.9$ and different system sizes indicated on top of each panel. The middle figure in each panel contains a close-up view of the spin-1 spectrum, while the bottom of each panel also shows the
momentum-resolved spin-1 gap in the moiré Brillouin zone. \label{fig:Neutral-vs.-spin-flip_ED_var_Sizes}}
\end{figure}

We now turn to the spin-1 excitation spectrum. We show the polarized and spin-1 neutral spectra for different system sizes in Fig.$\,$\ref{fig:Neutral-vs.-spin-flip_ED_var_Sizes}. In every case, one can identify a set of low-lying spin-1 modes separated from a continuum of excitations. In a separate work, we will provide a better understanding of the low lying modes through a generalization of the scattering matrix approach employed in Ref.$\,$\cite{TBGV}. Here we just make a couple of brief comments.
We find that the dispersion of these isolated modes strongly depends on system size. This can be better seen from the momentum-resolved plots of the spin neutral gap dependence, shown in the bottom row of Fig.$\,$\ref{fig:Neutral-vs.-spin-flip_ED_var_Sizes}. In particular, note that the dispersion of the lowest-lying mode as a function of momentum has either a minimum or maximum at $\bsl{k}=\bsl{\Gamma}_M$ depending on system size.
The origin of this size dependence, that will be detailed in a separate work, is tied to the underlying lattice geometry and whether it supports the FCI ground-states in the same momentum sector or not. 
When the FCI states live in the same momentum sector, as for the 27 unit cell lattice, the spin-1 excitation spectrum corresponds to creating magnon-like excitations on top of the three quasi-degenerate ground states. The resulting dispersion includes three magnon modes with almost the same energy and with a splitting that appears to be set by finite-size effects. In the calculation performed on a 27 unit cell lattice, we find that the minima of the quasi-degenerate magnon bands are located at $\bsl{K}_M$ and $\bsl{K}_M'$. We contrast this behavior to cases where FCI ground states are located at different momentum sectors, as for the 21- and 24 unit cell lattices. Like in the previous case, the spin-1 excitation spectrum can be understood as creating spin-1 magnon excitations on top of the FCI ground states. Nevertheless, even assuming negligible magnon scattering, we do not expect the magnon bands to remain quasi-degenerate throughout momentum space; instead, each band’s dispersion follows from the many-body momentum of its underlying ground state complicating the analysis.

\clearpage

\subsection{Occupation numbers}

\begin{figure}[h!]
\begin{centering}
\includegraphics[width=0.85\columnwidth]{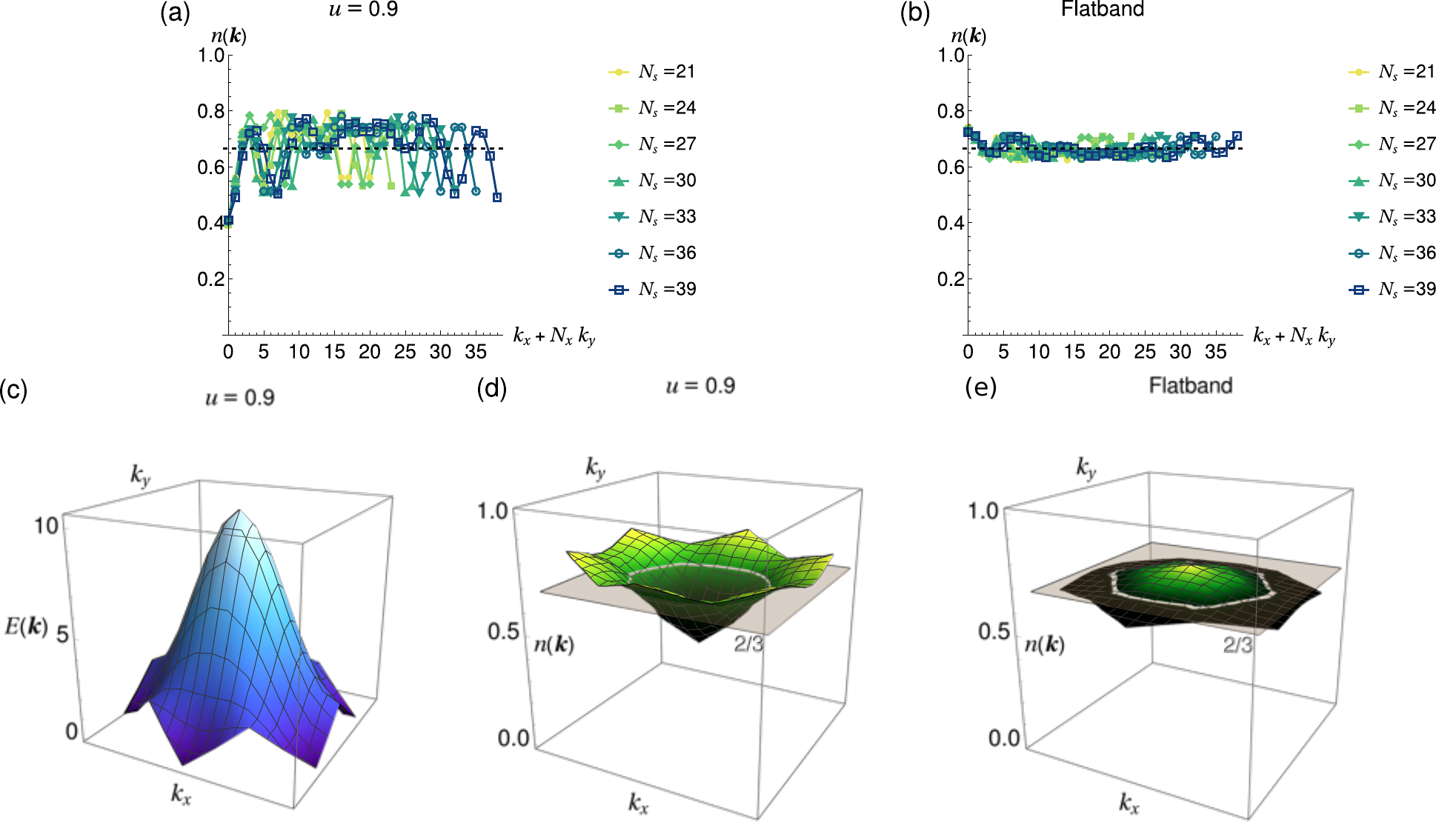}\caption{ 
Hole occupation number $n(\bf{k})$ at filling $\nu=-2/3$, averaged over the three-fold quasi-degenerate FCI ground-states, for $u=0.9$ (a,d) and in the flatband limit (b,e). In (c) we also show the single-particle energy dispersion for band 0 (topmost valence band for electrons) in the hole basis, for comparison with the $n(\bf{k})$ results. For the $n(\bf{k})$ results in (c,d), we interpolated the combined data for all the system sizes in (a,b). The excellent collapse of the data is a strong indication that the fluctuations in $n(\bf{k})$ are robust across the different system sizes. For the energy dispersion in (c), we used a momentum mesh defined by $(N_x,N_y,n_{11},n_{12},n_{21},n_{22})=(12,12,1,0,0,1)$ and interpolated the data using linear interpolation.
 \label{fig:nk_2o3}}
\par\end{centering}
\end{figure}

In this section, we study the hole occupation number defined as $n(\bf{k}) \equiv \langle \gamma^\dagger_{\up,0,\bf{k}}\gamma_{\up,0,\bf{k}} \rangle$, where the $\gamma^\dagger_{\up,0,\bf{k}}$ are defined in Eq.$\ref{eq:band_operators}$, in Appendix \ref{sec:1BPV}, and $\langle \rangle$ denotes the expectation value in the ground-state. For FCI ground-states, we will average over all the three-fold quasi-degenerate states.
We carry out this study in the fully polarized sector and at filling $\nu=-2/3$. The results are shown in Fig.$\,$\ref{fig:nk_2o3}. For $u=0.9$, $n({\bf k})$ shows fluctuations around the uniform distribution of $n({\bf k})=2/3$, that would be associated with a perfect FQH state, with a standard deviation of $0.1$ (considering the data for all the system sizes), see Fig.$\,$\ref{fig:nk_2o3}(a). In the flat band limit, defined in Sec.$\,$\ref{sec:1BPV}, the FCI is significantly more uniform, with $n({\bf k})$ having a standard deviation of only $0.03$, see Fig.$\,$\ref{fig:nk_2o3}(b).

To better understand the structure of the fluctuations in $n({\bf k})$, we combine the data for all the system sizes shown in Figs.$\,$\ref{fig:nk_2o3}(a,b) into a single function $n({\bf k})$ shown in Figs.$\,$\ref{fig:nk_2o3}(d,e). We find that fluctuations in $n({\bf k})$ are consistent among the different system sizes and thus the hole density, $n({\bf k})$, collapses into a smooth function on the moiré Brillouin zone. The collapse is obtained by combining the data across all the different $\bf{k}$ points obtained from different system sizes and performing a linear interpolation. For both $u=0.9$ and the flat-band case, the collapsed charge density has extremal points at $\bf{k}=\bf{0}$ and ${\bf k}={\bf K}_M,{\bf K}'_M$. Interestingly, while the calculations with dispersion and $u=0.9$ show a single minimum located at $\bf{k}=\bf{0}$ accompanied by maxima at ${\bf k}={\bf K}_M,{\bf K}'_M$, the opposite behavior is observed in the flat-band case where multiple minima occur at ${\bf k}={\bf K}_M,{\bf K}'_M$ accompanied by a single maximum at $\bf{k}=\bf{0}$. 
The $n({\bf k})$ results for $u=0.9$ can be understood by examining the  single-particle energy dispersion of band 0 (the topmost valence band for electrons, where interactions are projected in the 1BPV approximation). The single-particle dispersion in the hole basis is shown in Fig.$,$\ref{fig:nk_2o3}(c). Notably, the behavior of $n({\bf k})$ is set by the single-particle dispersion, reaching its maximum at $\bf{k}$ points where the single-particle energy is the smallest (${\bf k}={\bf K}_M,{\bf K}'_M$ ) and its minimum where the single-particle energy is the largest ($\bf{k}=\bf{0}$), as expected.

\clearpage

\subsection{Spectra of charged excitations}

In contrast to quasi-electron and quasi-hole excitations in the FQH effect, which are dispersionless due to the symmetry-enforced extensive center-of-mass degeneracy (that we will illustrate in detail below), we find that the analog excitations in the FCIs studied here are strongly dispersive, with this behavior persisting as system size increases. In this section, we take a closer look at the energy spectrum of the quasi-electron and quasi-hole charge excitations, in order to back this claim and investigate the energy dispersions defined in  Eq.$\,$\ref{eq:eps_general}. 
We will mostly focus at filling $\nu=-2/3$, the main target of this work, in Sec.$\,$\ref{sec:charge_excitations_2o3}. However, in order to further motivate the emergent magnetic translation symmetry discussed in the main text, we will also present results at filling $\nu=-3/5$ in Sec.$\,$\ref{sec:charge_excitations_3o5}.

\subsubsection{Charged excitations at filling $\nu=-2/3$}
\label{sec:charge_excitations_2o3}

\begin{figure}[h!]
\begin{centering}
\includegraphics[width=\columnwidth]{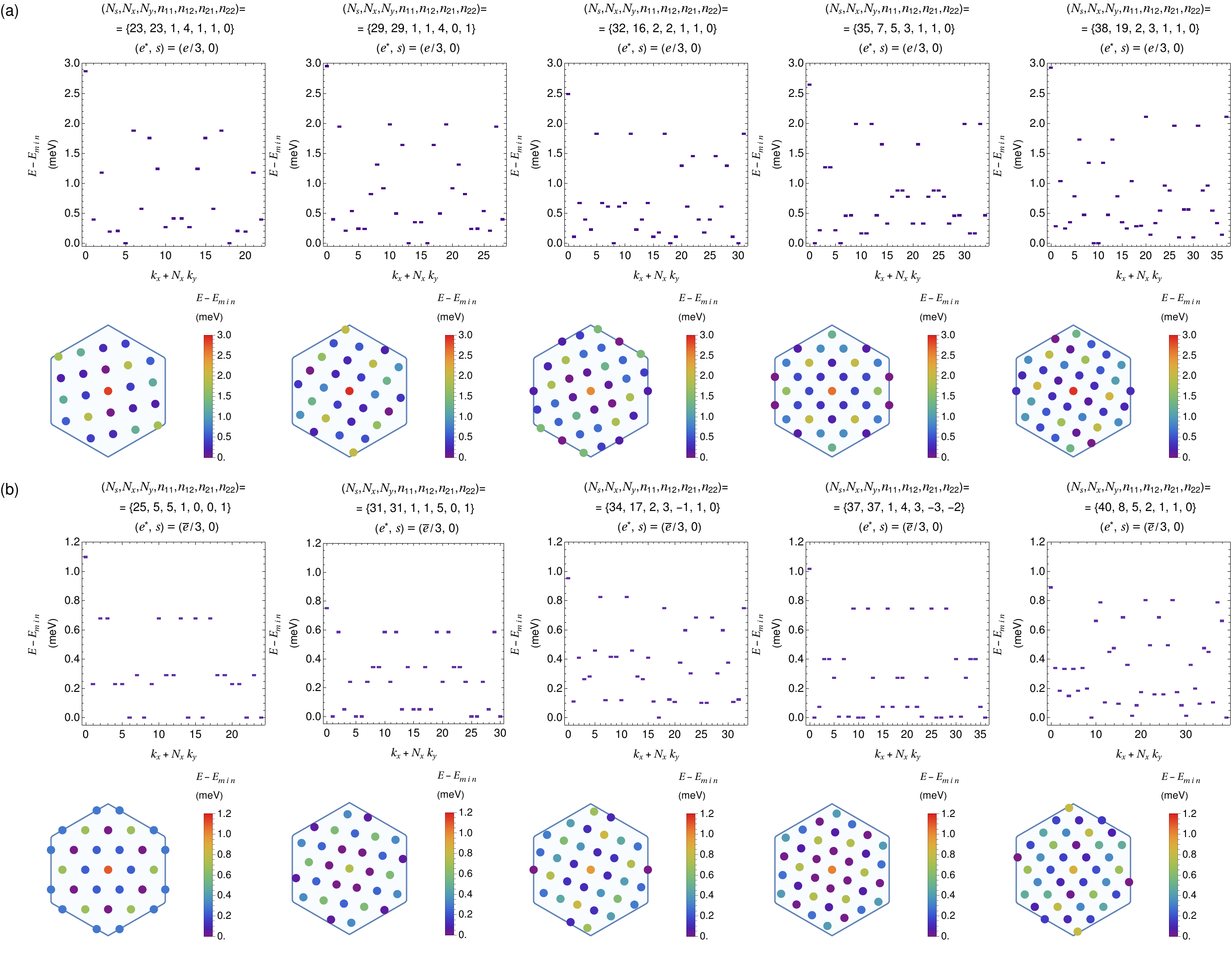}\caption{ 
Spectrum for $\nu=-2/3$, with $s=0$ and (a) charge-$e/3$ quasi-electron ($p'=-1,q'=-1$) and (b) charge-$\bar{e}/3$ quasi-hole ($p'=1,q'=1$), for $u=0.9$ and different system sizes, showing only the lowest energy state in each momentum sector. The $\bsl{k}$-resolved plots show the dispersion  as a function of many-body momentum, in the moiré Brillouin zone. The lowest energy is shifted to be zero for each system size, to make the comparison between the results for different system sizes easier.
 \label{fig:qh_qp_spectra_eo3_u0p9}}
\par\end{centering}
\end{figure}

\begin{figure}[h!]
\begin{centering}
\includegraphics[width=\columnwidth]{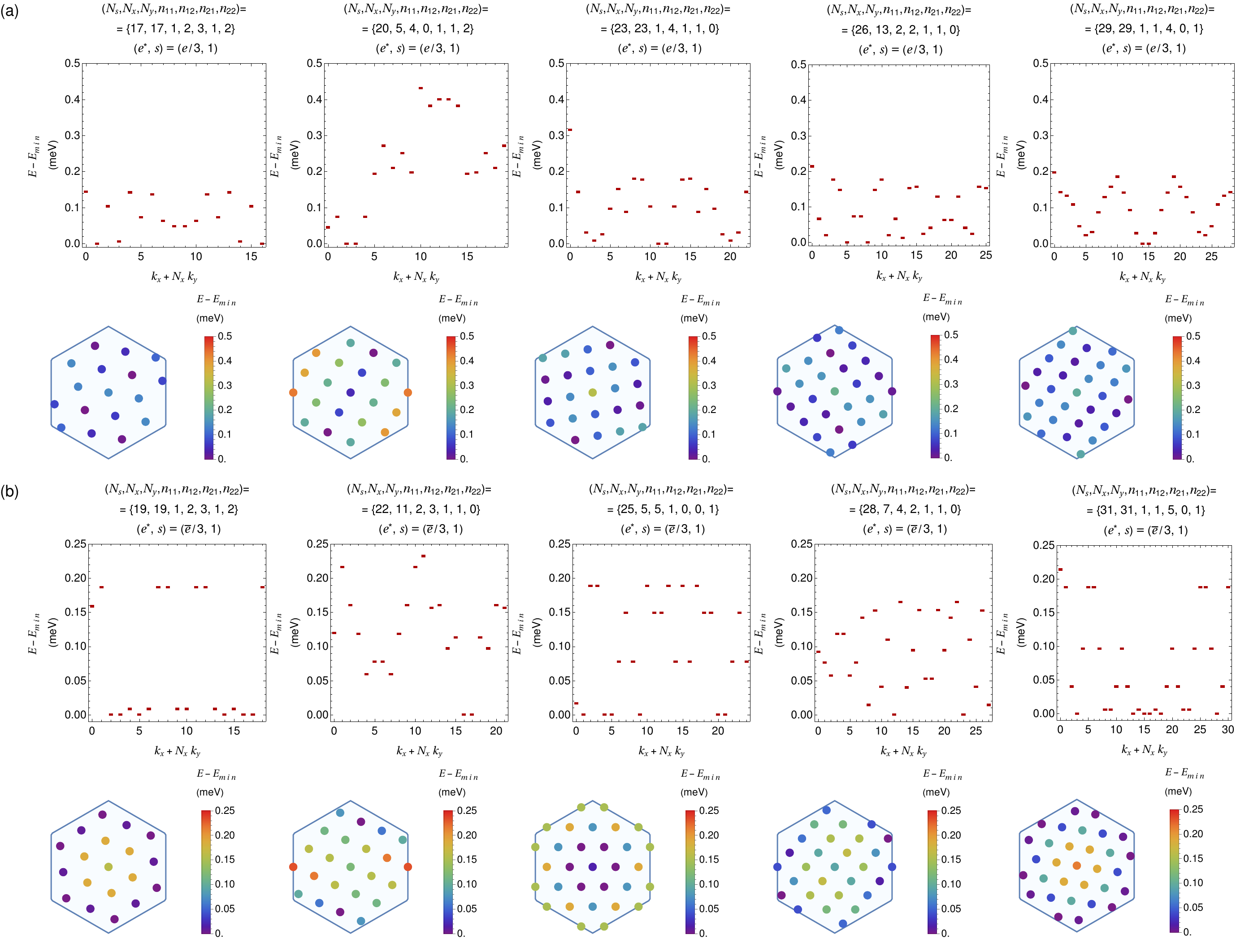}\caption{ 
Spectrum of magnon-dressed charge $e/3$ excitations ($s=1$) for $\nu=-2/3$. (a) charge-$e/3$ quasi-electron ($p'=-1,q'=-1$) and (b) charge-$\bar{e}/3$ quasi-hole ($p'=1,q'=1$) spectra for different system sizes, for $u=0.9$. The $\bsl{k}$-resolved plots show the dispersion of the lowest-energy excitations as a function of many-body momentum, in the moiré Brillouin zone. The lowest energy is shifted to be zero for each system size.
 \label{fig:qh_qp_spectra_eo3_u0p9_s_1}}
\par\end{centering}
\end{figure}

In Fig.$\,$\ref{fig:qh_qp_spectra_eo3_u0p9} we show the momentum-resolved energy spectrum for the pure charge-$e/3$ quasi-electron and quasi-hole excitations, defined respectively by $(e^*,s)=(e/3,0)$ and $(e^*,s)=(\bar{e}/3,0)$, for different system sizes. 
Both the quasi-electron and quasi-hole spectra have a very significant dispersion that is robust to increasing the system size. The dispersion of the quasi-electron has a bandwidth of $\Delta_B \approx 2.5$meV, while the quasi-hole is less dispersive, with $\Delta_B \approx 1$meV. 
The dispersion for the magnon-dressed quasi-electron and quasi-hole excitations,  defined by $(e^*,s)=(e/3,1)$ and $(e^*,s)=(\bar{e}/3,1)$ is also shown in Fig.$\,$\ref{fig:qh_qp_spectra_eo3_u0p9_s_1}. These excitations are much less dispersive, with a bandwidth $\Delta_B < 0.5$meV for the quasi-electron and $\Delta_B \approx 0.2$meV for the quasi-hole. While the bandwidths are approximately consistent for the different system sizes, we do not observe a common trend in the dispersion, in contrast to the $s=0$ excitations. This shows that finite-size effects are more pronounced for the dispersions of the $s=1$ excitations.

\begin{figure}[h!]
\begin{centering}
\includegraphics[width=\columnwidth]{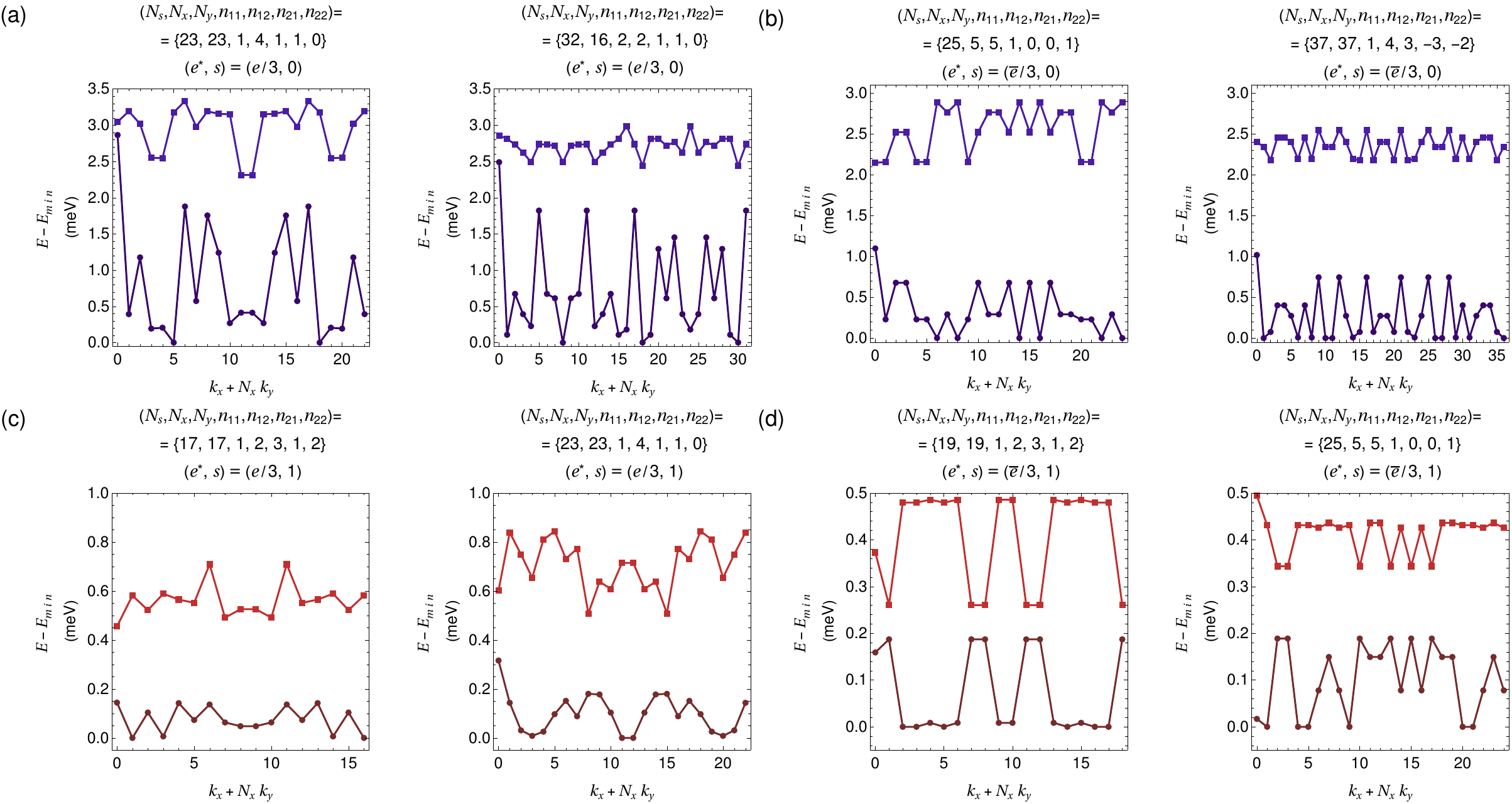}\caption{ 
Lowest two eigenenergies per momentum sector for different charge excitations, for $\nu=-2/3$ and $u=0.9$. In blue we show the spinless quasi-electron $(e^*,s;p',q')=(e/3,0,-1,-1)$  (a) and quasi-hole  $(e^*,s;p',q')=(\bar{e}/3,0,1,1)$ (b) excitations. In red, we show the spinful quasi-electron  $(e^*,s;p',q')=(e/3,1,-1,-1)$ (c) and quasi-hole $(e^*,s;p',q')=(\bar{e}/3,1,1,1)$ (d) excitations. These figures show that the lowest-energy modes are well isolated from higher energy modes by a gap in each momentum sector. 
\label{fig:qh_qp_spectra_multiple_evs}}
\par\end{centering}
\end{figure}

In order to understand if the $s=0$ and $s=1$ quasi-electron and quasi-hole excitations are well isolated from higher-energy modes, in Fig.$\,$\ref{fig:qh_qp_spectra_multiple_evs}, we show plots of at least the two lowest-energy eigenvalues. These results show that the excitations are in fact well isolated, with the largest gap occurring for the $s=0$ quasi-hole excitation. 

\begin{figure}[h!]
\begin{centering}
\includegraphics[width=\columnwidth]{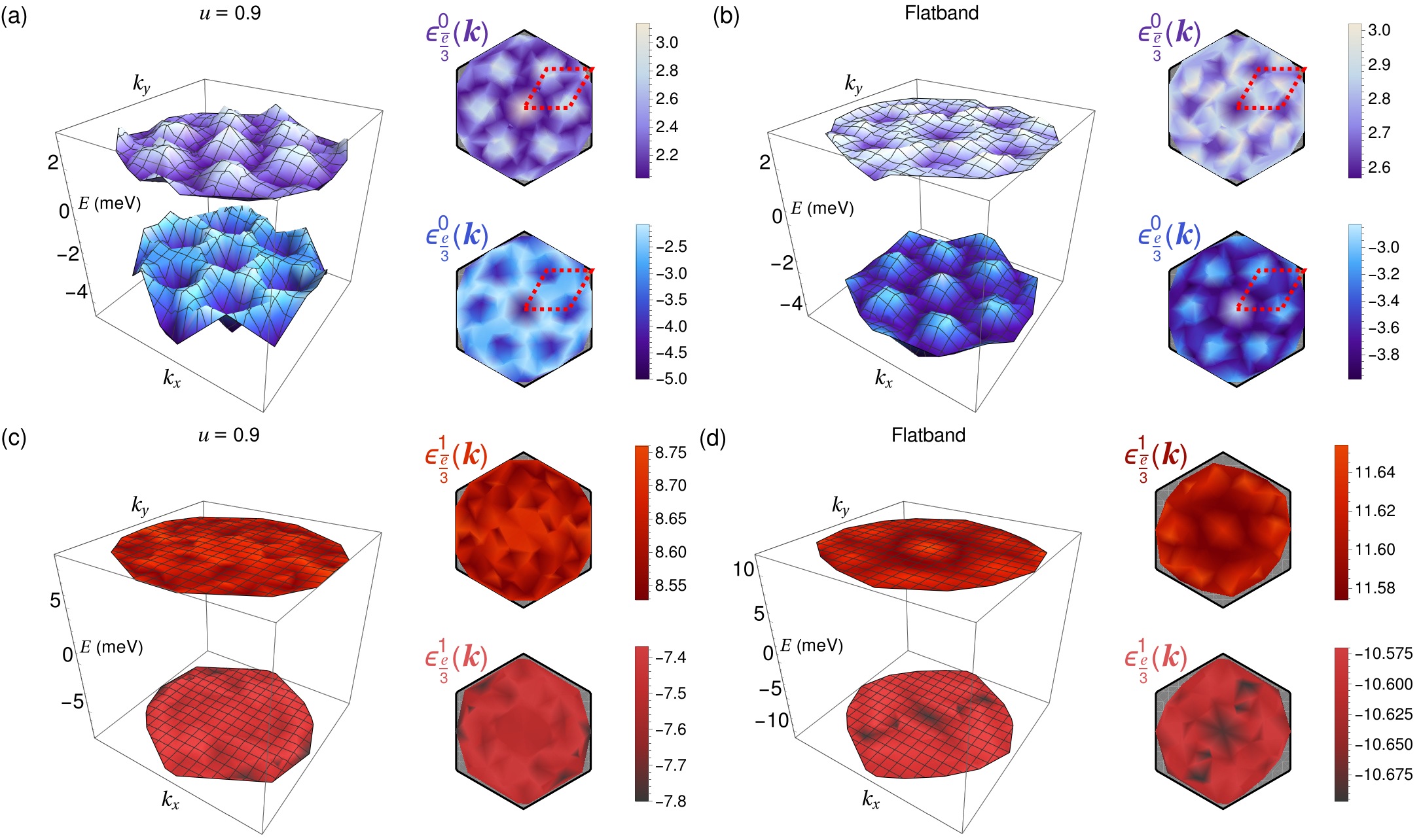}\caption{ 
 Energy dispersions for the charge-$e/3$ quasi-electron (bottom dispersion of each plot) and charge-$e/3$  quasi-hole (top dispersion of each plot), interpolating the combined data from all system sizes with $N_s \geq 20$. The dashed red lines indicate the approximate emergent Brillouin zone for the $s=0$ excitations. (a) $u=0.9,s=0$. (b) Flatband, $s=0$. (c) $u=0.9,s=1$. (d) Flatband, $s=1$. 
 \label{fig:qh_qp_dispersion_u0p9}}
\par\end{centering}
\end{figure}

Given that the lowest-energy excitations are well-defined, we inspect their emergent ``band structure" in Fig.$\,$\ref{fig:qh_qp_dispersion_u0p9}, combining the data from all the system sizes with $N_s \geq 20$. We observe a remarkable collapse of the data for the $s=0$ excitations into a smooth band structure, in Fig.$\,$\ref{fig:qh_qp_dispersion_u0p9}(a). This band structure exhibits an approximate emergent periodicity compatible with an enlarged moiré unit cell nine times larger than the original (corresponding to a
tripling along each moiré lattice vector direction). This corresponds to an emergent Brillouin zone nine times smaller than the moiré Brillouin zone, depicted by the red dashed lines in Fig.$\,$\ref{fig:qh_qp_dispersion_u0p9}(a).
The most natural interpretation of this data is as a single-particle Hofstadter spectrum, where magnetic translation symmetry at $\phi/2\pi = 1/3$ flux per unit cell enforces the enlarged $3\times 3$ real-space periodicity. Recall that, in actual non-interacting Hofstadter systems at $\phi/2\pi = 1/3$ \cite{PhysRevLett.125.236804,2023PhRvL.130w6601H}, the magnetic translation operators obey $[H_{hof},T_i] = 0$ and $T_1 T_2 = T_2 T_1 e^{i \phi N}$ where $N$ is the total particle number, and we take $N=1$ to describe an effective single-particle system. Then the quantum numbers of the system at $\phi/2\pi = 1/3$ (the general result easily follows) can be chosen as the eigenvalues of the commuting operators $T_1, T_2^3$  which we denote as $e^{i k_1}, e^{i 3 k_2}$. These magnetic Bloch momenta are non-redundantly defined in $k_1 \in [0,2\pi), k_2 \in [0,2\pi/3)$ which is called the magnetic Brillouin zone and is smaller than the crystalline Brillouin zone by a factor of 3. This means the dispersion relation trivially obeys $E(k_1, k_2)=E(k_1, k_2+2\pi/3)$. If we consider the spectrum in the \emph{crystalline} Brillouin zone, it has a trivial 3-fold periodicity along $k_2$ due to the reduced magnetic Brillouin zone.
It is easy to see that there is $3$-fold periodicity in the spectrum along $k_1$ as well. Consider an eigenstate $\ket{k_1,k_2}$ with energy $E(k_1,k_2)$. Then $T_2 \ket{k_1,k_2}$ is also an eigenstate since $[H,T_2] = 0$, but has momentum $(k_1+\frac{2\pi}{3},k_2)$ since
\bea
T_1 (T_2 \ket{k_1,k_2}) = T_2 T_1 e^{i \frac{2\pi}{3}}  \ket{k_1,k_2})= e^{i (k_1 +\frac{2\pi}{3})}  (T_2  \ket{k_1,k_2}) \ .
\eea
This means the spectrum is 3-fold periodic along $k_1$: $E(k_1, k_2)=E(k_1+2\pi/3, k_2)$. In total, there is a $3 \times 3$ periodicity of the spectrum when considered in the original crystalline Brillouin zone. In fact, this must be the case since our choice of simultaneously diagonalizing $T_1, T_2^3$ was arbitrary: we could have picked $T_{\mbf{R}_1}, T_{\mbf{R}_2}^3$ for any tilted lattice vectors $\mbf{R}_1,\mbf{R}_2$. Hence, the observation of an emergent $3 \times 3$ Brillouin zone periodicity in the quasi-particle spectrum is consistent with a single-particle Hofstadter description --- although there is no microscopic magnetic field in the Hamiltonian. In other words, the quasi-particles obey a projective representation of the microscopic translation group. This is consistent with composite fermion approaches \cite{PhysRevB.97.125131,PhysRevB.109.245125,RevModPhys.75.1101} that implement an exact projective representation within an enlarged parton Hilbert space.

We leave a derivation of the effective quasi-hole Hamiltonian for future work, but conjecture that it is defined by an effective mass, periodic potential, and periodic (generically non-uniform) magnetic field enclosing $2\pi/3$ flux per unit cell. Lastly, we note that there is no particle-hole symmetry relating the quasi-electron and quasi-hole, so they are expected to have different effective masses \cite{PhysRevB.42.212}. In the FQH problem where there is continuous magnetic translation symmetry, the quasi-particle bands are flat, and hence their kinetic energy is quenched and they are susceptible to disorder. We now briefly illustrate the flatness of the quasi-particle bands assuming a fully polarized FQH state at fillings $\nu=1/3$ and $\nu=2/3$. the many-body symmetries in a torus geometry allow to define a Brillouin zone with $N_k \times N_k$ many-body momenta, with  $N_k=\gcd(pN+p',N_{\phi})$,  where $pN+p'$ and $N_{\phi}$ are respectively the number of particles and number of fluxes \cite{PhysRevLett.55.2095,PhysRevB.85.075128}. At $\nu=1/3$, quasi-hole and quasi-electron excitations can be created by taking $p'=0$ and $N_{\phi}=3N \pm 1$ ($+$ for quasi-hole and $-$ for quasi-electron), where $N\rightarrow\infty$ in the thermodynamic limit. In both cases, we have $N_k=1$ for any $N$, directly implying an extensive $N_{\phi}$-fold center-of-mass degeneracy, and therefore, flat quasi-particle dispersions. At filling $\nu=2/3$, the same result is obtained by taking  $p'=\pm1$ and $N_{\phi}=3N \pm 1$ ($+$ for quasi-electron and $-$ for quasi-hole).  
It may therefore be interesting to design FCIs with highly dispersive quasi-particles to move farther away from the Landau level limit, where quasi-particles are easily localized by disorder.

For completeness, we also show results in the flat band limit in Fig.$\,$\ref{fig:qh_qp_dispersion_u0p9}(b), where we again observe the excellent collapse and the new emergent unit cell. Interestingly, the maxima/minima invert in the flatband limit, compared to $u=0.9$, correlating with the inversion in $n(\bf{k})$ observed in Fig.$\,$\ref{fig:nk_2o3}.
We also plot the $s=1$ excitations energies in Figs.$\,$\ref{fig:qh_qp_dispersion_u0p9}(c,d). In this case, the data collapse to a significantly flatter dispersion. It is not clear to see a $3 \times 3$ periodicity likely due to significantly stronger finite-size effects and the flatness of the dispersion, which is still not well converged for the available system sizes. We note that finite-size effects are also stronger for the spinful excitations in the context of FQH. In this case, many-body symmetries allow to define a Brillouin zone with $N_k \times N_k$ many-body momenta, with  $N_k=\gcd(N_{\up}-N_{\down},N_{\phi})$, where $N_{\up}$,$N_{\down}$ and $N_{\phi}$ are respectively the number of particles in the spin up and down sectors, and the number of fluxes (equivalent to $N_s$ in the FCI context). For the spinless charge-$e/3$ excitations ($N_{\down}=0,N_{\up}=2N\pm1,N_s=3N\pm1$), it is easy to see that $N_k=1$ for all $N$. In other words, symmetry enforces a flat dispersion for any number of fluxes/system size in this case. For the spinful quasi-hole excitation ($N_{\down}=1,N_{\up}=2N,N_s=3N+1$), however, we have $N_k=2$ ($N_k=1$) for $N$ odd (even). Since $N_k$ is non-extensive, the dispersion is also flat in the thermodynamic limit. However, stronger finite-size effects are expected due to the different behaviors for even and odd $N$. For the spinful quasi-electron excitation ($N_{\down}=1,N_{\up}=2N-3,N_s=3N-1$), we have $N_k=1$ for all $N$ except when $\mod(N-5,7)=0$, so that the finite-size effects are expected not to be as strong.

\clearpage

\subsubsection{Quasi-electron excitations at filling $\nu=-3/5$}

\label{sec:charge_excitations_3o5}

To further support the evidence of emergent magnetic translation symmetry observed for quasi-electron and quasi-hole excitations at $\nu = -2/3$, we extend our analysis to the nearby fractional filling $\nu = -3/5$, where FCI ground-states also exist in twisted MoTe$_2$ at $\theta=3.7^{\degree}$ \cite{Park2023,Zeng2023,Cai2023,PhysRevX.13.031037,Ji2024,Redekop2024,park2025observationhightemperaturedissipationlessfractional}. Employing the arguments discussed in Sec.~\ref{sec:charge_excitations_2o3}, and assuming that the lowest-energy excitations have fractional charge $\pm e/5$, as in FQH, one can anticipate the emergence of an enlarged magnetic unit cell $5 \times 5$ times larger than the moiré unit cell, corresponding to a Brillouin zone $25$ times smaller (i.e., reduced by a factor of $5$ along each moiré reciprocal lattice direction).

\begin{figure}[h!]
\begin{centering}
\includegraphics[width=0.75\columnwidth]{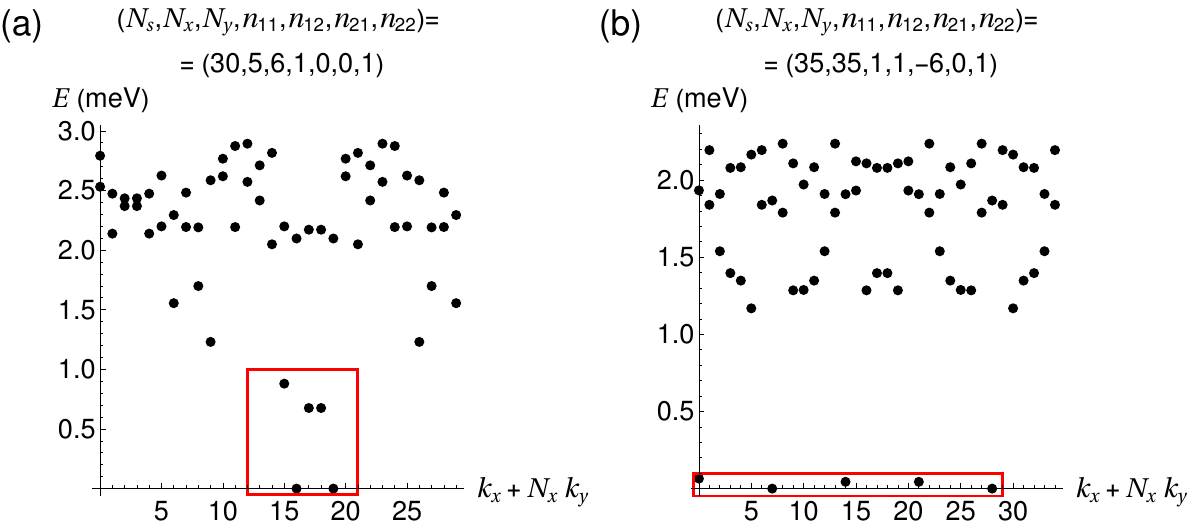}\caption{ 
Lowest two eigenvalues per momentum sector at filling $\nu=-3/5$ for $u=0.9$ and system sizes $N_s=30$ (a) and $N_s=35$ (b). The full specifications of the used momentum meshes are given in the figures. The red rectangles indicate the five FCI ground-states.
 \label{fig:3o5_FCI}}
\par
\end{centering}
\end{figure}

We start by verifying the robustness of the FCI ground state at $\nu = -3/5$. Figure~\ref{fig:3o5_FCI}(a) shows the many-body spectrum for $N_s = 30$ unit cells and $u = 0.9$. In this case, the energy spread of the FCI ground-states is larger than the spin-0 neutral gap, indicating strong finite-size effects. In contrast, for $N_s = 35$, shown in Fig.~\ref{fig:3o5_FCI}(b), the quality of the FCI is significantly better: the energy spread is reduced to only $6\%$ of the neutral gap.

Using the definitions introduced in Sec.$\,$\ref{sec:gaps}, a quasi-electron excitation with charge $e/5$ can be obtained by choosing $p'=2$ and $q'=1$ (i.e. adding two holes and one flux with respect to the ground-state). In what follows, we will obtain results for quasi-electron excitations with respect to the FCI ground-states at $N_s=30$ and $N_s=35$, involving respectively calculations on system sizes with $N_s=32$ and $N_s=37$ unit cells. We note that strong finite-size effects are expected at $N_s=32$ due to the poor quality FCI observed for $N_s=30$ in Fig.$\,$\ref{fig:3o5_FCI}(a). However, the system sizes with $N_s=32$ and $N_s=37$ are the largest we can reach. The calculation with $N_s=37$ involves an Hilbert space size of $2.5 \times 10^8$, and the next available size, with $N_s=42$, would involve an Hilbert space size of $6 \times 10^9$.

\begin{figure}[h!]
\begin{centering}
\includegraphics[width=\columnwidth]{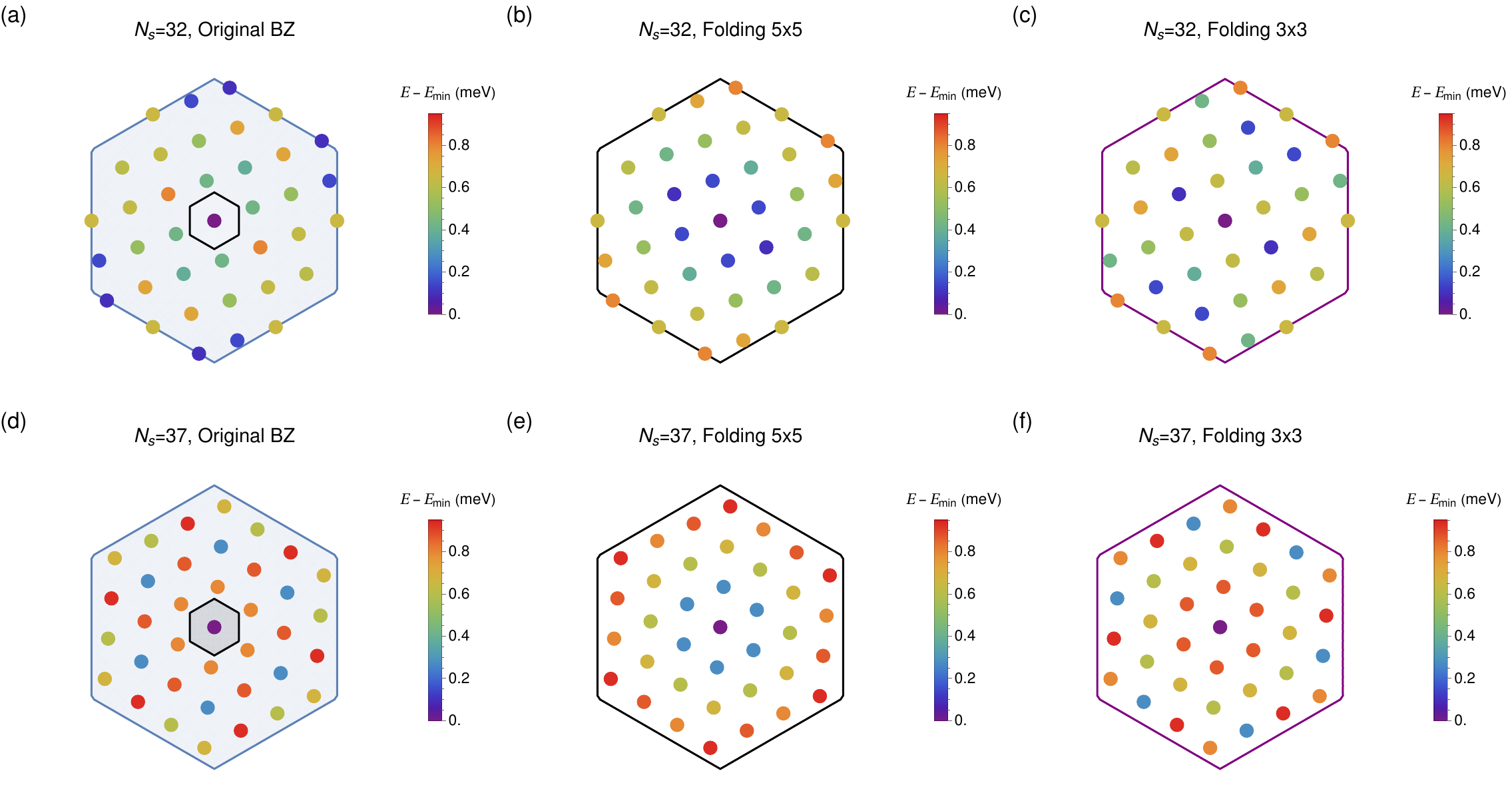}\caption{ 
 Charge $e/5$ quasi-electron excitation at filling $\nu=-3/5$, for $u=0.9$. Results for $N_s=32$ with 19 holes (a-c), and $N_s=37$ with 22 holes (d-f). 
(a,d) shows the lowest-energy per momentum-sector in the moiré Brillouin zone subtracted by $E_{\textrm{min}}$, the lowest-energy among all the momentum-sectors.  The expected emergent Brillouin zone, defined by the reciprocal lattice vectors ${\bf b}_{M,1}/5$ and ${\bf b}_{M,2}/5$, is also shown in black. 
 In (b,c) and (e,f), the same data is shown, assuming an emergent unit cell $n\times n$ larger, giving rise to a new Brillouin zone defined by the reciprocal lattice vectors ${\bf b}_{M,1}/n$ and ${\bf b}_{M,2}/n$, with $n=5$ (b,e) and $n=3$ (c,f). 
 \label{fig:3o5_quasielectron}}
\par\end{centering}
\end{figure}

We show the results for the quasi-electron spectra in Fig.$\,$\ref{fig:3o5_quasielectron}. In Figs.$\,$\ref{fig:3o5_quasielectron}(a,d) we show the spectra in the original moiré Brillouin zone. It is difficult to draw any conclusions with the data represented in this way, since only one or two ${\bf k}$ points fall within each emergent mini-Brillouin zone, when represented in the original moiré Brillouin zone.
A clearer picture is obtained by folding the energy dispersion into the emergent Brillouin zone, as shown in Figs.$\,$\ref{fig:3o5_quasielectron}(b,e). In this case, a smooth dispersion is obtained, showing global minima near ${\bf k} = {\bf 0}$ and maxima close to the moiré Dirac points for both $N_s = 32$ and $N_s = 37$.

For comparison, in Figs.~\ref{fig:3o5_quasielectron}(c,f), we also fold the energy dispersion into a Brillouin zone nine times smaller than the original (i.e., defined by the reciprocal vectors ${\bf b}_{M,i}/3$), corresponding to the emergent Brillouin zone at $\nu = -2/3$. In this case, it is clear that the obtained energy dispersion is a less smooth function of momentum, compared with the $5\times 5$ folding. 

While these results do not constitute definitive proof of an emergent $5\times 5$ larger unit cell at $\nu=-3/5$, they provide strong supporting evidence. Together with the results for $\nu = -2/3$ presented in Sec.~\ref{sec:charge_excitations_2o3}, they strengthen the Hofstadter interpretation presented in Sec.~\ref{sec:charge_excitations_2o3}.

\clearpage

\section{2BPV results: Impact of band mixing}

\subsection{Neutral and charge gaps}

\label{sec:2BPV}

In this section, we investigate the effect of band mixing on the neutral and charge excitation gaps at filling $\nu = -2/3$, carrying out 2BPV ED calculations, i.e. using the Hamiltonian in Eq.$\,$\ref{eq:H_full_band_basis} projected into the topmost (band 0) and second-topmost (band 1) valence bands per valley. 

Full 2BPV calculations are highly demanding and we are restricted to very small system sizes. As an example, to compute all the charge-$e/3$ excitation gaps from a ground-state with $N_s=18$ (recall that this involves also calculations for systems with $N_s\pm1$ unit cells), we would need to deal with Hilbert space sizes up to $5 \times 10^9$. 
To make these calculations tractable and reach the largest system sizes possible, we implement the \emph{band maximum} technique introduced in Ref.~\cite{MFCI_IV}. This method amounts to restrict the number of holes allowed in band 1, denoted $N_{\text{band1}}$, while keeping the occupation of band 0 unrestricted. It offers a controlled interpolation between the 1BPV ($N_{\text{band1}} = 0$) and full 2BPV ($N_{\text{band1}} =2 N_s$) limits, while expanding the range of accessible system sizes. Going back to the example of $N_s=18$ mentioned above, restricting $N_{\text{band1}} = 3$ (which already provides a very good convergence for this system size, as we will see), reduces the maximum required Hilbert space dimension to $10^8$,  more than an order of magnitude compared to the full (unrestricted) Hilbert space.

\begin{figure}[h!]
\centering{}%
\includegraphics[width=0.85\columnwidth]{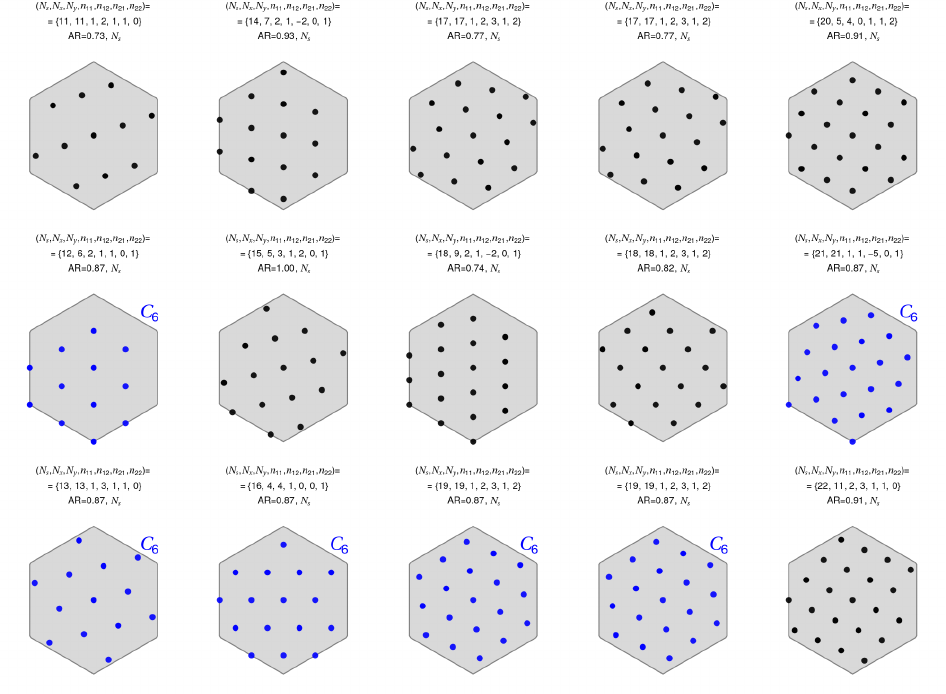}
\caption{Momentum meshes used for the 2BPV with "bandmax" calculations. We show all the geometries used for the neutral and charge gap calculations, and corresponding aspect ratio (AR). The calculations for the charge $e/3$ and $2e/3$ charge gaps involve three different system sizes, with $N_s-1,N_s$ and $N_s+1$ unit cells, see Eqs.$\,$\ref{eq:charge_gaps_eOver3_1}-\ref{eq:charge_gaps_2eOver3_4}. These are respectively shown from the top to bottom rows, while each column corresponds to the sizes used in a given calculation. For the charge-$e$ gaps, only the number of particles is varied (see Eqs.$\,$\ref{eq:charge_gaps_e1}-\ref{eq:charge_gaps_e4}) and therefore only the sizes in the middle row were used. The $C_6$ symmetric meshes are highlighted in blue.}
\label{fig:meshes_2BPV}
\end{figure}

In Tab.$\,$\ref{tab:dimH_2BPV} we show the maximum system sizes and Hilbert space sizes used for the
main calculations shown in the main text and in this section. All the momentum-space meshes we used for the calculations in this Appendix are shown in Fig.$\,$\ref{fig:meshes_2BPV}.

\begin{table}[h]
\centering
\begin{tabular}{|c|c|c|c|c|c|c|}
\hline
& $\Delta_N$ & $\Delta_N^s$ & $(e/3,0)$ &   $(\bar{e}/3,0)$ &   $(e/3,1)$ & $(\bar{e}/3,1)$ \\
\hline
dimH & $1.3\times10^8$ &   $8.7 \times 10^7$    &    $5.3\times10^7$      &    $2.9\times 10^8$                &   $ 10^9$       &     $1.7 \times 10^8$     \\
\hline
$N_s$ & 21 &    21    &   20       &    22      &      20    &   22                \\
\hline
$N_{\text{band1}}$ & 4 &    2    &   4       &    4      &      4    &   2                \\
\hline
\end{tabular}
\caption{Largest Hilbert space sizes (dimH) and corresponding  $N_s$ and $N_{\text{band1}}$ used in the 2BPV calculations, for the neutral gaps and different types of charge excitations defined by $(e^*,s)$ (see Eq.$\,$\ref{eq:eps_general}).}
\label{tab:dimH_2BPV}
\end{table}

\begin{figure}[h!]
\centering{}%
\includegraphics[width=\columnwidth]{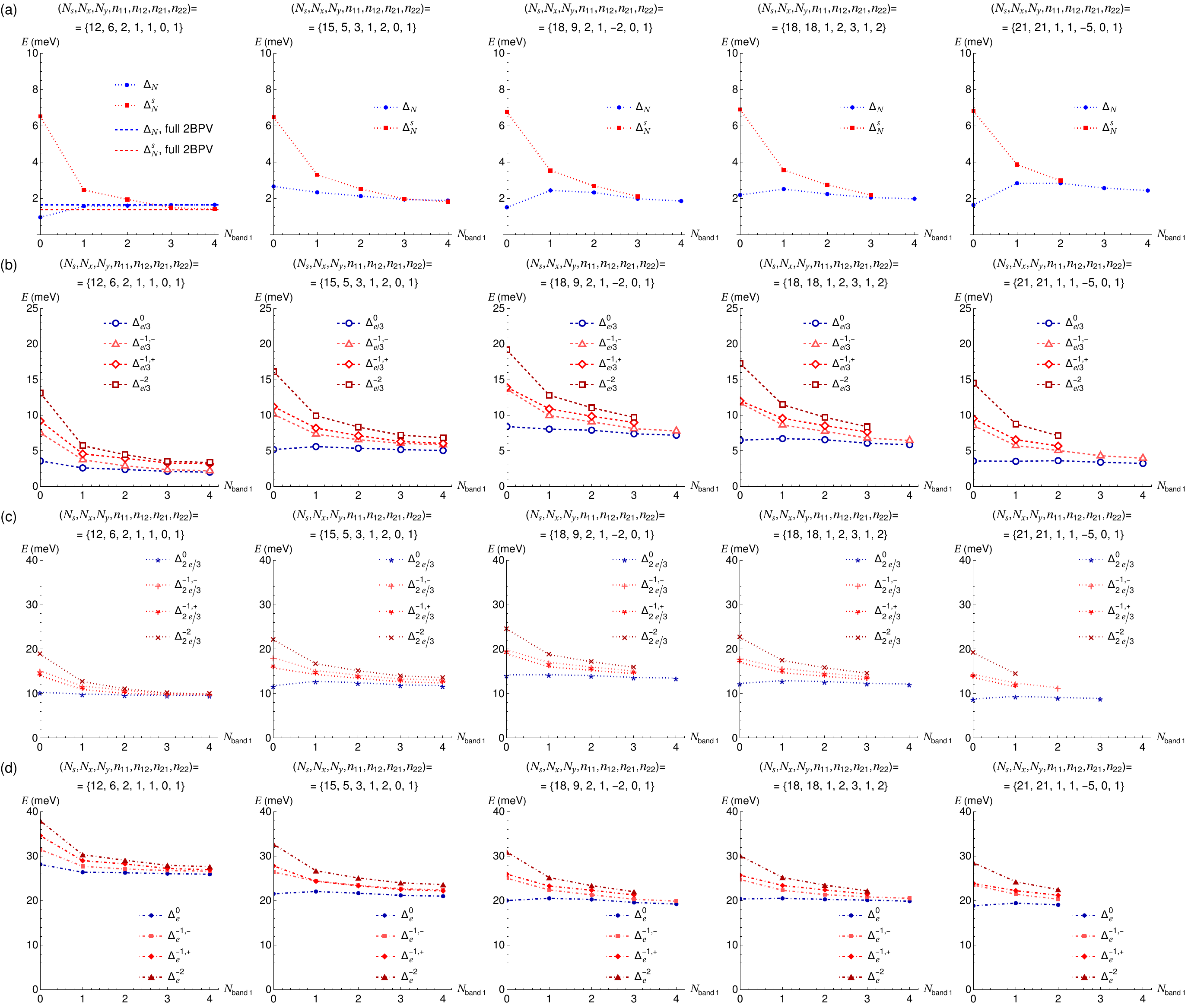}\caption{Effect of band mixing at $\nu=-2/3$ on the neutral gaps (a) and charge-$e/3$ (b), charge-$2e/3$ (c) and charge-$e$ (d) charge gaps, for $u=0.9$ and different system sizes specified in each figure. Note that for the charge gaps with fractional charge, additional system sizes with $N_s\pm1$ unit cells were used for the calculation (see Eqs.$\,$\ref{eq:charge_gaps_eOver3_1} -\ref{eq:charge_gaps_2eOver3_4}). All the momentum meshes used for these calculations are shown in Fig.$\,$\ref{fig:meshes_2BPV}}.
\label{fig:gaps_bmixing_all_u0p9}
\end{figure}

In Fig~\ref{fig:gaps_bmixing_all_u0p9} we show the dependence of the neutral and charge excitation gaps on $N_{\text{band1}}$ for several system sizes at $u  = 0.9$. 
The key trends are:

\vspace{0.5em}
\noindent\textbf{Neutral gaps:} The spin-1 neutral gap $\Delta_N^s$ is very sensitive to band mixing. It decreases rapidly with increasing $N_{\text{band1}}$ and becomes comparable to the spin-0 neutral gap $\Delta_N$. In comparison, $\Delta_N$ is relatively insensitive to band mixing.

\vspace{0.5em}
\noindent\textbf{Charge gaps:} The spin-0 charge gaps remain very stable as $N_{\text{band1}}$ increases. However, similarly to their neutral counterparts, the spinful charge gaps also decrease substantially, approaching the spin-0 gap (but always remaining larger). Taking the example of the charge-$e/3$ gaps, the ratio between the spin-1 gaps and the spin-0 gap can decrease by almost a factor of 2 (or even more for the spin-2 gap). 

\vspace{0.5em}

Focusing on the spin-0 neutral gap first, in Fig.$\,$\ref{fig:gaps_bmixing_all_u0p9}(a) we can see that there are still some finite-size effects. In particular, the relative differences for the largest $N_{\text{band1}}$ calculations among the three largest system sizes, $N_s=15,18$ and $21$ reaches $24\%$. We note that we considered two different momentum meshes with $N_s=18$ to evaluate the impact of changing the aspect ratio for a fixed system size. Changing the aspect ratio from 0.74 ($9 \times 2$ mesh) to 0.82 ($18 \times 1$ mesh) induces a mild relative increase of $7\%$ for the largest $N_{\text{band1}}$  calculations of $\Delta_N$, approaching the calculation for $N_s=21$ that has the optimal aspect ratio $\sqrt{3}/2$.

\begin{figure}[h!]
\centering{}%
\includegraphics[width=0.7\columnwidth]{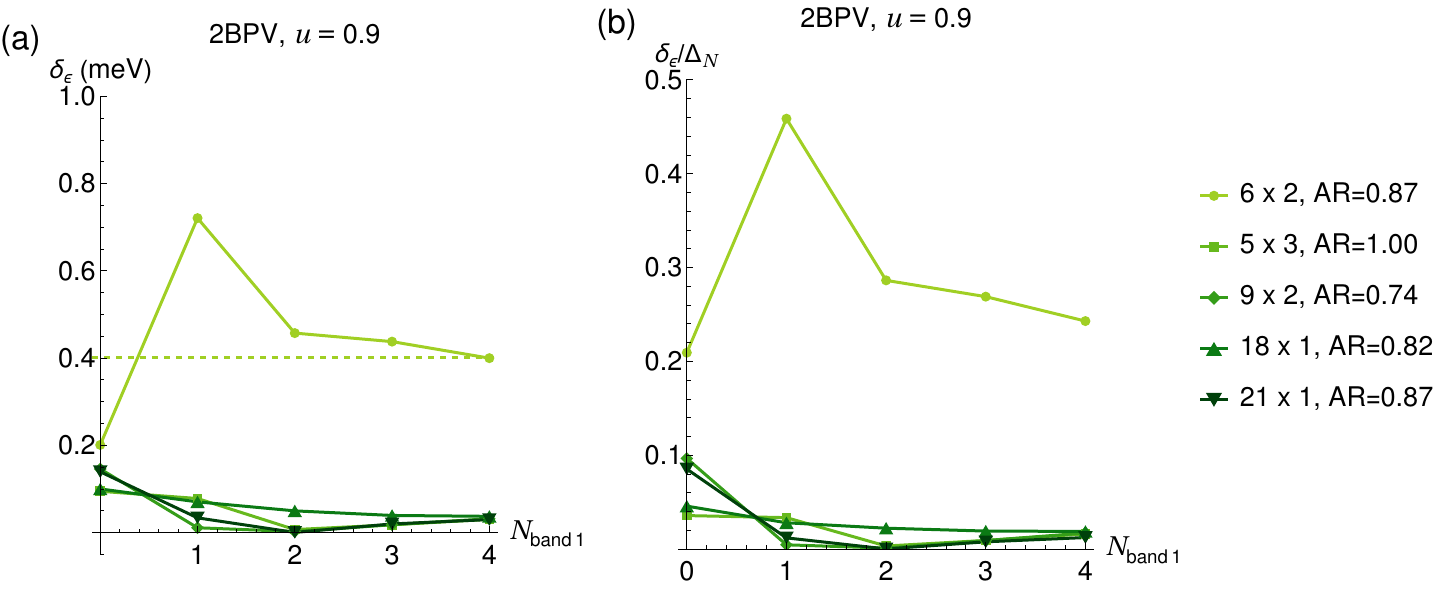}\caption{Finite-size spread of the quasi-degenerate FCI ground-states at $\nu=-2/3$, defined in Eq.$\,$\ref{eq:FCI_spread} (a) and ratio with spin-0 neutral gap (b). Results for 2BPV and $u=0.9$, for different  $N_{\textrm{band}1}$ and system sizes labeled by $N_x \times N_y$. The dashed line in (a) corresponds to the spread for the full 2BPV calculation, for $N_s=12$. 'AR' is the aspect ratio. The momentum meshes used in these calculations are shown in Fig.$\,$\ref{fig:meshes_2BPV} (middle row).}\label{fig:spread_2BPV}
\end{figure}

In order to inspect the quality of the FCI ground-states, in Fig.$\,$\ref{fig:spread_2BPV}(a) we show the finite-size spread defined in Eq.$\,$\ref{eq:FCI_spread}. Excepting the smallest system size ($N_s=12$) where the spread can reach $20 \%$ of the FCI neutral gap, the ratio $\delta_{\epsilon}/\Delta_N$ never exceeds $10\%$ ($\delta_{\epsilon}/\Delta_N=1.2\%$ for the largest $N_s$ and $N_{\textrm{band}1}$), see Fig.$\,$\ref{fig:spread_2BPV}(b). 

\begin{figure}[h!]
\centering{}%
\includegraphics[width=0.8\columnwidth]{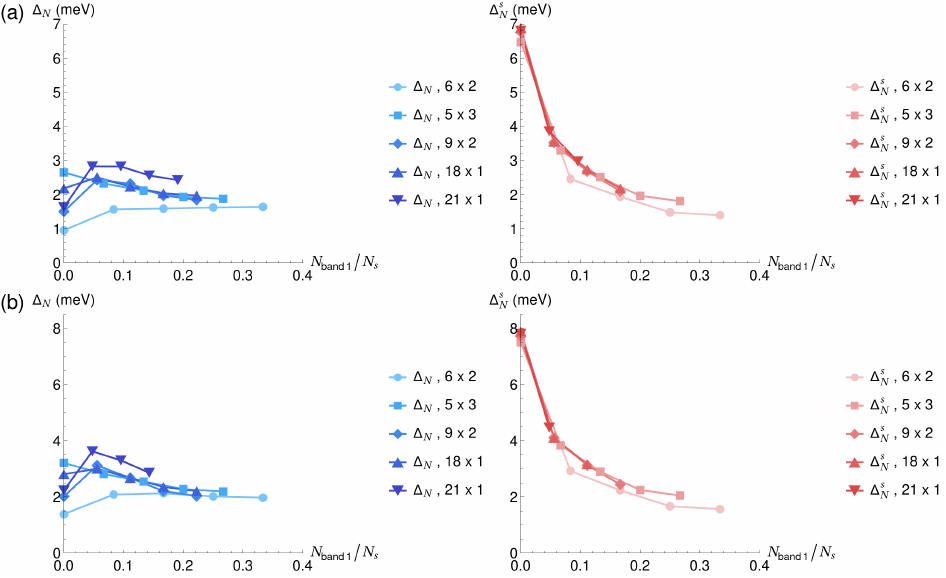}\caption{Scaling collapses of the spin-0 (blue) and spin-1 (red) neutral gap at $\nu=-2/3$ for different system sizes as a function of the density of holes in band 1, $\rho_1=N_{\textrm{band}1}/N_s$, for $u=0.9$ (a) and $u=1$ (b). The momentum meshes used in these calculations are shown in Fig.$\,$\ref{fig:meshes_2BPV} (middle row).  }\label{fig:gaps_bmixing_collapse_u0p9}
\end{figure}

\begin{figure}[h!]
\centering{}%
\includegraphics[width=\columnwidth]{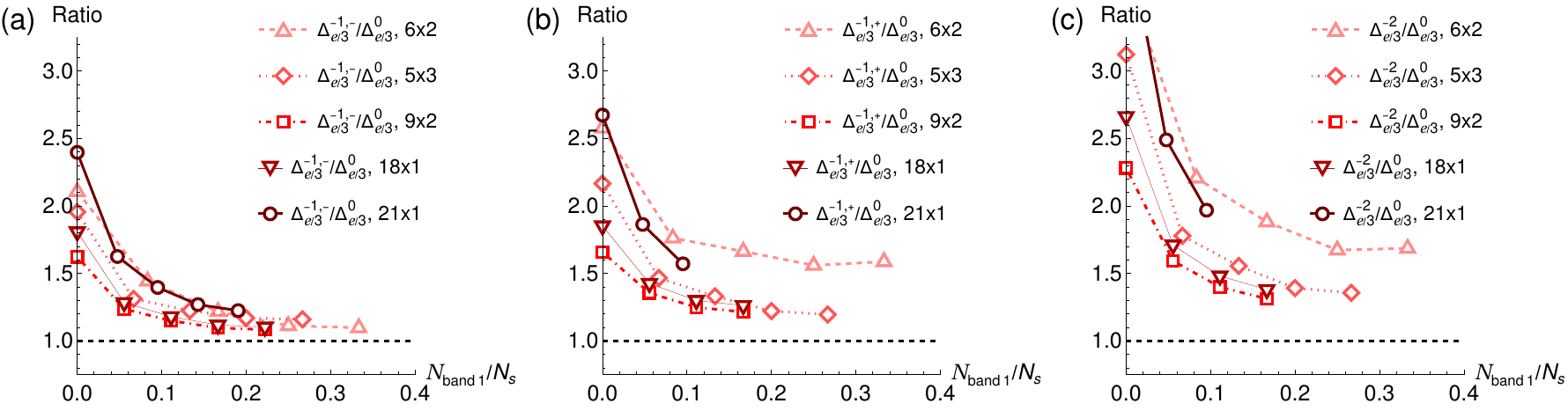}\caption{Ratios between all the $s\geq1$ charge-$e/3$ gaps and $\Delta_{e/3}^0$, as a function of the density of holes in band 1, $\rho_1=N_{\textrm{band}1}/N_s$, for $\nu=-2/3,u=0.9$ and different system sizes. (a) $\Delta_{e/3}^{-1,-}/\Delta_{e/3}^0$. (b) $\Delta_{e/3}^{-1,+}/\Delta_{e/3}^0$. (c) $\Delta_{e/3}^{-2}/\Delta_{e/3}^0$. All the momentum meshes used in these calculations are shown in Fig.$\,$\ref{fig:meshes_2BPV}.  }
\label{fig:gaps_bmixing_all_u0p9_ratios_collapse}
\end{figure}

Defining $\rho_1=N_{\textrm{band}1}/N_s$ as the density of occupied holes in band 1, the results at fixed density should converge for large enough sizes. We therefore expect the energy gaps to be scaling functions of $\rho_1$. With this in mind, we attempt a scaling collapse of the neutral gaps as a function of $\rho_1$, for the different system sizes. 
The results are shown in Fig.$\,$\ref{fig:gaps_bmixing_collapse_u0p9}(a). The data for $\Delta_N$ shows a good collapse for $15 \leq N_s \leq 18$. However, the results for $N_s=21$ deviate significantly, showing that there are still important finite-size effects. On the other hand, the spin-1 gaps show an excellent scaling collapse, suggesting that the asymptotic value of the corresponding scaling curve (at large $\rho_1$) provides a reliable estimate of the thermodynamic-limit $\Delta_N$. For completeness, in  Fig.$\,$\ref{fig:gaps_bmixing_all_u0p9_ratios_collapse}(b) we also show results for a different interaction strength, $u=1$, to show that the collapses are robust to changing the interaction strength.

Turning to the charge gaps, we can see in Figs.$\,$\ref{fig:gaps_bmixing_all_u0p9}(b-d) that there are stronger finite-size fluctuations than for the neutral gaps. The existence of finite-size fluctuations for the fractional-charge gaps is expected since their calculation depends on three different lattices, with $N_s$ and $N_s\pm1$ unit cells. In addition, even for the same system size fluctuations can occur for lattices with different geometry/aspect ratios. We illustrate this by computing the charge gaps using the two different momentum meshes for $N_s=18$, while keeping the geometries for the additional $17$ unit cell and $19$ unit cell systems involved in the calculation fixed (see geometries in Fig.$\,$\ref{fig:meshes_2BPV}). The relative differences for the $e/3$ charge gaps computed in these two cases  can reach $23\%$ for $N_{\text{band1}}=0$. The significant finite-size fluctuations show the importance of choosing lattices with aspect ratios close to $\sqrt{3}/2$ (recall this is the aspect ratio in the thermodynamic limit, enforced by $C_6$ symmetry), which is not always possible for small system sizes.
It is important to note that while for the charge-$e$ gaps there is a clear monotonic decrease of the gaps with system size, due to the residual repulsion between quasi-particles, the finite-size fluctuations for the charge-$e/3$ gaps are non-monotonic. This is expected from the 1BPV calculations in  Fig.$\,$\ref{fig:gaps_charge_u0p9_1BPV}, where we can see that the finite-size fluctuations only get under control for $N_s \gtrsim 20$. Crucially, we can see in Fig.$\,$\ref{fig:gaps_charge_u0p9_1BPV} that in the 1BPV case, the charge gap $\Delta_{e/3}^0$ computed for $N_s=21$ only differs by $13\%$ from the calculation using the largest system size  ($N_s=39$). This very good agreement is expected to hold with band mixing, given its small impact on $\Delta_{e/3}^0$.  With this in mind, we expect the results for $N_s=21$ to provide the most reliable  approximations for the thermodynamic-limit gaps. 

\begin{figure}[h!]
\centering{}%
\includegraphics[width=\columnwidth]{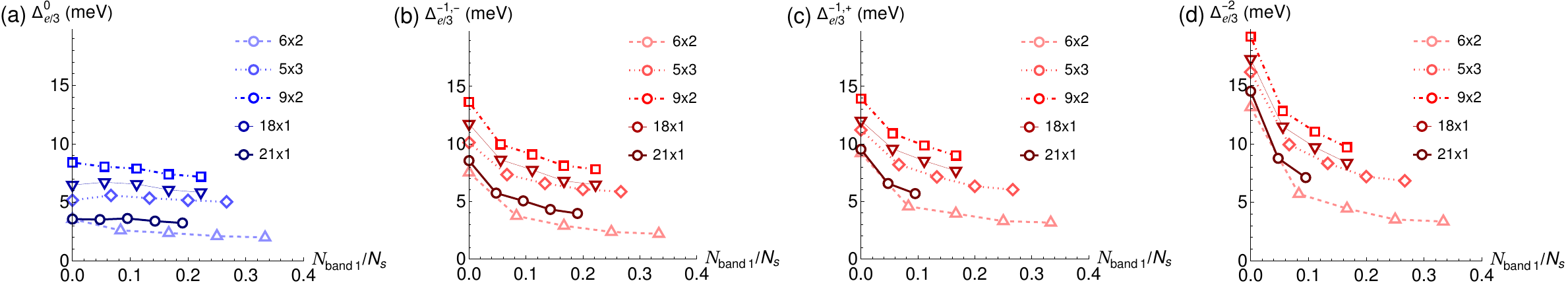}\caption{Charge-$e/3$ gap as a function of the density of holes in band 1, $\rho_1=N_{\textrm{band}1}/N_s$, at $\nu=-2/3$ and for different system sizes. (a) $\Delta_{e/3}^{0}$. (b) $\Delta_{e/3}^{-1,-}$. (c) $\Delta_{e/3}^{-1,+}$. (d) $\Delta_{e/3}^{-2}$. All the momentum meshes used in these calculations are shown in Fig.$\,$\ref{fig:meshes_2BPV}.  }
\label{fig:gaps_bmixing_all_u0p9_collapse_bare}
\end{figure}

\begin{figure}[h!]
\centering{}%
\includegraphics[width=\columnwidth]{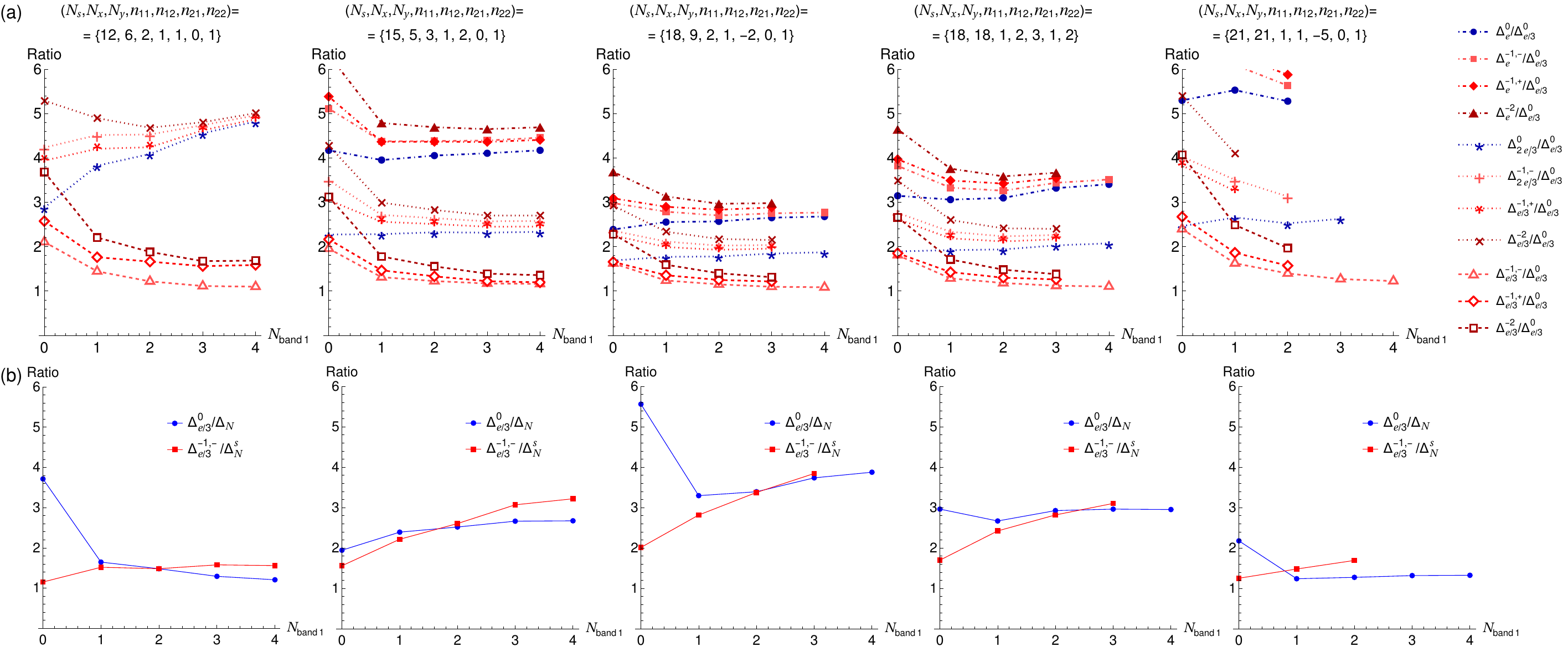}\caption{Impact of band mixing on ratios between excitation gaps, for $\nu=-2/3$ and $u=0.9$. (a) Ratios between all the charge gaps and the smallest charge gap $\Delta_{e/3}^0$. (b) Ratio between the spin-0 charge gap $\Delta_{e/3}^0$ and the corresponding spin-0 neutral gap  $\Delta_N$, together with the ratio between the smallest spin-1 charge gap $\Delta_{e/3}^{-1,-}$ and the corresponding spin-1 neutral gap $\Delta_N^s$. All the momentum meshes used in these calculations are shown in Fig.$\,$\ref{fig:meshes_2BPV}. }
\label{fig:gaps_bmixing_all_u0p9_ratios}
\end{figure}

The strong finite-size fluctuations of the charge-$e/3$ gaps can be more clearly seen in Fig.$\,$\ref{fig:gaps_bmixing_all_u0p9_collapse_bare}, where we plot the results for the different system sizes in the same figure. The ratio between the spinful and spin-0 charge-$e/3$ gaps is, however, less sensitive, as shown in Fig.$\,$\ref{fig:gaps_bmixing_all_u0p9_ratios}(a). For the charge-$2e/3$ and charge-$e$ gaps, the ratio is much more sensitive, which is expected from the 1BPV results and discussion in Appendix \ref{sec:1BPV}. In addition, for the largest considered size with  $N_s=21$, we have $\Delta_{2e/3}^0/\Delta_{e/3}^0\approx 2.6$  and $\Delta_{e}^0/\Delta_{e/3}^0\approx 5.3$ for the largest $N_{\textrm{band}1}$ calculations, still significantly different from the expected thermodynamic limit values $\Delta_{2e/3}^0/\Delta_{e/3}^0 =  2$  and $\Delta_{e}^0/\Delta_{e/3}^0 =  3$, yet another hint of the finite-size effects. With this in mind, we attempt a scaling collapse for the ratios in  Fig.$\,$\ref{fig:gaps_bmixing_all_u0p9_ratios_collapse}. For the ratio $\Delta_{e/3}^{-1,-}/\Delta_{e/3}^0$, although a good collapse cannot be made for all the system sizes, they extrapolate to approximately the same value. The remaining ratios, shown in Figs.$\,$\ref{fig:gaps_bmixing_all_u0p9_ratios_collapse}(b,c) also do not show a good-quality collapse. The ratio extrapolates to similar values for the system sizes with $N_s=15,18$, substantially different from the extrapolated value with the smallest size $N_s=12$. For $N_s=21$, we only reached convergence in $N_{\textrm{band1}}$ for the gaps $\Delta_{e/3}^0$ and $\Delta_{e/3}^{-1,-}$, with the latter requiring a Hilbert space size of $10^9$. The relative error between the calculations using $N_{\textrm{band1}}=3$ and $N_{\textrm{band1}}=4$ was respectively $5\%$ and $8\%$ for $\Delta_{e/3}^{0}$ and $\Delta_{e/3}^{-1,-}$.

In Fig.$\,$\ref{fig:gaps_bmixing_all_u0p9_ratios}(b) we also show the ratio between the spin-0 charge gap $\Delta_{e/3}^0$ and the corresponding spin-0 neutral gap  $\Delta_N$, as well as the ratio between the smallest spin-1 charge gap $\Delta_{e/3}^{-1,-}$ and the corresponding spin-1 neutral gap $\Delta_N^s$.  Although the strong finite-size effects are clear in this figure, the general trend is that $\Delta_{e/3}^0/\Delta_N$ decreases with band mixing or remains roughly constant, while $\Delta_{e/3}^{-1,-}/\Delta_N^s$ increases.

\begin{figure}[h!]
\centering{}%
\includegraphics[width=\columnwidth]{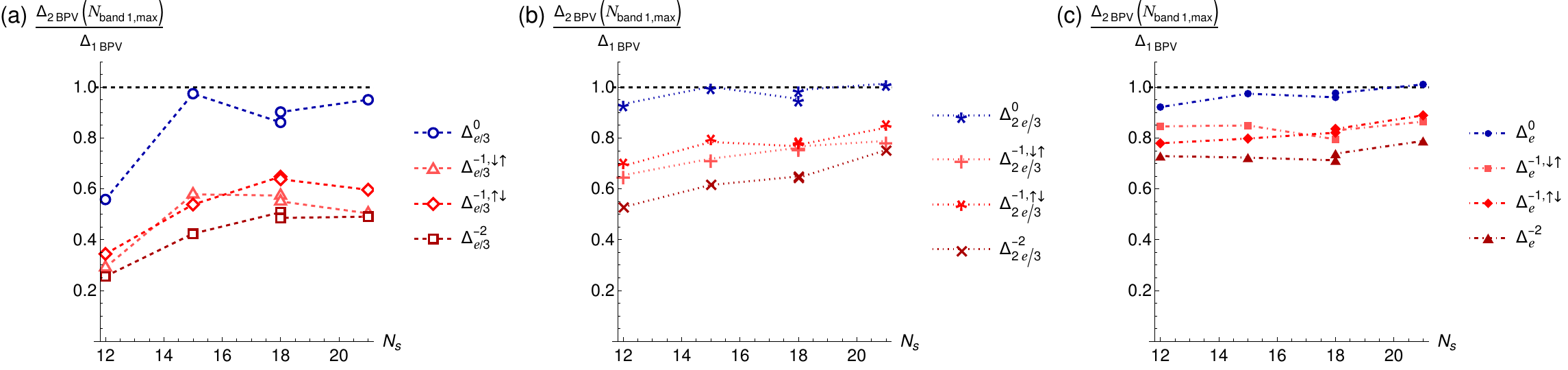}\caption{ Ratios between the charge gaps estimated from 2BPV and 1BPV calculations for $\nu=-2/3,u=0.9$, where in the former case we used the maximum computed $N_{\textrm{band}1}$. (a) Charge-$e/3$ gaps. (b) Charge-$2e/3$ gaps. (c) Charge-$e$ gaps.    }
\label{fig:ratios_2BPV_1BPV}
\end{figure}

Finally, in Fig.$\,$\ref{fig:ratios_2BPV_1BPV}, we show the ratios between the 2BPV calculation obtained for the largest computed $N_{\textrm{band}1}$ and the 1BPV calculation, for all the gaps. The ratios for the spinless gaps are always close to 1, reinforcing that band mixing has little effect on these gaps. On the other hand, the spinful gaps are significantly affected, with the charge-$e/3$ gaps being the most sensitive, in which case the ratios can get close to $1/2$.

\clearpage

\subsection{Spectra of spin-0 and spin-1 neutral excitations}

\begin{figure}[h!]
\centering{}%
\includegraphics[width=\columnwidth]{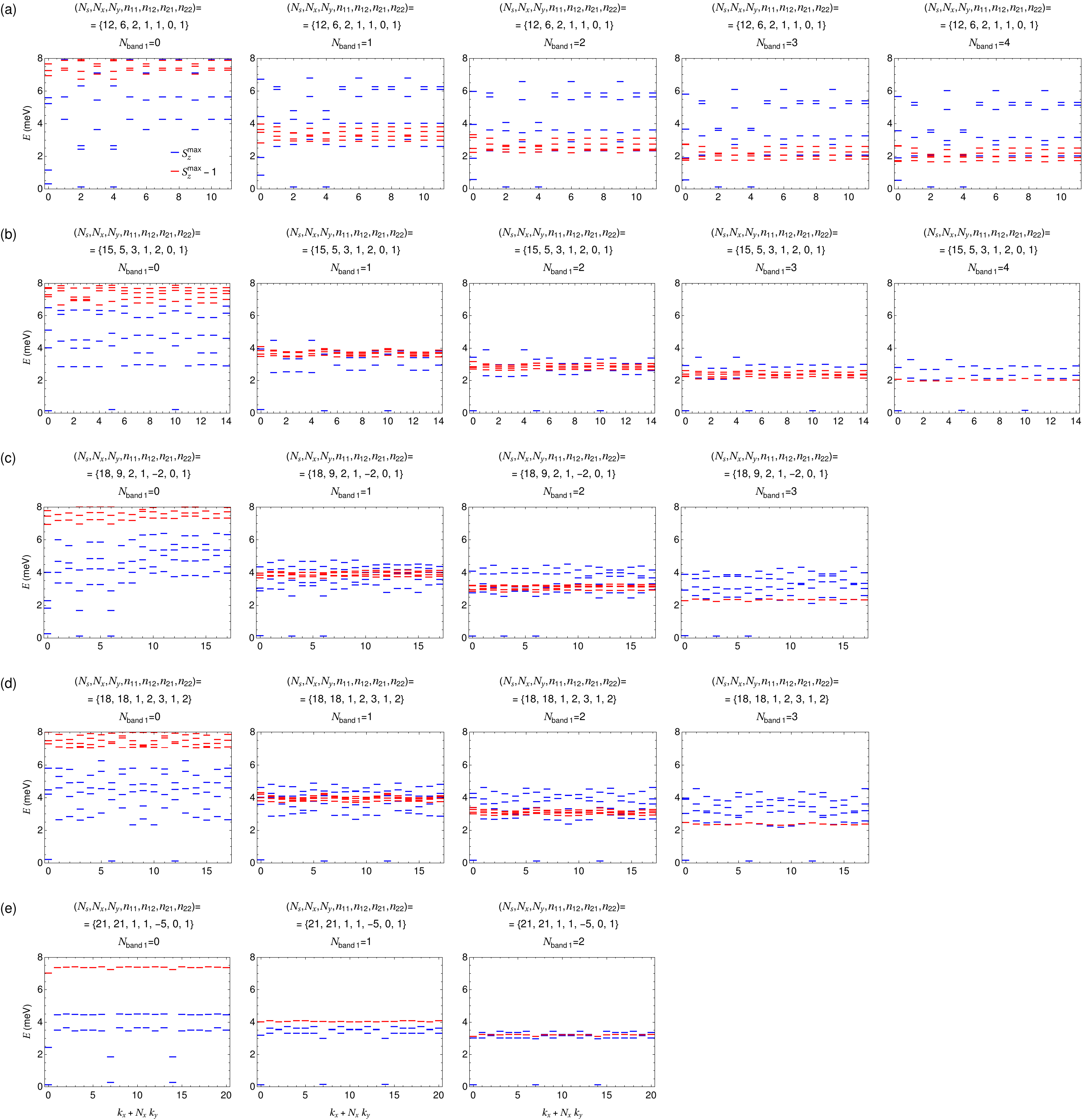}\caption{ Influence of band mixing in the neutral spectra at $\nu=-2/3$ for different system sizes shown in the different panels (a-e), for $u=0.9$. We show between 1 and 4 lowest eigenstates per momentum.}\label{fig:neutral_spec_bmax}
\end{figure}

In Fig.$\,$\ref{fig:neutral_spec_bmax} we study in more detail the influence of band mixing in the neutral spectra. Very significant changes with respect to the 1BPV calculations occur already for $N_{\textrm{band}1}=1$, for the studied system sizes, being much more prominent for the spin-1 excitation spectrum. 
A clear convergence of the spectra with increasing $N_{\textrm{band}1}$ can be seen for $N_s \leq 18$.
Comparing the results for the maximum $N_{\textrm{band}1}$ and $N_{\textrm{band}1}=0$ (equivalent to 1BPV), we see can that while there is a very significant change in the energy gap of the spin-1 excitations, the energy dispersion of the lowest energy modes (both in the spin-polarized and spin-1 sector) shows similarities between 1BPV and 2BPV for $N_s=18$ and $N_s=21$.

\clearpage

\subsection{Spectra of charge excitations}

\label{sec:spectra_charge_excitations_bmixing}

\begin{figure}[h!]
\begin{centering}
\includegraphics[width=\columnwidth]{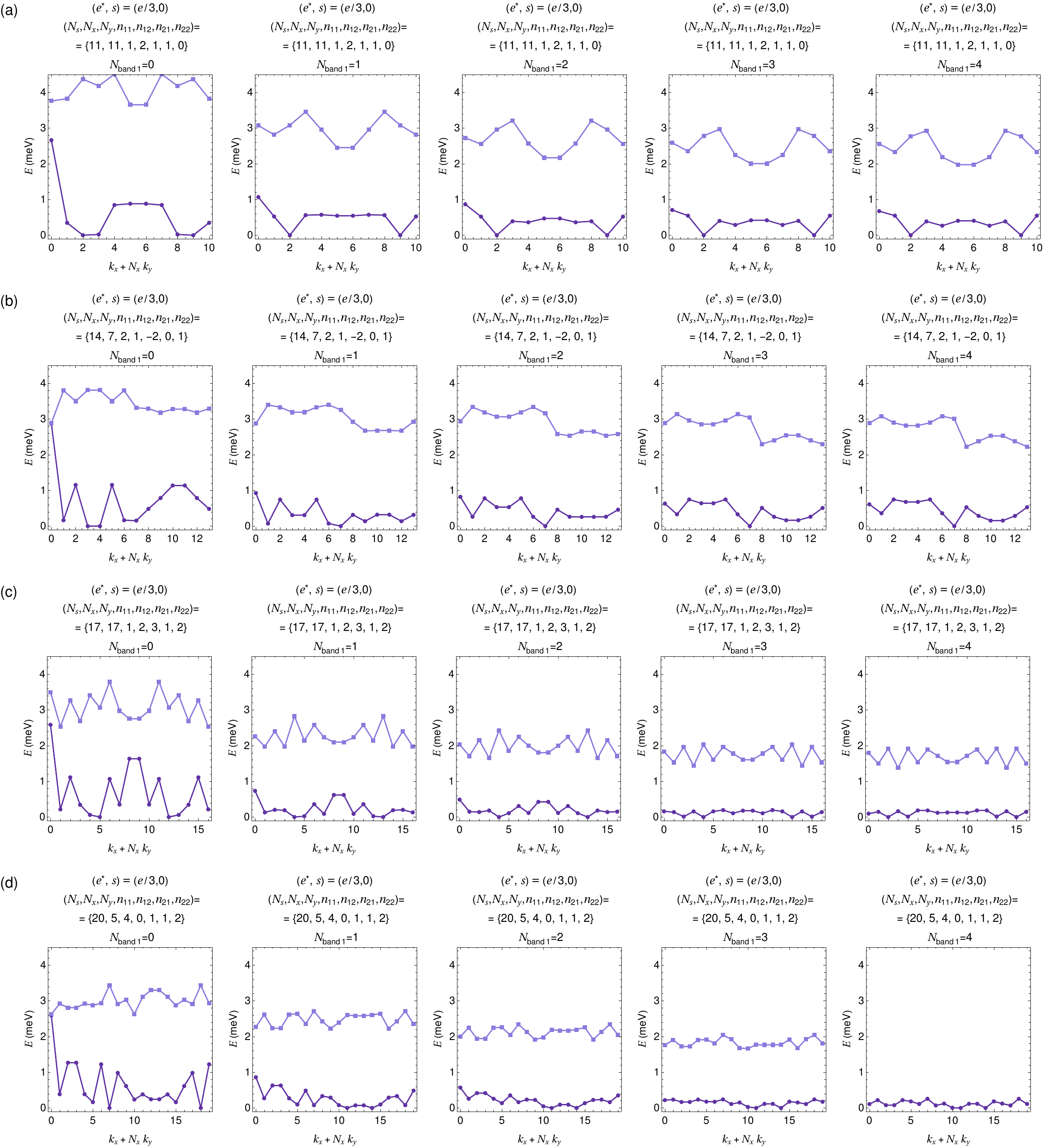}\caption{ 
Effect of band mixing at $\nu=-2/3$ on the two lowest-energy charge excitations specified by $(e^*,s;p',q')=(e/3,0,-1,-1)$, per momentum sector, and for $u=0.9$. $N_{\textrm{band1}}$ increases from left to right. Different system sizes, specified in the figures, are shown in (a-d).
 \label{fig:bmixing_eOver3_s0}}
\par\end{centering}
\end{figure}

\begin{figure}[h!]
\begin{centering}
\includegraphics[width=\columnwidth]{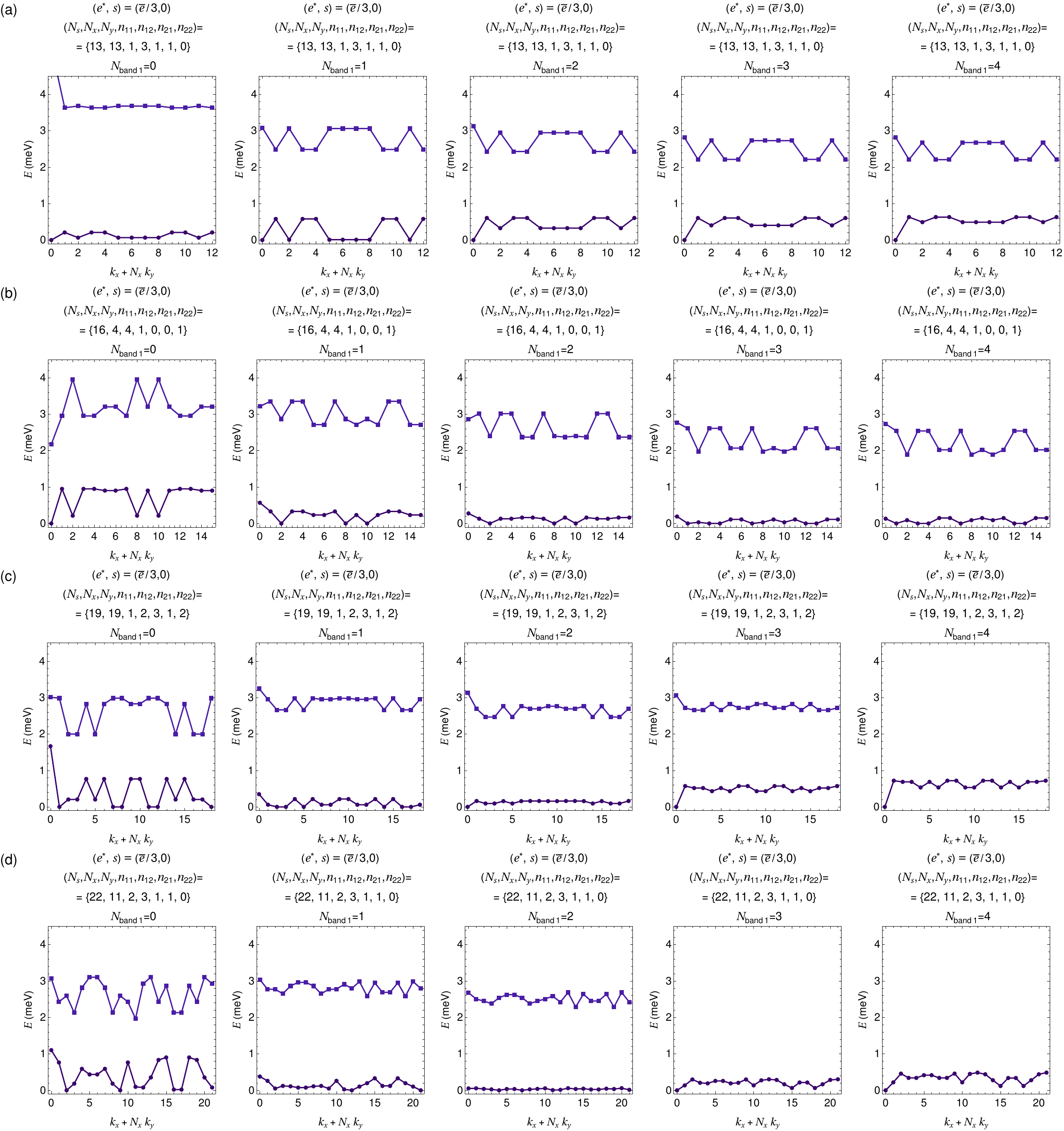}\caption{ 
Effect of band mixing at $\nu=-2/3$ on the two lowest-energy charge excitations specified by $(e^*,s;p',q')=(\bar{e}/3,0;1,1)$, per momentum sector, and for $u=0.9$. $N_{\textrm{band1}}$ increases from left to right. Different system sizes, specified in the figures, are shown in (a-d).
 \label{fig:bmixing_ebarOver3_s0}}
\par\end{centering}
\end{figure}

\begin{figure}[h!]
\begin{centering}
\includegraphics[width=\columnwidth]{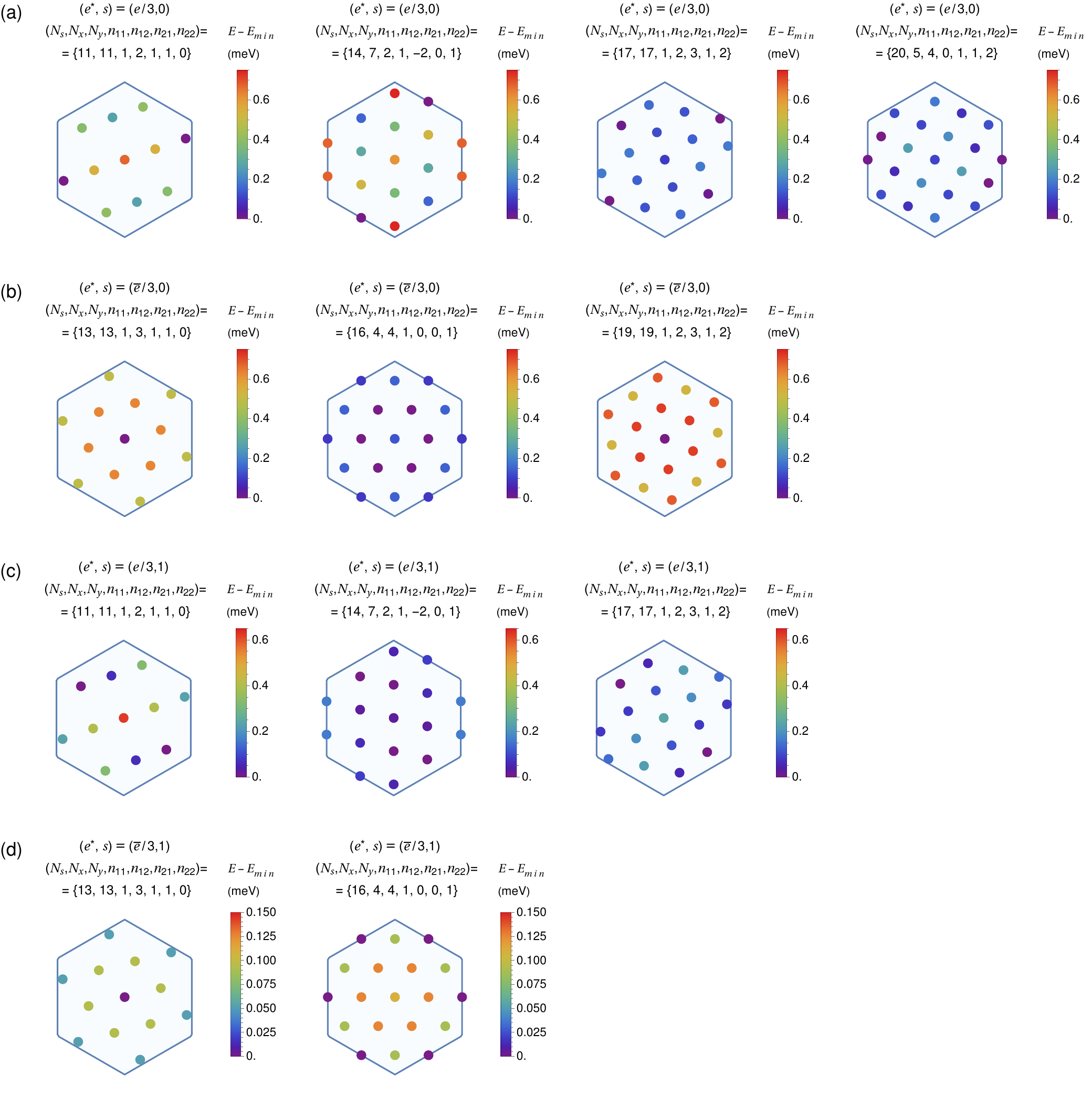}\caption{ 
Dispersions of the lowest-energy charge-$e/3$ excitations in the moiré Brillouin zone, for $\nu=-2/3,u=0.9$ and $N_{\textrm{band1}}=4$. (a) $(e^*,s;p',q')=(e/3,0,-1,-1)$. (b) $(e^*,s,p',q')=(\bar{e}/3,0,1,1)$. (c) $(e^*,s;p',q')=(e/3,1,-1,-1)$. (d) $(e^*,s;p',q')=(\bar{e}/3,1,1,1)$. 
The system sizes are indicated in the figures. The lowest energy is
shifted to be zero for each system size to make the comparison between different system sizes easier.
 \label{fig:bmixing_All_BZ}}
\par\end{centering}
\end{figure}

In this section we study the impact of band mixing on the spectra of charge excitations. The results are shown in Figs.$\,$\ref{fig:bmixing_eOver3_s0}-\ref{fig:bmixing_All_BZ}.

In Figs.$\,$\ref{fig:bmixing_eOver3_s0} and \ref{fig:bmixing_ebarOver3_s0} we show the evolution of respectively the charge-$e/3$ quasi-electron and quasi-hole spectra ($s=0$ excitations) with $N_{\textrm{band1}}$. The common trend observed for the different system sizes is that in both cases, band mixing narrows the dispersion of the lowest-energy excitation, and as a consequence increases the gap to higher-energy excitations. By comparing the results for the different system sizes, we can infer that there are still strong finite-size effects, e.g. the bandwidth for the quasi-electron varies between $0.18$ and $0.74$ for the quasi-electron and between $0.15$ and $0.72$ for the quasi-hole dispersions. It is however clear that although  band mixing does not have a significant impact on the spinless charge-$e/3$ gap, it substantially affects the dispersion of the lowest-energy excitation compared to the 1BPV calculations. 

\begin{figure}[h!]
\begin{centering}
\includegraphics[width=\columnwidth]{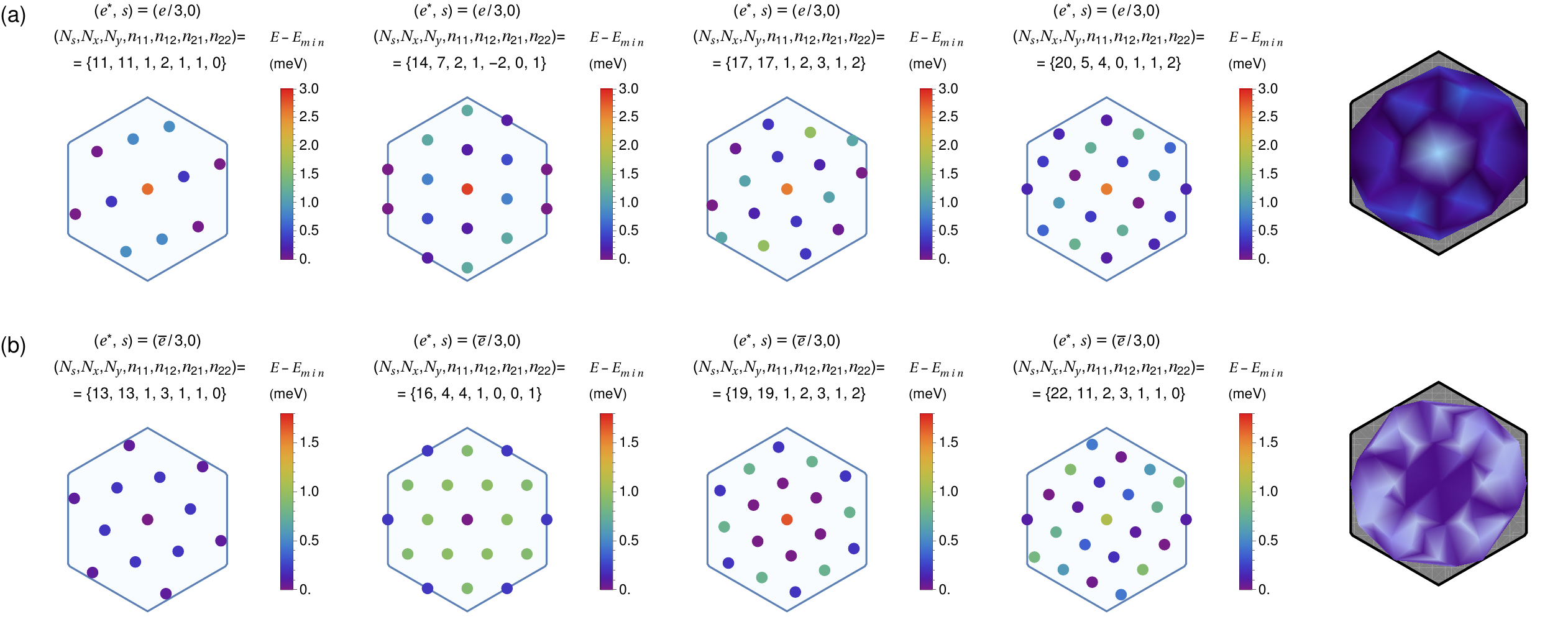}\caption{ 
Dispersions of the lowest-energy spinless charge-$e/3$ excitations in the moiré Brillouin zone, for $\nu=-2/3,u=0.9$ and $N_{\textrm{band1}}=0$ (equivalent to 1BPV), restricting to the system sizes for which it is possible to converge the 2BPV calculations in $N_{\textrm{band1}}$. (a) $(e^*,s;p',q')=(e/3,0,-1,-1)$. (b) $(e^*,s;p',q')=(\bar{e}/3,0,1,1)$.
The system sizes are indicated in the figures. The last columns show density plots obtained by combining the data for all the system sizes shown in the figure. The lowest energy is
shifted to be zero for each system size to make the comparison between different system sizes easier.
 \label{fig:bmixing_All_BZ_1BPV}}
\par\end{centering}
\end{figure}

In Figs.$\,$\ref{fig:bmixing_All_BZ} (a,b), we show the quasi-electron and quasi-hole dispersions in the moiré Brillouin zone for the largest $N_{\textrm{band1}}(=4)$ we considered (in which case we observe a good convergence of the results). It is clear from these results that the dispersions are not consistent between different system sizes, and therefore it is not possible to do a meaningful collapse of the data as it was done in Fig.$\,$\ref{fig:qh_qp_dispersion_u0p9}(a) for the 1BPV calculations. This is expected since we can only obtain  convergence in $N_{\textrm{band1}}$ for system sizes with $N_s < 20$, smaller than the smallest size used for the collapse in the 1BPV calculations, in Fig.$\,$\ref{fig:qh_qp_dispersion_u0p9}(a). To further back this claim, we also show the energy dispersions for $N_{\textrm{band1}}=0$ (equivalent to 1BPV) in Fig.$\,$\ref{fig:bmixing_All_BZ_1BPV}, for $N_s<20$, where we can also see significant variability among the different system sizes and almost featureless interpolations when we combine all the data.

\begin{figure}[h!]
\begin{centering}
\includegraphics[width=\columnwidth]{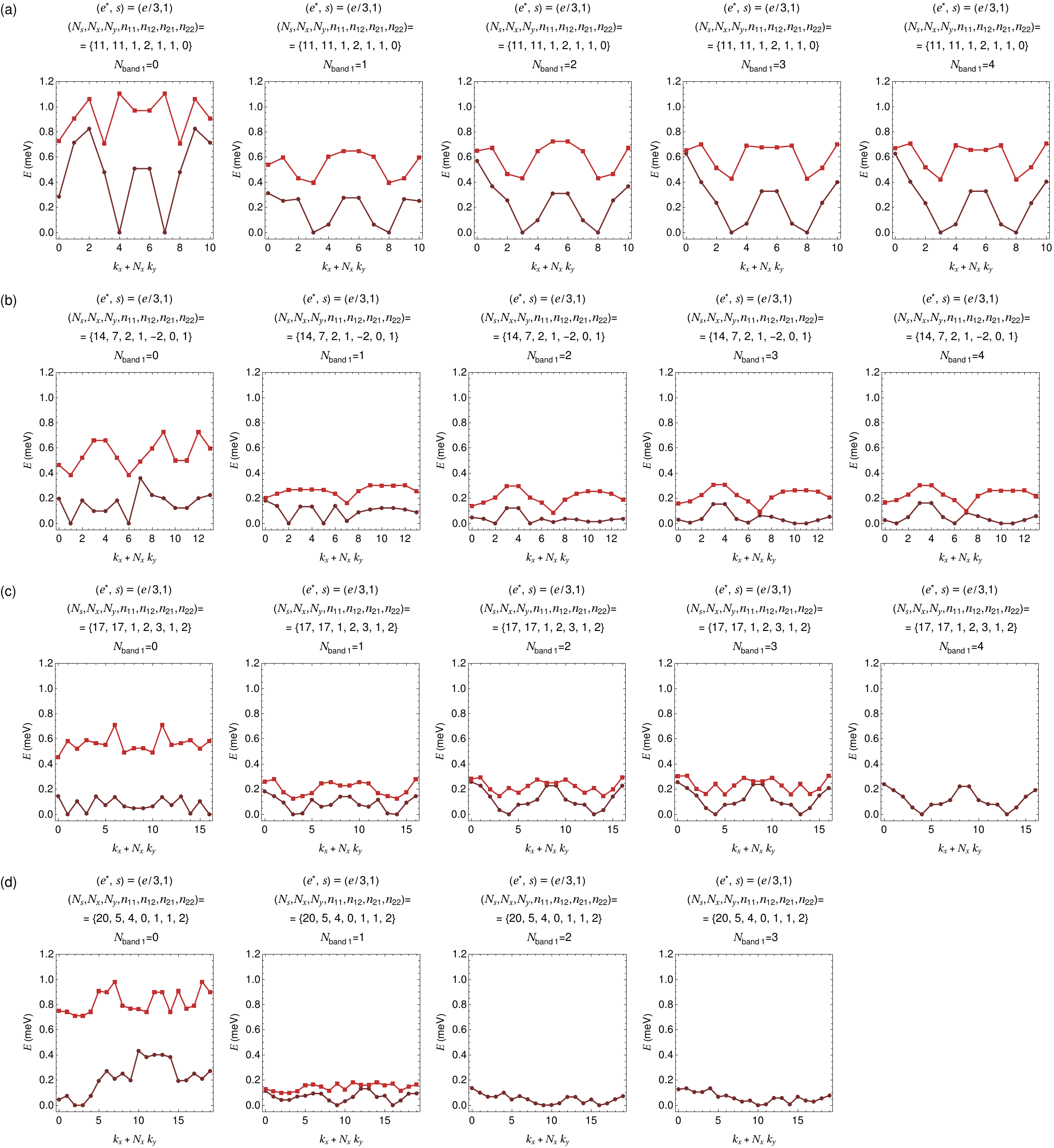}\caption{ 
Effect of band mixing at $\nu=-2/3$ on the two lowest-energy charge excitations specified by $(e^*,s;p',q')=(e/3,1,-1,-1)$, per momentum sector, and for $u=0.9$. $N_{\textrm{band1}}$ increases from left to right. Different system sizes, specified in the figures, are shown in (a-d).
 \label{fig:bmixing_eOver3_s1}}
\par\end{centering}
\end{figure}

\begin{figure}[h!]
\begin{centering}
\includegraphics[width=\columnwidth]{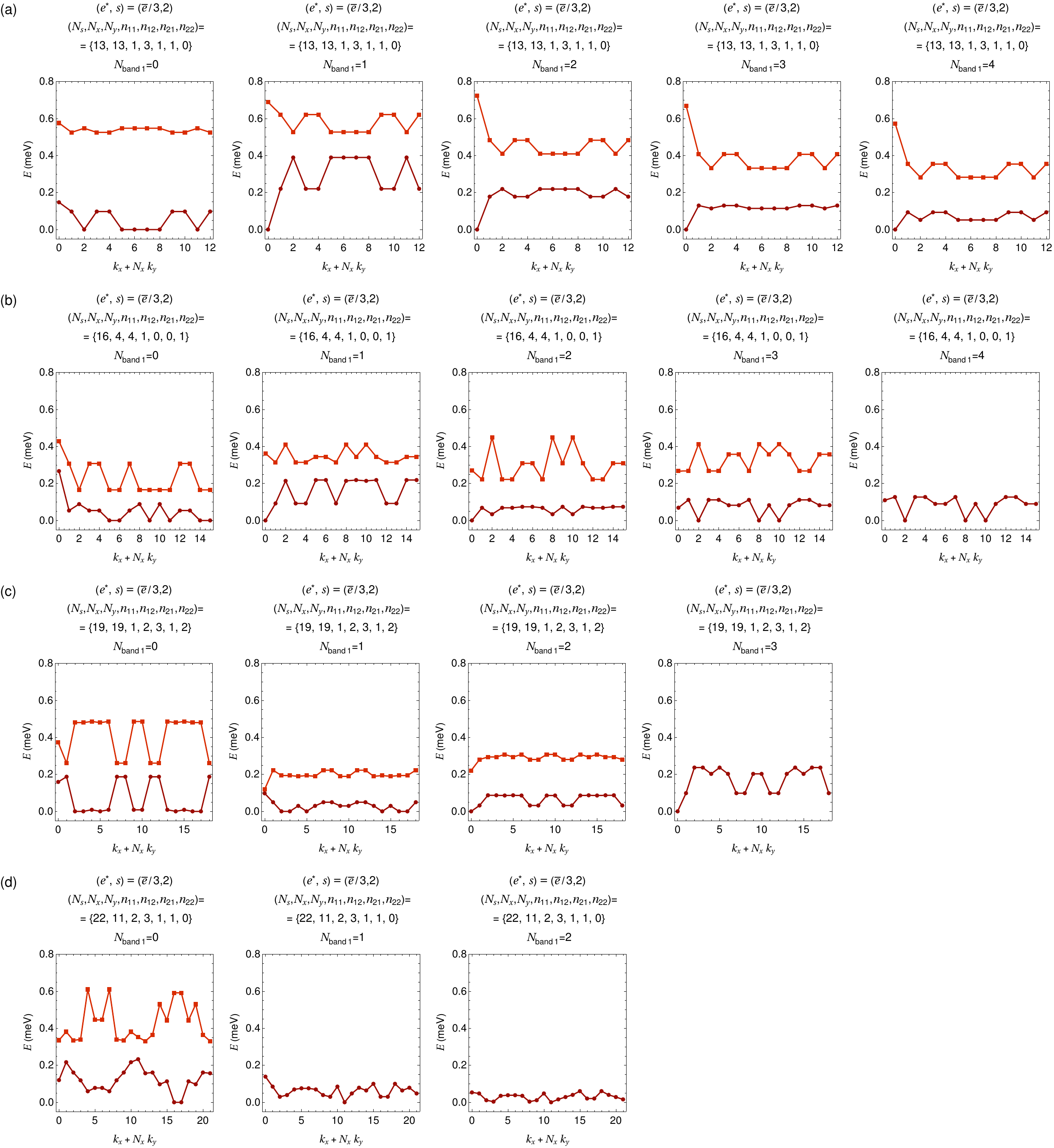}\caption{ 
Effect of band mixing at $\nu=-2/3$ on the two lowest-energy charge excitations specified by $(e^*,s;p',q')=(\bar{e}/3,1,1,1)$, per momentum sector, and for $u=0.9$. $N_{\textrm{band1}}$ increases from left to right. Different system sizes, specified in the figures, are shown in (a-d).
 \label{fig:bmixing_ebarOver3_s1}}
\par\end{centering}
\end{figure}

For the magnon-dressed quasi-electron and quasi-hole spectra ($s=1$ excitations), there are also significant changes, as can be seen in Figs.$\,$\ref{fig:bmixing_eOver3_s1} and \ref{fig:bmixing_ebarOver3_s1}. The dispersion remains very narrow even with band mixing. For the quasi-electron,  Fig.$\,$\ref{fig:bmixing_eOver3_s1}, band mixing shows a clear trend of closing the energy gap between the lowest and higher energy excitations. Note that while there is no clear gap with $N_{\textrm{band1}}=0$ for the smaller sizes $ N_s \leq 14$,  it develops for $N_s>17$, see also the 1BPV calculations for larger sizes in Fig.$\,$\ref{fig:qh_qp_spectra_multiple_evs}(a) in Appendix \ref{sec:1BPV}. The same trend is not observed for the magnon-dressed quasi-hole, see Fig.$\,$\ref{fig:bmixing_ebarOver3_s1}, with a (small) gap still being present for the largest accessible $N_{\textrm{band1}}$. As expected from the results in the 1BPV case, where we did not observe a consistent dispersion for the different system sizes, this is also the case here where the sizes are substantially smaller, see  Figs.$\,$\ref{fig:bmixing_All_BZ}(c,d). 

\clearpage

\subsection{Single-particle correlation functions}

\begin{figure}[h!]
\centering{}%
\includegraphics[width=\columnwidth]{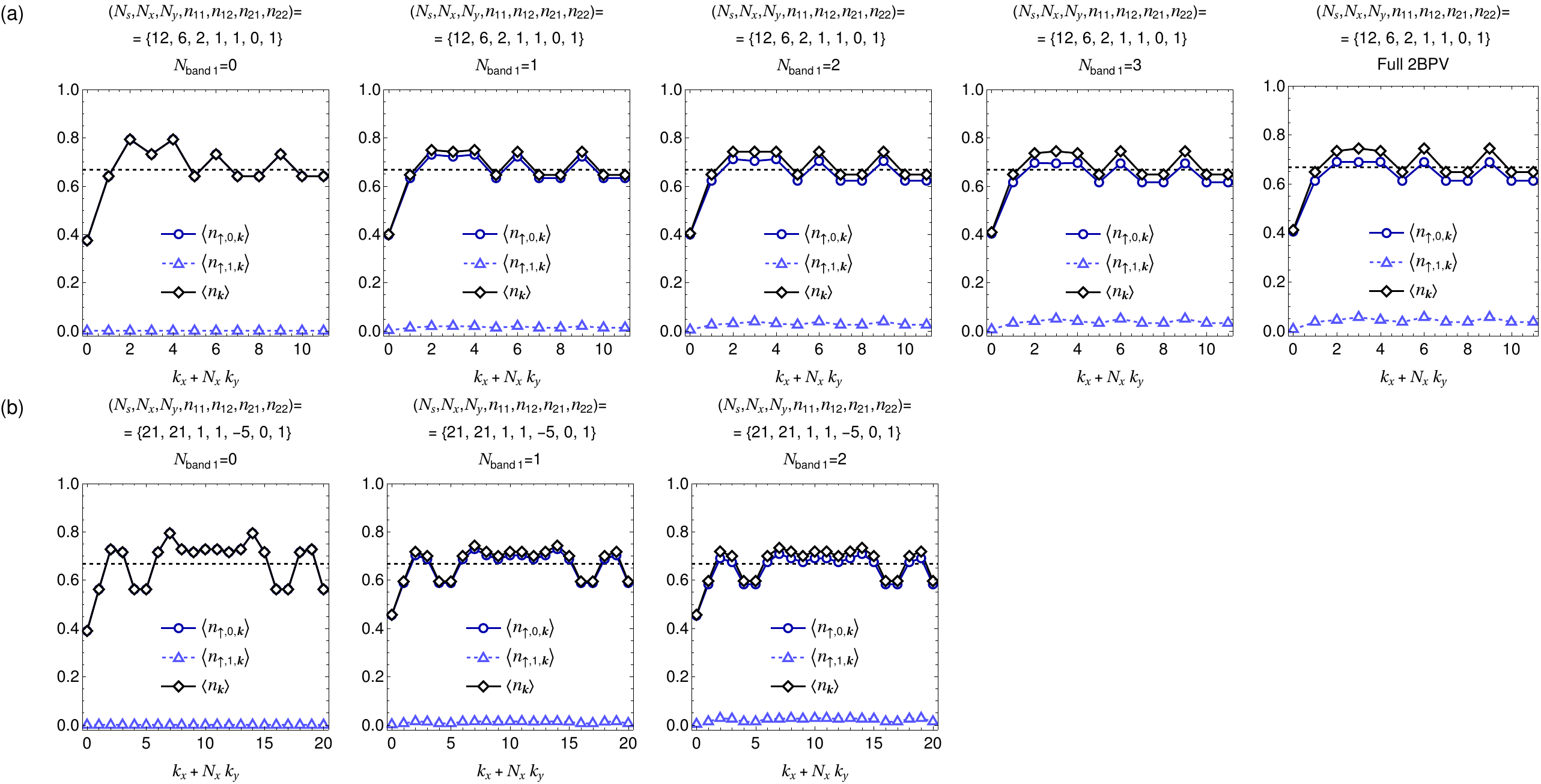}\caption{Hole band occupation numbers and full hole occupation number $\langle n_{{\bf k}} \rangle = \langle n_{\up,0,{\bf k}} \rangle + \langle n_{\up,1,{\bf k}} \rangle$, for $\nu=-2/3$, $u=0.9$ and the 12 unit cell (a) and 21 unit cell (b) systems specified in the figures. Different columns correspond to  different $N_{\textrm{band1}}$. The results were averaged over the three-fold quasi-degenerate FCI ground-states.}\label{fig:nk_12_21_bandmixing}
\end{figure}

We finish our analysis of band mixing by studying its impact on one-body correlation functions. 
We start by analyzing the impact on the band occupation numbers of the FCI ground-states at filling $\nu=-2/3$, shown in Fig.$\,$\ref{fig:nk_12_21_bandmixing}. Band mixing has minimal impact, with the occupation in band 1 being $\langle n_{{\bf k},1,\up} \rangle < 0.06$. In addition, maximum relative differences in  $\langle n_{{\bf k}} \rangle$ are smaller than $16 \%$, comparing the 1BPV and maximum $N_{\textrm{band1}}$ calculations.

\begin{figure}[h!]
\centering{}%
\includegraphics[width=0.75\columnwidth]{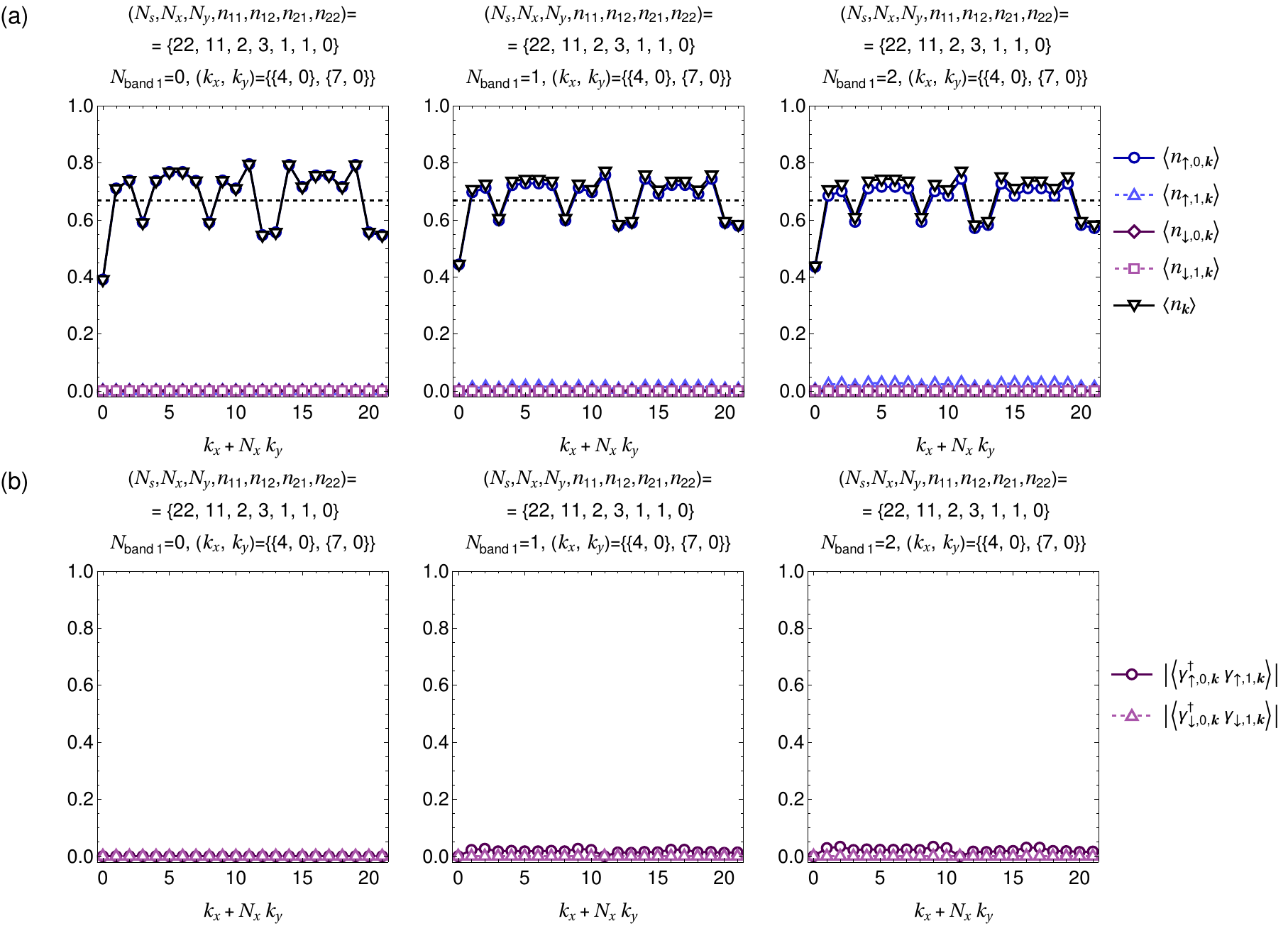}\caption{Impact of band mixing on single-particle correlation functions, for the lowest-energy many-body states containing the quasi-hole defined by $(e^*,s)=(\bar{e}/3,0)$.  (a) Hole band occupations. (b) Interband correlation functions. All the results are for $u=0.9$ and system size defined by $(N_x,N_y,n_{11},n_{12},n_{21},n_{22})=(11,2,3,1,1,0)$. Different columns correspond to different $N_{\textrm{band1}}$. Note that, as can be seen in Fig.$\,$\ref{fig:bmixing_ebarOver3_s0}, the momentum-sectors of the lowest-energy states can change with $N_{\textrm{band1}}$. For all the calculations in this figure, we considered the momentum sectors of the lowest-energy states obtained for the largest shown $N_{\textrm{band1}}$, which correspond to $(k_x,k_y)=(4,0)$ and $(k_x,k_y)=(7,0)$.}\label{fig:nk_bandmixing_ebarOver3_s0}
\end{figure}

\begin{figure}[h!]
\centering{}%
\includegraphics[width=\columnwidth]{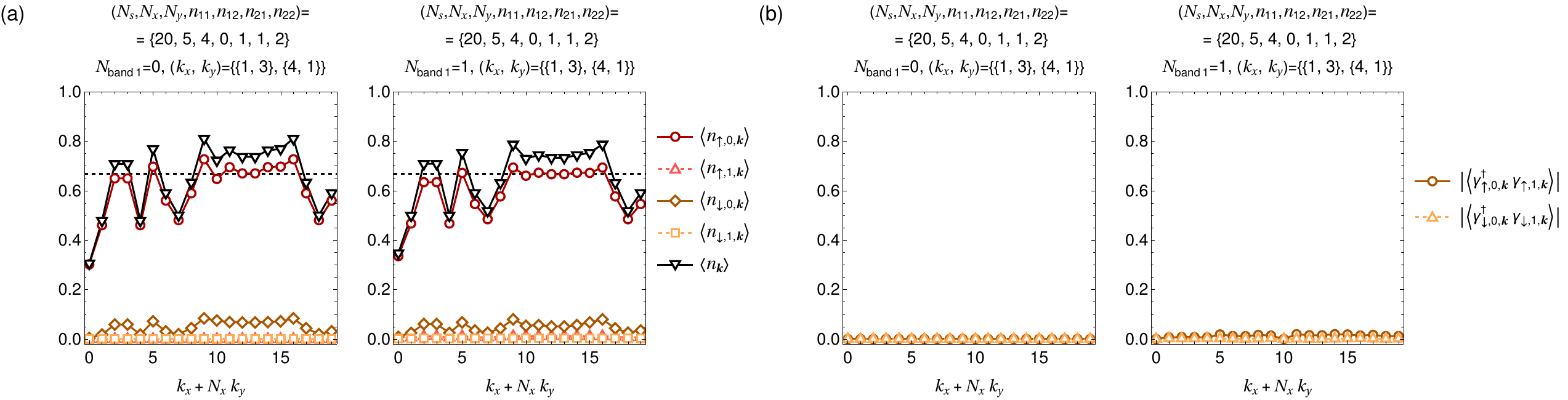}\caption{
Impact of band mixing on single-particle correlation functions, for the lowest-energy many-body states containing the magnon-dressed quasi-electron defined by $(e^*,s)=(e/3,1)$.  (a) Hole band occupations. (b) Interband correlation functions. All the results are for $u=0.9$ and system size defined by $(N_x,N_y,n_{11},n_{12},n_{21},n_{22})=(5,4,0,1,1,2)$.
Different columns correspond to different $N_{\textrm{band1}}$. Note that, as can be seen in Fig.$\,$\ref{fig:bmixing_eOver3_s1}, the momentum-sectors of the lowest-energy states can change with $N_{\textrm{band1}}$. For all the calculations in this figure, we considered the momentum sectors of the lowest-energy states obtained for the largest shown $N_{\textrm{band1}}$, which correspond to $(k_x,k_y)=(1,3)$ and $(k_x,k_y)=(4,1)$.}\label{fig:nk_bandmixing_eOver3_s1}
\end{figure}

The impact of band mixing on the single-particle correlation functions for the $s=0$ quasi-electron and quasi-hole excitations is also minimal. Some representative results are shown for the latter case in Fig.$\,$\ref{fig:nk_bandmixing_ebarOver3_s0}. We compute both the band-1 occupations and the interband correlations defined as $|\langle \gamma^{\dagger}_{\up,0,{\bf k}} \gamma_{\up,1,{\bf k}} \rangle|$ and $ |\langle \gamma^{\dagger}_{\down,0,{\bf k}} \gamma_{\down,1,{\bf k}} \rangle|$, where $ \gamma^{\dagger}_{\eta,m,{\bf k}}$ are the band operators defined in Eq.$\,$\ref{eq:band_operators}. They become finite with band mixing, being however always very small, with $\langle n_{\up,1,{\bf k}} \rangle, |\langle \gamma^{\dagger}_{\up,0,{\bf k}} \gamma_{\up,1,{\bf k}} \rangle|, |\langle \gamma^{\dagger}_{\down,0,{\bf k}} \gamma_{\down,1,{\bf k}} \rangle| < 0.05$. The same holds true for $s=1$ 
 quasi-electron and quasi-hole excitations, with some representative results for the former shown in Fig.$\,$\ref{fig:nk_bandmixing_eOver3_s1}.

\clearpage

\section{Atomic-scale Hubbard interaction}

\label{sec:Hubbard}

In this section we provide details on the phenomenological  atomic Hubbard interaction used in the main text, deriving its contribution to the interacting Hamiltonian in Eq.$\,$\ref{eq:H_full_band_basis} and presenting some additional results. 

\subsection{Contribution of atomic Hubbard interaction to Hamiltonian }
It has been established by\textit{ab-initio} calculations that the valence band maximum
around ${\bf K}$ (${\bf K}'$) is spanned by the $d_{x^{2}-y^{2}}+id_{xy}$ ($d_{x^{2}-y^{2}}-id_{xy}$) orbitals 
located around the Mo sites \cite{MFCI_0}. However, these calculations tend to underestimate the correlation effects of residual atomic-scale interactions between tightly localized orbitals --- like the $d$ orbitals in MoTe$_2$ \cite{PhysRevB.44.943, pavarini2012correlated}. As we show below on this appendix, including these additional short-range repulsive interactions can further stabilize magnetism on MoTe$_2$. To explore these additional contributions to the interaction, we define the operator $c_{{\bf R}_{\ell},\ell,\sigma}^{\dagger}$
that creates an electron with spin $\sigma$ in the orbitals at site
${\bf R}_{\ell}$ in layer $\ell=b,t$. We will assume that the short-range interactions are not well captured by DFT and will model them by an additional atomistic Hubbard interaction given by 

\begin{equation}
H_{U}=U_H\sum_{{\bf R}_{\ell},\ell}c_{{\bf R}_{\ell},\ell,\up}^{\dagger}c_{{\bf R}_{\ell},\ell,\up}c_{{\bf R}_{\ell},\ell,\down}^{\dagger}c_{{\bf R}_{\ell},\ell,\down}
\label{eq:Hubbard_init}
\end{equation}

Applying the particle-hole transformation $c_{{\bf R}_{\ell},\ell,\sigma}^{\dagger}\rightarrow \tilde{c}_{{\bf R}_{\ell},\ell,\sigma}$ and $c_{{\bf R}_{\ell},\ell,\sigma}\rightarrow \tilde{c}_{{\bf R}_{\ell},\ell,\sigma}^{\dagger}$, in order to work with the hole operators $\tilde{c}_{{\bf R}_{\ell},\ell,\sigma}^{\dagger}$, we get

\begin{equation}
\begin{array}{cc}
H_{U}= & U_{H}\sum_{{\bf R}_{\ell},\ell}\tilde{c}_{{\bf R}_{\ell},\ell,\up}^{\dagger}\tilde{c}_{{\bf R}_{\ell},\ell,\up}\tilde{c}_{{\bf R}_{\ell},\ell,\down}^{\dagger}\tilde{c}_{{\bf R}_{\ell},\ell,\down}-U_{H}\sum_{{\bf R}_{\ell},\ell,\sigma}\tilde{c}_{{\bf R}_{\ell},\ell,\sigma}^{\dagger}\tilde{c}_{{\bf R}_{\ell},\ell,\sigma}+U_H N_{\textrm{sites}}\\
 & =U_{H}\sum_{{\bf R}_{\ell},\ell}\tilde{c}_{{\bf R}_{\ell},\ell,\up}^{\dagger}\tilde{c}_{{\bf R}_{\ell},\ell,\up}\tilde{c}_{{\bf R}_{\ell},\ell,\down}^{\dagger}\tilde{c}_{{\bf R}_{\ell},\ell,\down}-U_{H}N+U_H N_{\textrm{sites}}
\end{array}
\label{eq:Hubbard_hole}
\end{equation}

\noindent where $N_{\textrm{sites}}$ is the total number of sites and $N$ is the total number of holes, that we take to be a constant since we work at a fixed number of holes. In what follows, we will drop the constants in Eq.$\,$\ref{eq:Hubbard_hole}, therefore only keep the first term. We define the basis of plane-waves in the single-layer Brillouin zone as

\begin{equation}
a_{{\bf p},\ell,\sigma}^{\dagger}=\frac{1}{\sqrt{N_{SL}}}\sum_{{\bf R}_{\ell}}e^{i{\bf R}_{\ell}\cdot{\bf p}}\tilde{c}_{{\bf R}_{\ell},\ell,\sigma}^{\dagger} \, ,
\label{eq:plane_wave_SL}
\end{equation}
where ${\bf p}\in\textrm{SLBZ}$, ``SLBZ'' stands for single-layer
Brillouin zone and $N_{SL}$ is the number of single-layer unit cells. Replacing Eq.$\,$\ref{eq:plane_wave_SL} in Eq.$\,$\ref{eq:Hubbard_hole}, we obtain 

\begin{equation}
H_{U}=\frac{U_H}{N_{SL}}\sum_{{\bf p},{\bf p}',{\bf q}\in\textrm{SLBZ}}a_{{\bf p}+{\bf q},\ell,\up}^{\dagger}a_{{\bf p},\ell,\up}a_{{\bf p}'-{\bf q},\ell,\down}^{\dagger}a_{{\bf p}',\ell,\down}=\frac{U_H}{N_{SL}}\sum_{{\bf q}\in\textrm{SLBZ},\ell}n_{{\bf q},\ell,\up}n_{-{\bf q},\ell,\down}
\end{equation}
where we defined the density operators $n_{{\bf q},\ell,\sigma}=\sum_{{\bf p}\in\textrm{SLBZ}}a_{{\bf p}+{\bf q},\ell,\sigma}^{\dagger}a_{{\bf p},\ell,\sigma}$.
Considering only momenta close to $\pm{\bf K}_{\ell}$, we choose
${\bf p}=\pm{\bf K}_{\ell}+\delta{\bf p}$ and ${\bf p}'=\pm{\bf K}_{\ell}+\delta{\bf p}'$,
to get

\begin{equation}
H_{U}=\frac{U_H}{N_{SL}}\sum_{\delta{\bf p},\delta{\bf p}',{\bf q}\in\textrm{SLBZ},\ell,\eta,\eta'}a_{\eta{\bf K}_{\ell}+\delta{\bf p}+{\bf q},\ell,\up}^{\dagger}a_{\eta{\bf K}_{\ell}+\delta{\bf p},\ell,\up}a_{\eta'{\bf K}_{\ell}+\delta{\bf p}'-{\bf q},\ell,\down}^{\dagger}a_{\eta'{\bf K}_{\ell}+\delta{\bf p}',\ell,\down}
\end{equation}

Due to the spin-valley locking in MoTe2, we assume that this term should only have a relevant
contribution when valley ${\bf K}_{\ell}$ ($-{\bf K}_{\ell}$) is
locked to spin $\up(\down)$. This fixes $\eta$ and $\eta'$. As a consequence, ${\bf q}$ should also be small so that the momenta
$\pm{\bf K}_{\ell}+\delta{\bf p}\pm{\bf q}$ is close to $\pm{\bf K}_{\ell}$.
We then get

\begin{equation}
H_{U}=\frac{U_H}{N_{SL}}\sum_{\delta{\bf p},\delta{\bf p}',{\bf q}\in\textrm{SLBZ},\ell}a_{{\bf K}_{\ell}+\delta{\bf p}+{\bf q},\ell,\up}^{\dagger}a_{{\bf K}_{\ell}+\delta{\bf p},\ell,\up}a_{-{\bf K}_{\ell}+\delta{\bf p}'-{\bf q},\ell,\down}^{\dagger}a_{-{\bf K}_{\ell}+\delta{\bf p}',\ell,\down} \, .
\label{eq:U_H_aOperators}
\end{equation}

We now define the coarse-grained real-space operators $c_{\eta,\ell,{\bf r}}^{\dagger}$ in terms of the plane-wave operators $a^\dagger_{\bf{p},\ell,\sigma}$ as 

\begin{equation}
a_{\eta{\bf K}_{l}+\delta{\bf k},\ell,\eta}^{\dagger}=\frac{1}{\sqrt{\Omega_{tot}}}\int d{\bf r}\,e^{i\delta{\bf k}\cdot{\bf r}}c_{\eta,\ell,{\bf r}}^{\dagger} \, ,
\end{equation}
where $\Omega_{tot}$ is the total system's area. Replacing in Eq.$\,$\ref{eq:U_H_aOperators}, we obtain 

\begin{equation}
\def\arraystretch{2.2}
\begin{array}{c}
H_{U}=\frac{U_H}{N_{SL}}\sum_{{\bf q},\delta{\bf p},\delta{\bf p}',\ell}a_{{\bf K}_{\ell}+\delta{\bf p}+{\bf q},\ell,\up}^{\dagger}a_{{\bf K}_{\ell}+\delta{\bf p},\ell,\up}a_{-{\bf K}_{\ell}+\delta{\bf p}'-{\bf q},\ell,\down}^{\dagger}a_{-{\bf K}_{\ell}+\delta{\bf p}',\ell,\down}\\
=\frac{U_H}{N_{SL}\Omega_{tot}^{2}}\int d{\bf r}\,d{\bf r}'\,d{\bf r}''\,d{\bf r}'''\sum_{{\bf q},\delta{\bf p},\delta{\bf p}',\ell}e^{i(\delta{\bf p}+{\bf q})\cdot{\bf r}}e^{-i\delta{\bf p}\cdot{\bf r}'}e^{i(\delta{\bf p}'-{\bf q})\cdot{\bf r}''}e^{-i\delta{\bf p}'\cdot{\bf r}'''}c_{\up,\ell,{\bf r}}^{\dagger}c_{\up,\ell,{\bf r}'}c_{\down,\ell,{\bf r}''}^{\dagger}c_{\down,\ell,{\bf r}'''}\\
=\frac{\Omega_{tot} U_H}{N_{SL}}\sum_{\ell}\int d{\bf r}\,c_{\up,\ell,{\bf r}}^{\dagger}c_{\up,\ell,{\bf r}}c_{\down,\ell,{\bf r}}^{\dagger}c_{\down,\ell,{\bf r}}=UA_{SL}\sum_{\ell}\int d{\bf r}\,c_{\up,\ell,{\bf r}}^{\dagger}c_{\down,\ell,{\bf r}}^{\dagger}c_{\down,\ell,{\bf r}}c_{\up,\ell,{\bf r}}
\end{array}\label{eq:H_U_intermediate}
\end{equation}
where $A_{SL}=\sqrt{3}a_0/2$ is the area of the monolayer unit cell
and $a_0=3.52A^{\circ}$ is the lattice constant of MoTe2. We now Fourier transform 
into the moiré plane-wave basis through

\begin{equation}
c_{\eta,\ell,{\bf r}}^{\dagger}=\frac{1}{\sqrt{\Omega_{tot}}}\sum_{{\bf k}\in\textrm{MBZ},{\bf Q}\in\mathcal{Q}_{\ell}^{\eta}}e^{i({\bf k}-{\bf Q}^{\eta\ell})\cdot{\bf r}}c_{\eta,{\bf k},{\bf Q}^{\eta\ell}}^{\dagger} \, ,
\label{eq:plane_wave_basis:HU}
\end{equation}
where ${\bf Q}^{\eta\ell}=\{{\bf G}_{M}+\eta(-)^{\ell}{\bf q}_{1}\}$,
with ${\bf G}_{M}$ moiré reciprocal lattice vectors, $(-)^{t}=1,(-)^{b}=-1$
and ${\bf q}_{1}={\bf K}_{b}-{\bf K}_{t}$.

Inserting Eq.$\,$\ref{eq:plane_wave_basis:HU} into Eq.$\,$\ref{eq:H_U_intermediate},
we get

\begin{equation}
\def\arraystretch{2}
\begin{array}{c}
H_{U}=U_H\frac{A_{SL}}{\Omega_{tot}^{2}}\sum_{\ell}\sum_{{\bf k}_{1},{\bf k}_{2},{\bf k}_{3},{\bf k}_{4}\in\textrm{MBZ}}\sum_{{\bf Q}_{1}^{\up\ell},{\bf Q}_{4}^{\up\ell}\in\mathcal{Q}_{\ell}^{\up}}\sum_{{\bf Q}_{2}^{\down\ell},{\bf Q}_{3}^{\down\ell}\in\mathcal{Q}_{\ell}^{\down}}\\
\int d{\bf r}\,e^{i({\bf k}_{1}+{\bf k}_{2}-{\bf k}_{3}-{\bf k}_{4}-{\bf Q}_{1}^{\up\ell}-{\bf Q}_{2}^{\down\ell}+{\bf Q}_{3}^{\down\ell}+{\bf Q}_{4}^{\up\ell})\cdot{\bf r}}c_{\up,{\bf k}_{1},{\bf Q}_{1}^{\up\ell}}^{\dagger}c_{\down,{\bf k}_{2},{\bf Q}_{2}^{\down\ell}}^{\dagger}c_{\down,{\bf k}_{3},{\bf Q}_{3}^{\down\ell}}c_{\up,{\bf k}_{4},{\bf Q}_{4}^{\up\ell}}\\
=U_H\frac{A_{SL}}{\Omega_{tot}}\sum_{{\bf k}_{1},{\bf k}_{2},{\bf k}_{3},{\bf k}_{4}\in\textrm{MBZ}}\sum_{\ell}\sum_{{\bf Q}_{1}^{\up\ell},{\bf Q}_{2}^{\down\ell},{\bf Q}_{3}^{\down\ell},{\bf Q}_{4}^{\up\ell}}\delta({\bf k}_{1}+{\bf k}_{2}-{\bf k}_{3}-{\bf k}_{4}-{\bf Q}_{1}^{\up\ell}-{\bf Q}_{2}^{\down\ell}+{\bf Q}_{3}^{\down\ell}+{\bf Q}_{4}^{\up\ell})\\
c_{\up,{\bf k}_{1},{\bf Q}_{1}^{\up\ell}}^{\dagger}c_{\down,{\bf k}_{2},{\bf Q}_{2}^{\down\ell}}^{\dagger}c_{\down,{\bf k}_{3},{\bf Q}_{3}^{\down\ell}}c_{\up,{\bf k}_{4},{\bf Q}_{4}^{\up\ell}}\\
\textrm{Taking }{\bf k}_{1}={\bf k}+{\bf q},{\bf k}_{2}={\bf k}'-{\bf q},{\bf k}_{3}={\bf k}',{\bf k}_{4}={\bf k},{\bf Q}_{1}^{\up\ell}={\bf Q}^{\up\ell}+{\bf G},{\bf Q}_{2}^{\down\ell}={\bf Q}'^{\down\ell}-{\bf {\bf G}},{\bf Q}_{3}^{\down\ell}={\bf Q}'^{\down\ell},{\bf Q}_{4}^{\up\ell}={\bf Q}^{\up\ell}:\\
=\frac{U_{\textrm{eff}}}{N_{M}}\sum_{\ell}\sum_{{\bf k},{\bf k}',{\bf q}\in\textrm{MBZ},{\bf Q}^{\up\ell},{\bf Q}'^{\down\ell},{\bf {\bf G}}}c_{\up,{\bf k}+{\bf q},{\bf Q}^{\up\ell}+{\bf {\bf G}}}^{\dagger}c_{\down,{\bf k}'-{\bf q},{\bf Q}'^{\down\ell}-{\bf {\bf G}}}^{\dagger}c_{\down,{\bf k}',{\bf Q}'^{\down\ell}}c_{\up,{\bf k},{\bf Q}^{\up\ell}}
\end{array}
\end{equation}
where ${\bf G}$ are moiré reciprocal lattice vectors, we defined
$U_{\textrm{eff}}=U_H A_{SL}/A_M$ and $N_{M}$ and $A_{M}$
are respectively the number of moiré unit cells and the area of a
moiré unit cell. We therefore conclude that the effective interaction
strength at the moiré scale is rescaled by $A_{SL}/A_{M}\sim a_0^{2}/a_{M}^{2}$.
Using $a_{M}=\frac{a_0}{2\sin(\theta/2)}$, we have that $a_{M}\approx55 \mathring{\textrm{A}}$
at $\theta=3.7^{\circ}$, which implies that $U_{\textrm{eff}}\approx U_H/244$.
Taking $U_H\sim10\,$eV, we obtain $U_{\textrm{eff}}\sim40\,$meV.

We finally project into the band basis, using $c_{\eta,{\bf k},{\bf Q}^{\eta\ell}}^{\dagger}=\sum_{m}[U_{\eta,m,{\bf k}}^{*}]_{{\bf Q}^{\eta\ell}}\gamma_{\eta,m,{\bf k}}^{\dagger}$, 
to obtain 

\begin{equation}
\def\arraystretch{2}
\begin{array}{c}
H_{U}=\frac{U_{\textrm{eff}}}{N_{M}}\sum_{m,m',n',n}\sum_{{\bf k},{\bf k}',{\bf q}\in\textrm{MBZ},{\bf Q}^{\up\ell},{\bf Q}'^{\down\ell},{\bf {\bf G}}}[U_{\up,m,{\bf k}+{\bf q}}^{*}]_{{\bf Q}^{\up\ell}+{\bf {\bf G}}}[U_{\down,m',{\bf k}'-{\bf q}}^{*}]_{{\bf Q}'^{\down\ell}-{\bf {\bf G}}}[U_{\down,n',{\bf k}'}]_{{\bf Q}'^{\down\ell}}[U_{\up,n,{\bf k}}]_{{\bf Q}^{\up\ell}}\\
\gamma_{\up,m,{\bf k}+{\bf q}}^{\dagger}\gamma_{\down,m',{\bf k}'-{\bf q}}^{\dagger}\gamma_{\down,n',{\bf k}'}\gamma_{\up,n,{\bf k}}\\
=\frac{U_{\textrm{eff}}}{N_{M}}\sum_{m,m',n',n}\sum_{{\bf k},{\bf k}',{\bf q}\in\textrm{MBZ},{\bf {\bf G}}}\sum_{\ell}M_{\up,mn}^{\ell}({\bf k},{\bf q+{\bf {\bf G}}})M_{\down,m'n'}^{\ell}({\bf k}',-{\bf q-{\bf {\bf G}}})\gamma_{\up,m,{\bf k}+{\bf q}}^{\dagger}\gamma_{\down,m',{\bf k}'-{\bf q}}^{\dagger}\gamma_{\down,n',{\bf k}'}\gamma_{\up,n,{\bf k}}\\
=\frac{U_{\textrm{eff}}}{N_{M}}\sum_{m,m',n',n}\sum_{{\bf k},{\bf k}',{\bf q}}\sum_{\ell}V_{\up\down,mm'n'n}^{\ell}({\bf k},{\bf k}',{\bf q})\gamma_{\up,m,{\bf k}+{\bf q}}^{\dagger}\gamma_{\down,m',{\bf k}'-{\bf q}}^{\dagger}\gamma_{\down,n',{\bf k}'}\gamma_{\up,n,{\bf k}}
\end{array}
\label{eq:Hubbard_ff}
\end{equation}
where we defined the layer-resolved form factors $M_{\eta,mn}^{\ell}({\bf k},{\bf q+{\bf {\bf G}}})=\sum_{{\bf Q}^{\eta\ell}}[U_{\eta,m,{\bf k}+{\bf q}}^{*}]_{{\bf Q}^{\eta\ell}+{\bf {\bf G}}}[U_{\eta,n,{\bf k}}]_{{\bf Q}^{\eta\ell}}$
and $V_{\up\down,mm'n'n}^{\ell}({\bf k},{\bf k}',{\bf q})=\sum_{{\bf {\bf G}}}M_{\up,mn}^{\ell}({\bf k},{\bf q+{\bf {\bf G}}})M_{\down,m'n'}^{\ell}({\bf k}',-{\bf q-{\bf {\bf G}}})$. 
 Note that the effective on-site Hubbard interaction is not expected to strongly affect the polarized state but rather provides a mechanism to stabilize it against other competing states. In particular, the effective Hubbard term is analogous to the $V_0$ Haldane pseudopotential, which has no effect for the intra-valley term in the 1BPV calculations, thus not influencing the spin-0 gaps. An on-site Hubbard interaction was also considered in Ref.$\,$\cite{FTI_Yves} for twisted MoTe2. We note however, that in contrast with the $V_0$ pseudo-potential, the Hubbard interaction here considered in Eq.$\,$\ref{eq:Hubbard_ff} is purely inter-valley, and thus does not contribute to the intra-valley interaction. 
 This implies that the Hubbard term here considered selectively increases the spinful gaps, while leaving the spinless gaps unaffected.

We point out that $U_{\textrm{eff}}$ will be treated as a tunable parameter in our calculations. However, it is not physically sensible to increase its value arbitrarily. Experiments constraint $U_{\mathrm{eff}}$, which must remain low enough such that it does not drive spin polarization at other fillings. A particular example is filling $\nu=-4/3$, where unpolarized states are observed experimentally \cite{Zeng2023,Cai2023,Redekop2024}. Ref.$\,$\cite{FTI_Yves} showed that band mixing is essential to obtain an unpolarized ground-state at  $\nu=-4/3$ and $\theta=3.7^{\degree}$. In the next section, we will study the polarization of the ground-state in this case for $U_{\textrm{eff}}\neq 0$, to inspect the range of values of $U_{\textrm{eff}}$ that is compatible with the experimental results.

\vspace{0.1cm}

\subsection{Ground-state polarization at $\nu=-4/3$}

\begin{figure}[h!]
\centering{}\includegraphics[width=0.4\columnwidth]{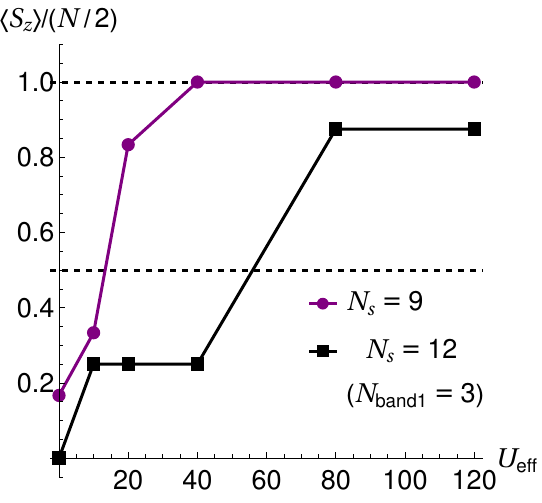}\caption{ Polarization at  $\nu=-4/3$ and $\theta=3.7^{\degree}$ as function of  $U_{\textrm{eff}}$ for system sizes with  $N_s=9$ and $N_s=12$ unit cells, respectively specified by  $(N_x,N_y,n_{11},n_{12},n_{21},n_{22})=(3,3,1,0,0,1)$ and $(N_x,N_y,n_{11},n_{12},n_{21},n_{22})=(6,2,1,1,0,1)$. 
For $N_s = 9$, fully unrestricted 2BPV results are shown. For $N_s = 12$, results are restricted to $N_{\textrm{band1}} = 3$, the maximal $N_{\textrm{band1}}$ for which calculations were performed down to the $S_z=0$ spin sector. The spin polarization is normalized to the maximal $S_z$ value given by $S_z^{\textrm{max}}=N/2$, where $N$ is the number of particles.  The dashed lines correspond to the situation where all particles are in one valley/spin ($\langle S_z \rangle/(N/2)=1 $), or filling 1 for the up spin and filling 1/3 for the down spin ($\langle S_z \rangle/(N/2)=1/2 $).\label{fig:polarization_4o3}}
\end{figure}

We start by studying the spin polarization of the ground state at filling $\nu = -4/3$ and twist angle $\theta = 3.7^{\degree}$ as a function of the on-site Hubbard interaction strength $U_{\textrm{eff}}$. Results for two system sizes, with $N_s = 9$ and $N_s = 12$, are summarized in Fig.~\ref{fig:polarization_4o3}. All calculations were constrained to Hilbert space dimensions below $2 \times 10^8$. For $N_s = 9$, fully unrestricted 2BPV calculations were feasible across all spin sectors. For $N_s = 12$, however, all spin sectors could only be accessed using the band maximum technique with $N_{\textrm{band1}} \leq 3$.

\begin{figure}[h!]
\centering{}\includegraphics[width=\columnwidth]{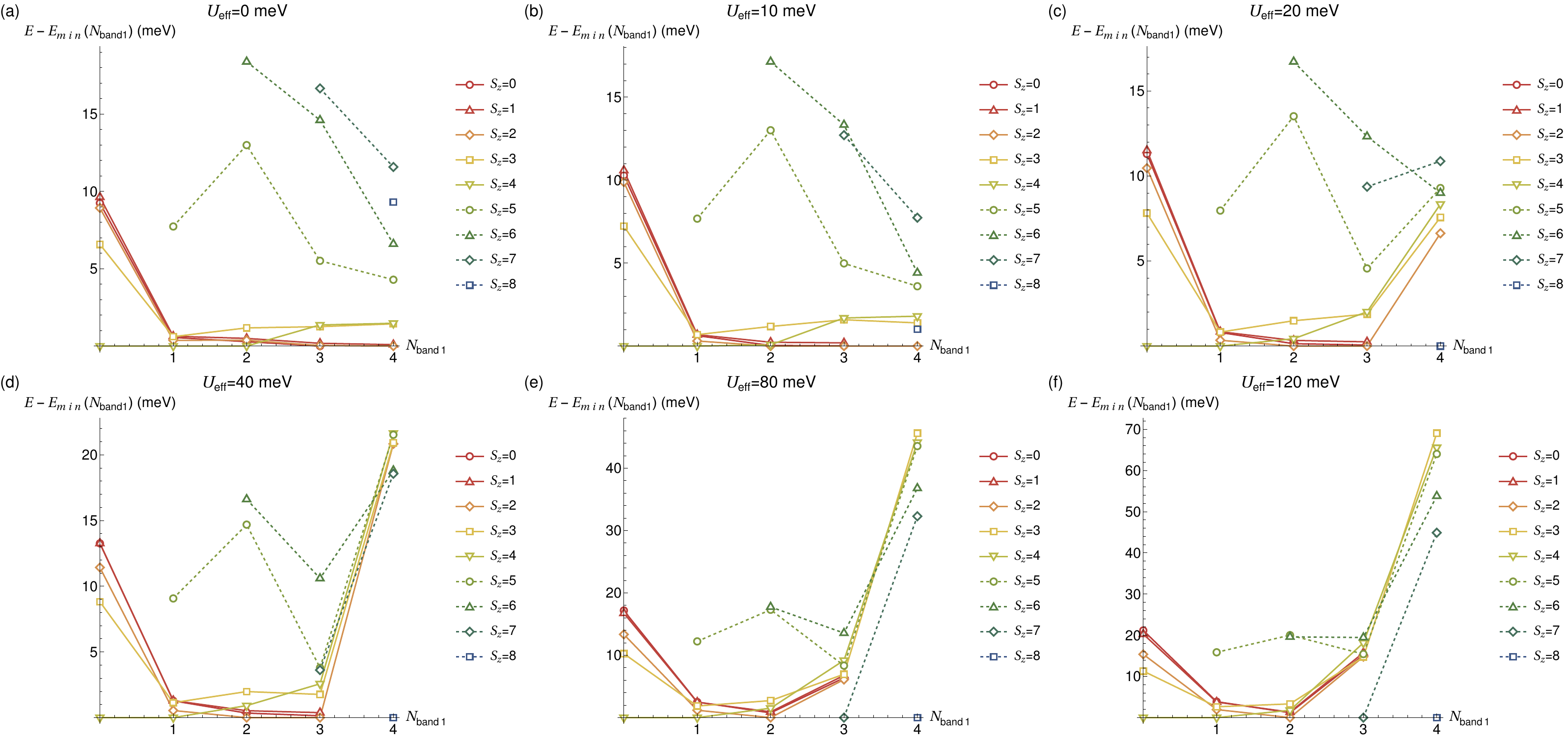}\caption{ 
Ground-state energies for different  spin sectors $S_z$ as a function of $N_{\textrm{band1}}$, at $\nu = -4/3$ and for the $N_s = 12$ system defined by $(N_x,N_y,n_{11},n_{12},n_{21},n_{22}) = (6,2,1,1,0,1)$. Data is plotted for different values of $U_{\textrm{eff}}$ in panels (a–f). $E_{\textrm{min}}(N_{\textrm{band1}})$ denotes the minimum energy among all the spin sectors for each $N_{\textrm{band1}}$. Dashed lines indicate spin sectors only accessible at  $N_{\textrm{band1}}>1$.
\label{fig:4o3_varUeff}}
\end{figure}

\begin{figure}[h!]
\centering{}\includegraphics[width=\columnwidth]{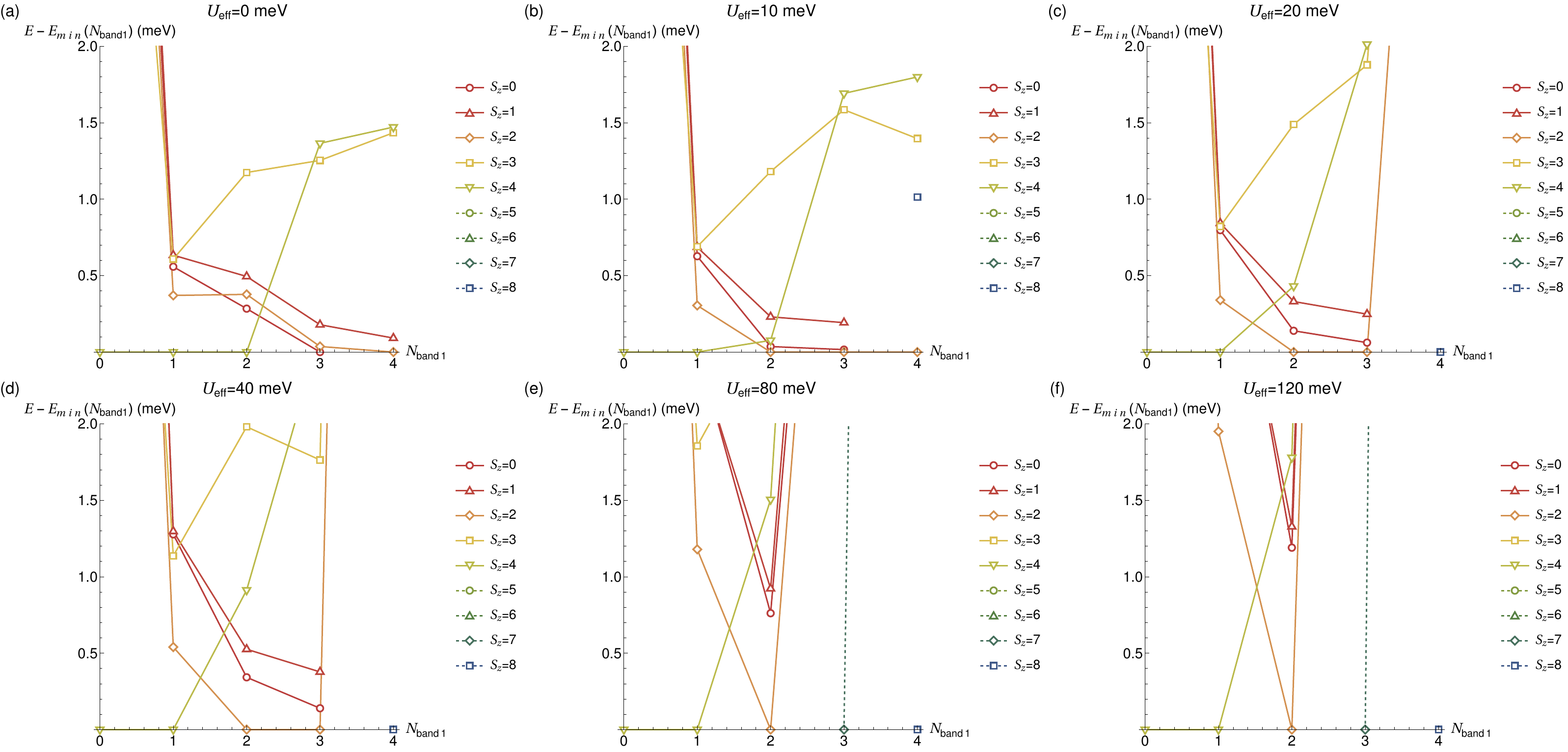}\caption{ Same data as Fig.~\ref{fig:4o3_varUeff}, plotted over a reduced energy window to better resolve the low-energy hierarchy between different spin sectors.
\label{fig:4o3_varUeff_zoom}}
\end{figure}

At $U_{\textrm{eff}} = 0$, the ground state becomes unpolarized for $N_s = 12$ and $N_{\textrm{band1}} = 3$, consistent with previous results in Ref.~\cite{FTI_Yves}. Compared to the smaller size with $N_s = 9$, the polarization decreases. For $ 10\, \textrm{meV}\leq U_{\textrm{eff}} \leq 40\,$meV, a small but finite spin polarization is observed for $N_s = 12$. However, the results are not converged with system size: within this range of $U_{\textrm{eff}}$, the polarization for $N_s = 12$ decreases compared to  $N_s = 9$. It is important to note, however, that  the fully polarized sector for $N_s = 12$ ($S_z = 8$) is only accessible for $N_{\textrm{band1}} \geq 4$, and thus we cannot definitively rule out full spin polarization of the ground state for $ 10\, \textrm{meV}\leq U_{\textrm{eff}} \leq 40\,$meV.
To clarify this point, Figs.~\ref{fig:4o3_varUeff} and~\ref{fig:4o3_varUeff_zoom} show the ground-state energy in each $S_z$ sector for the $N_s = 12$ system as a function of $N_{\textrm{band1}}$, up to $N_{\textrm{band1}} = 4$. For $N_{\textrm{band1}} = 4$, only spin sectors down to $S_z = 3$ were accessible, corresponding to Hilbert spaces dimensions up to $1.15 \times 10^8$. For $N_{\textrm{band1}} = 4$, the fully polarized state --- that is only accessible for $N_{\textrm{band1}} \geq 4$ --- has the lowest energy among the spin sectors $3 \leq S_z \leq 8$, for $U_{\textrm{eff}} \geq 10$ meV. However, the spin sectors with $S_z<3$ have lower energy than $S_z=3$ for $N_{\textrm{band1}}=3$ within this range of interaction strengths. A definite answer on the polarization of the ground-state would require accessing all the spin sectors 
for at least $N_{\textrm{band1}}=4$.

At larger interaction strengths, $U_{\textrm{eff}} \geq 80$ meV, spin polarization is more clearly established. The ground state is fully polarized for $N_s = 9$, and for $N_s = 12$, it resides in the maximally polarized sector accessible for $N_{\textrm{band1}} = 3$ ($S_z = 7$), as shown in Fig.~\ref{fig:polarization_4o3}. Moreover, for $N_{\textrm{band1}} = 4$, the energy of the fully polarized ground-state is more than $40$ meV lower than the ground-state energy for the  lowest accessible spin sector $S_z = 3$.

\clearpage

\subsection{Spinful gaps at $\nu=-2/3$ vs. Hubbard interaction}

In this section, we study the spinful gaps at finite $U_{\textrm{eff}}$. Most of the calculations we perform are for $U_{\mathrm{eff}} = 40$ meV, the largest interaction strength for which the results in the previous section are compatible with an unpolarized ground-state in the thermodynamic limit. This corresponds to $U_H = 10\,$ eV, which is also an upper bound on the atomic Hubbard interaction from prior estimates \cite{PhysRevB.105.195153,PhysRevMaterials.8.014409}.

\begin{figure}[h!]
\centering{}\includegraphics[width=0.55\columnwidth]{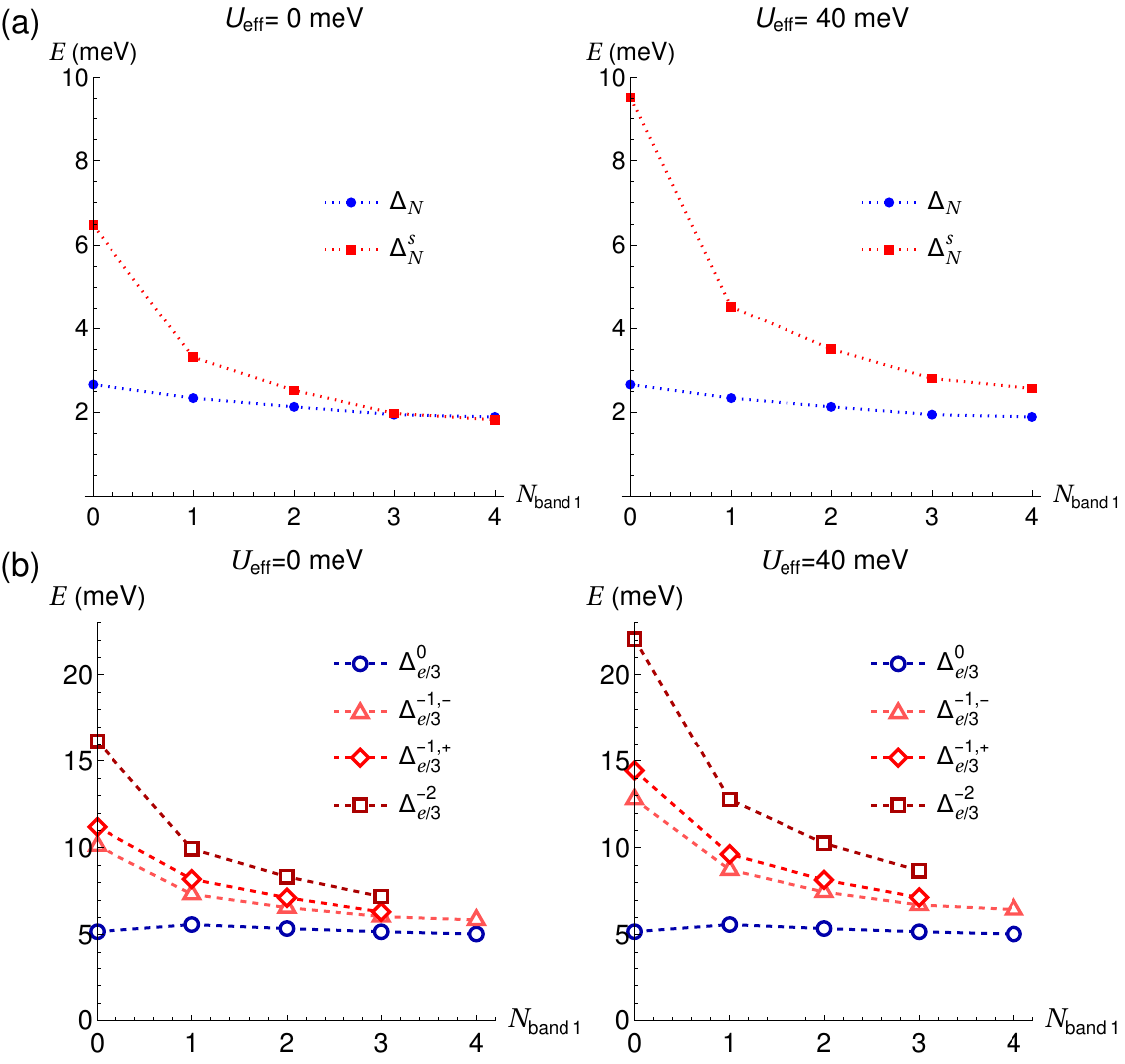}\caption{ Comparison of energy gaps without (left) and with (right) the Hubbard interaction in Eq.$\,$\ref{eq:Hubbard_ff}. Results for $\nu=-2/3, u=0.9$ and the 15 unit cell system specified by $(N_x,N_y,n_{11},n_{12},n_{21},n_{22})=(5,3,1,2,0,1)$. For the Hubbard interaction, we take $U_{\textrm{eff}}=40$meV. (a) Neutral gaps. (b) Charge-$e/3$ gaps. \label{fig:Hubbard_varU_15site}}
\end{figure}

Figure \ref{fig:Hubbard_varU_15site} compares the neutral and charge‑$e/3$ gaps computed with and without the on‑site Hubbard term at fixed system size ($N_s=15$), where all gaps converge in $N_{\rm band1}$ for the largest value of $N_{\rm band1}$ shown. Both the spinful neutral gap $\Delta_N^s$ and the spinful charge‑$e/3$ gap $\Delta_{e/3}^{-1,-}$ increase with $U_{\rm eff}$, while the spinless neutral and charge gaps remain unchanged—since $U_{\rm eff}$ does not contribute to the intra-valley interaction.  Crucially, band mixing substantially reduces these enhancements: at $U_{\rm eff}=40,$meV and taking the largest $N_{\rm band1}$, $\Delta_N^s$ grows by only 29$\%$, and $\Delta_{e/3}^{-1,-}$ by 9$\%$, relative to the 1BPV results. As we will show below, the percentage increase in $\Delta_N^s$ is largely insensitive to system size, whereas that in $\Delta_{e/3}^{-1,-}$ exhibits a much stronger size dependence.

\begin{figure}[h!]
\centering{}\includegraphics[width=0.7\columnwidth]{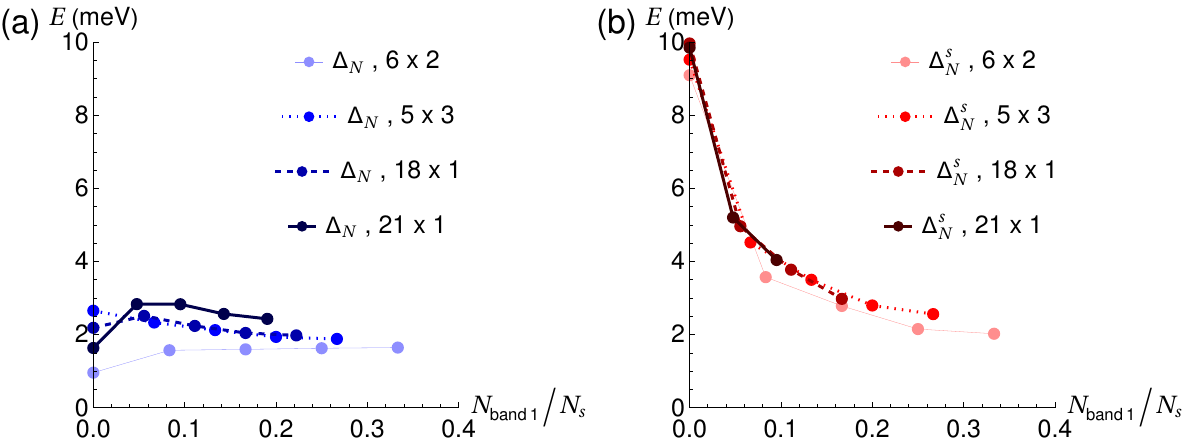}\caption{ Scaling collapse as a function of the density of holes in band 1 for the spin-0 (a) and spin-1 (b) neutral gaps, for $\nu=-2/3, u=0.9$ and including the additional Hubbard interaction in Eq.$\,$\ref{eq:Hubbard_ff} with  $U_{\textrm{eff}}=40$meV. All the momentum meshes used for these calculations are shown in Fig.$\,$\ref{fig:meshes_2BPV} (middle figures).\label{fig:Hubbard_neutral_collapse}}
\end{figure}

Figure \ref{fig:Hubbard_neutral_collapse} shows a scaling collapse of the neutral gaps versus the number of holes in band 1, in direct analogy to Fig. \ref{fig:gaps_bmixing_collapse_u0p9} for $U_{\mathrm{eff}}=0$. As before, the spin‑1 gap collapses cleanly for all $N_s>12$, reaffirming the robustness of the $N_s=15$ results in Fig. \ref{fig:Hubbard_varU_15site}(a). By contrast, the full neutral gap $\Delta_N$ fails to collapse across every system size. Most notably, the $N_s=21$ curve clearly deviates from the apparent trend established by the smaller lattices. Such finite-size effects were already present at $U_{\textrm{eff}}=0$meV, see Appendix \ref{sec:2BPV}.

\begin{figure}[h!]
\centering{}\includegraphics[width=\columnwidth]{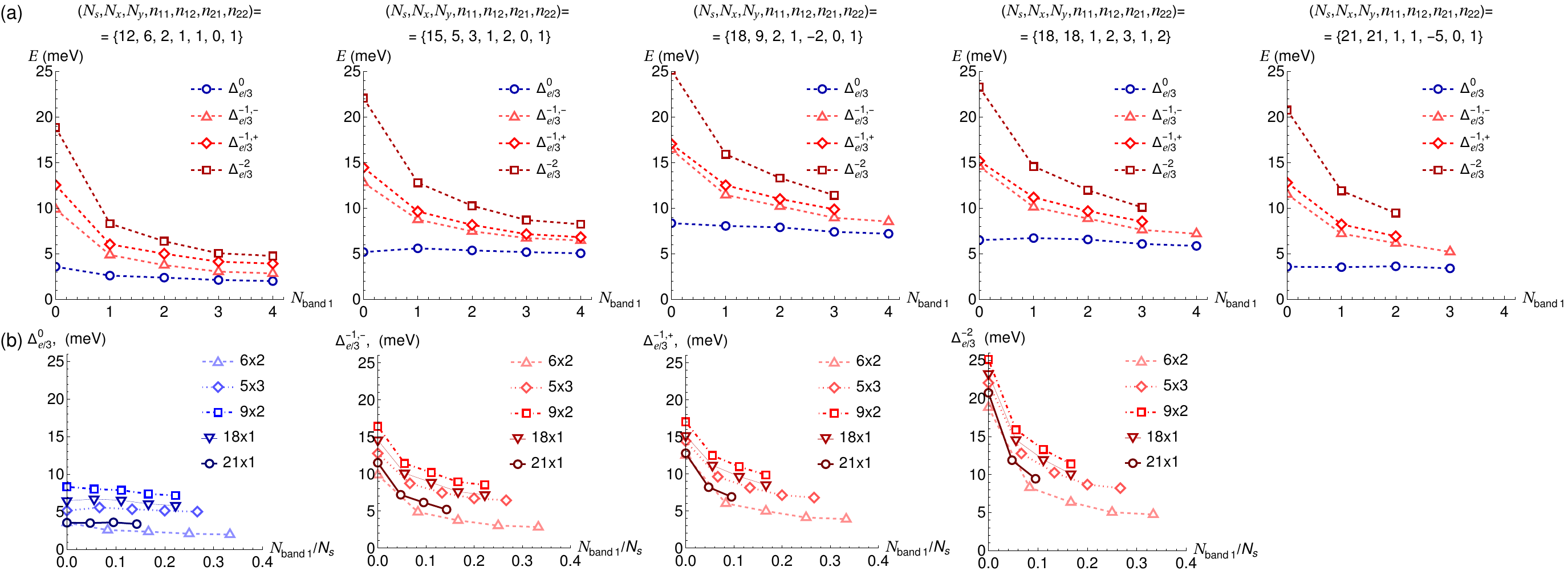}\caption{  Charge-$e/3$ gaps for $\nu=-2/3, u=0.9$ and the additional Hubbard interaction in Eq.$\,$\ref{eq:Hubbard_ff}, with  and  $U_{\textrm{eff}}=40$meV. (a) Gaps as a function of $N_{\textrm{band1}}$, where each figure contains data for a specified system size. (b) Gaps as a function of  $N_{\textrm{band1}}/N_s$, where each figure corresponds to a different gap type. All the momentum meshes used for these calculations are shown in Fig.$\,$\ref{fig:meshes_2BPV}. \label{fig:bare_gaps_Hubbard_All}}
\end{figure}

\begin{figure}[h!]
\centering{}\includegraphics[width=\columnwidth]{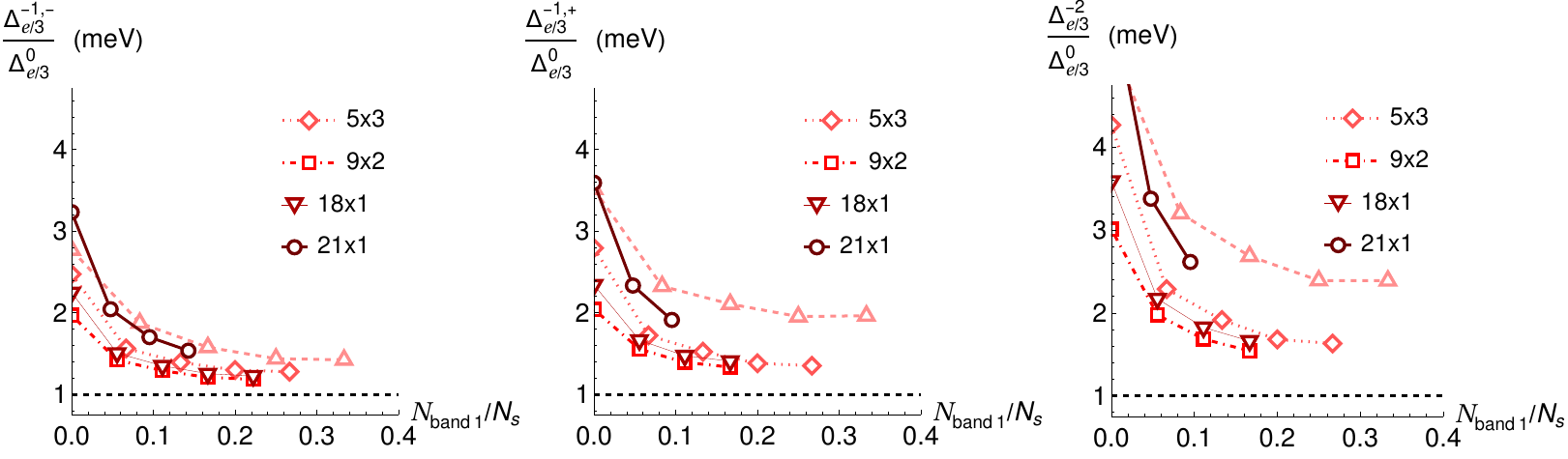}\caption{ Ratios between all the charge-$e/3$ spinful gaps and the charge-$e/3$ spin-0 gap, for $\nu=-2/3, u=0.9$ and the additional Hubbard interaction in Eq.$\,$\ref{eq:Hubbard_ff} with  $U_{\textrm{eff}}=40$meV.  All the momentum meshes used for these calculations are shown in Fig.$\,$\ref{fig:meshes_2BPV}. All the momentum meshes used for these calculations are shown in Fig.$\,$\ref{fig:meshes_2BPV}. \label{fig:ratios_Hubbard_All}}
\end{figure}

\begin{figure}[h!]
\centering{}\includegraphics[width=0.4\columnwidth]{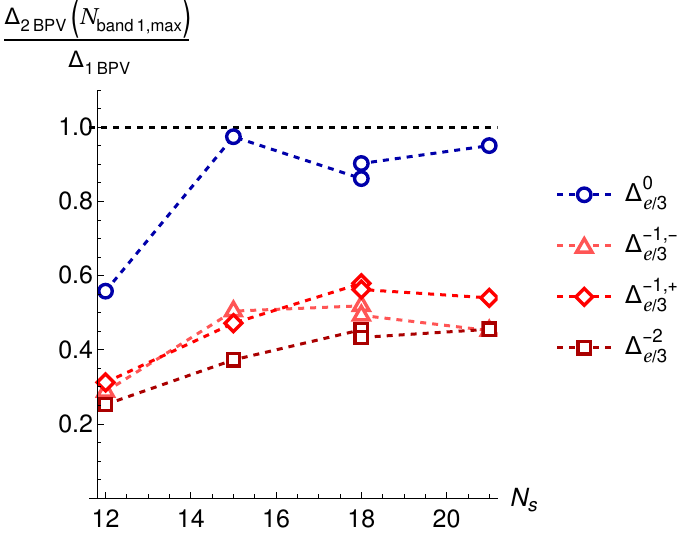}\caption{
Ratios between the gaps estimated from 2BPV and 1BPV calculations for the additional Hubbard interaction in Eq.$\,$\ref{eq:Hubbard_ff}, for $\nu=-2/3, u=0.9$ and  $U_{\textrm{eff}}=40$meV. For the 2BPV gaps, we used the calculation for the maximum computed $N_{\textrm{band}1}$ (see Fig.$\,$\ref{fig:bare_gaps_Hubbard_All}). All the momentum meshes used for these calculations are shown in Fig.$\,$\ref{fig:meshes_2BPV}.
 \label{fig:ratios_2BPV_1BPV_Hubbard}}
\end{figure}

We now report on the impact of the Hubbard interaction on the charge-$e/3$ gaps. Figure $\,$\ref{fig:bare_gaps_Hubbard_All} shows all the charge-$e/3$ energy gaps at fixed $U_{\textrm{eff}}=40$ meV, across different system sizes as a function of $N_{\textrm{band1}}$. The trends as a function of $N_{\textrm{band1}}$ closely mirror those for $U_{\textrm{eff}}=0$ meV case in Figs.$\,$\ref{fig:gaps_bmixing_all_u0p9}(b) and \ref{fig:gaps_bmixing_all_u0p9_collapse_bare}. We again observe that band mixing strongly suppresses the spinful gaps and leaves the spinless gaps almost unchanged. 
The ratios between all spinful charge gaps and $\Delta_{e/3}^0$ are shown in Fig.$\,$\ref{fig:ratios_Hubbard_All}. The value of the ratio is in better agreement between the different system sizes for the largest computed $N_{\textrm{band1}}/N_s$ than the value of the bare gaps. However, no clear scaling collapse is observed as in the $U_{\textrm{eff}}=0$ case (see Fig.$\,$\ref{fig:gaps_bmixing_all_u0p9_ratios_collapse}).  In Fig.$\,$\ref{fig:ratios_2BPV_1BPV_Hubbard} we also show the ratios between the gaps estimated from 2BPV (using the largest computed $N_{\textrm{band1}}$, see Fig.$\,$\ref{fig:bare_gaps_Hubbard_All}) and 1BPV calculations. This ratio is slightly reduced compared to the analogous calculation for $U_{\textrm{eff}}=0$ in Fig.$\,$\ref{fig:ratios_2BPV_1BPV}(a) - as a reference, for $N_s=21$, the ratio $\Delta_{e/3}^{-1,-}/\Delta_{e/3}^0$ only decreases by $10 \%$ compared to its value at $U_{\textrm{eff}}=0$. 

\begin{figure}[h!]
\centering{}\includegraphics[width=\columnwidth]{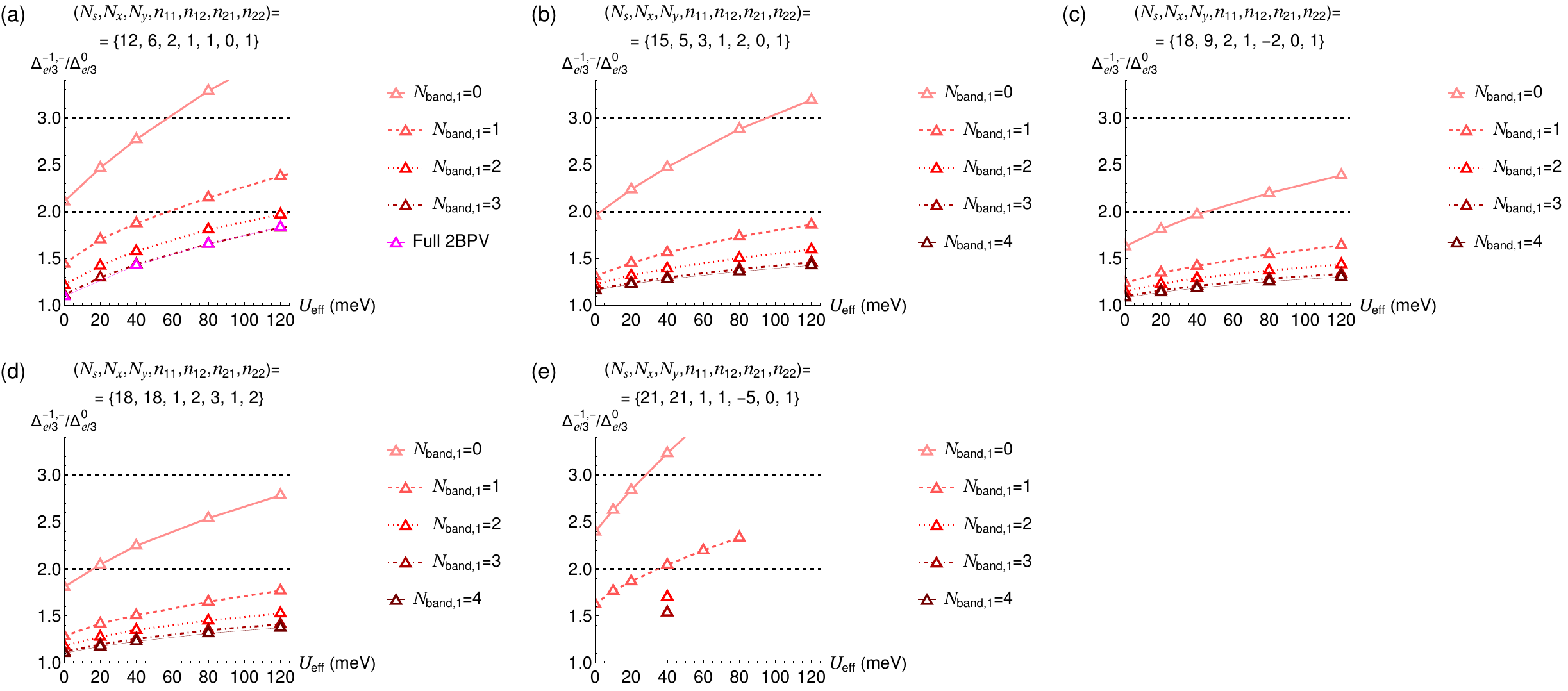}\caption{ Ratio between the charge-$e/3$ spin-1 and spin-0 gaps, $\Delta_{e/3}^{-1,-}/\Delta_{e/3}^0$, for $\nu=-2/3, u=0.9$  as a function of $U_{\textrm{eff}}$ (see Eq.$\,$\ref{eq:Hubbard_ff}), for different system sizes shown in the different panels (a-e).\label{fig:Hubbard_varU_AllSizes}}
\end{figure}

Finally, Fig.$\,$\ref{fig:Hubbard_varU_AllSizes} shows the evolution of the ratio $\Delta_{e/3}^{-1,-}/\Delta_{e/3}^0$ as a function of $U_{\textrm{eff}}$, for different system sizes and $N_{\textrm{band1}}$. Even though we treat $U_{\textrm{eff}}$ as a tunable parameter in this figure to get an idea of the trend, we reemphasize that our results in the previous section imply that reasonable values of $U_{\textrm{eff}}$ should not exceed $40\,$meV. We find that the ratio,  $\Delta_{e/3}^{-1,-}/\Delta_{e/3}^0$, increases consistently with increasing $U_{\textrm{eff}}$ for all $N_{\textrm{band1}}$. However, band mixing suppresses both the value of the gap ratio and its rate of growth with increasing $U_{\textrm{eff}}$. This trend highlights the competition between $U_{\textrm{eff}}$ and strong demagnetizing tendency induced by band mixing.

\clearpage

\end{document}